\documentclass[12pt,preprint]{aastex}

\shorttitle{A model of interstellar gaz}
\shortauthors{F. Le Petit et al.}

\begin{document}

\title{A model for atomic and molecular interstellar gas:\\
The Meudon PDR code}

\author{Franck Le Petit\altaffilmark{1}, Cyrine Nehm\'e, Jacques Le Bourlot and Evelyne Roueff}
\affil{LUTH, Observatoire de Paris et Universit\'e Paris 7, 5 place Jules
Janssen, F-92190 Meudon, France}

\altaffiltext{1}{Onsala Space Observatory, 439 92 Onsala, Sweden}

\begin{abstract}
We present the revised ``Meudon'' model of Photon Dominated Region (PDR code), presently available
on the web under the Gnu Public Licence at: \url{http://aristote.obspm.fr/MIS}.
General organisation of the code is described down to a level that
should allow most observers to use it as an interpretation tool with
minimal help from our part. Two grids of models, one for low excitation
diffuse clouds and one for dense highly illuminated clouds, are discussed,
and some new results on PDR modelisation highlighted.
\end{abstract}



\section{Introduction}

The recognition of the importance of photodissociation processes and
their relation to atomic to molecular transitions in interstellar clouds
started with \cite{Bates51}. However, the first numerical quantitative
model computing the various atomic and molecular abundances was
presented by \cite{BD77}. These authors focused their study on the
diffuse cloud toward the bright star $\zeta$ Oph described as a ``standard''
diffuse cloud for which extensive observational information was available,
including the first detection of molecular hydrogen in various rotational
levels derived from absorption transitions with the Copernicus satellite
\citep{Morton75}. Further impetus to study such environments has
been provided on the one hand by the detection of infrared emission
of molecular hydrogen, ionised and neutral carbon, neutral oxygen
and the so called aromatic bands at 3.3, 7.6 and 11 $\micron$.
The availability of the facility cryogenic grating spectrometer with 
the 91cm telescope of the Kuiper Airborne Observatory has allowed 
to detect for the first time, the far infrared emission lines of [OI] 
(63, 146 $\micron$), [CII] (158 $\micron$), [SiII] (35 $\micron$), 
and excited rotational lines of CO in luminous galaxies, planetary 
and reflection nebulae. These observations have emphasized the 
importance of Photon Dominated Regions and were interpreted in 
the context of theoretical models such as those of \cite{Tielens85a,Tielens85b}
 and \cite{LPRF93}. The ISO (Infrared Space Observatory)
 observed bright nebulae such as in Orion \citep{Abergel03}. 

These almost neutral astrophysical regions where ultraviolet photons
penetrate and affect the physical and chemical properties are described
now as Photon Dominated Regions (PDRs). These include hot gas close
to HII regions, diffuse and translucent clouds (Galactic and inter-galactic),
and the envelopes of dark clouds where star formation takes place.
They have been extensively described in the review article of \cite{HT99}.
Additional impulse came after the launch of the FUSE (Far Ultraviolet
Spectroscopic Explorer) satellite in 1999 which can detect $\rm{H}_{2}$
and HD in absorption in a variety of Galactic and extragalactic environments
\citep{Rachford02,Tumlinson02,Bluhm03,Boisse05,Lacour05}. The perspective of Spitzer and Herschel reinforces the need to describe as accurately as possible the physical conditions of photon-dominated environments and clarify the possible diagnostics. Various
groups have developed PDR codes at different stages of sophistication
and a workshop has been held in the Lorentz Center in Leiden in March
2004 to make detailed comparisons between the different numerical
codes. \cite{roellig05} describe the results obtained for different
test cases.

The present paper delineates the recent implementations performed
in our PDR model \citep{LPRF93}, labelled as the ``Meudon PDR
code'', which is available at the address \url{http://aristote.obspm.fr/MIS}.
The aim is to gather in a single paper the physical questions and
the numerical answers which have been implemented so far and to provide
to the observers a numerical tool to interpret their observations.
We also want to emphasise remaining issues to consider, such as the
inclusion of data and/or processes which are poorly known. The
organisation of the paper is the following: Sects.~\ref{Sec_strategy}
to \ref{Sec_chemistry} describe the physics and give technical details
on the organisation of the code (\ref{Sec_strategy}: general features,
\ref{Sec_grains}: grains, \ref{Sec_transfer}: radiative transfer,
\ref{Sec_excitation}: excitation, \ref{Sec_Ther_Bal}: thermal balance, 
\ref{Sec_chemistry}: chemistry). Some representative examples
are found in Sect.~\ref{Sec_diffuse} and \ref{Sec_Dense}, and Sect.~\ref{Sec_conclusion}
is our conclusion. Various appendices highlight more specific points.

\section{General features\label{Sec_strategy}}

\subsection{Overview}

The model considers a stationary plane-parallel
slab of gas and dust illuminated by an ultraviolet radiation field
(described in Appendix \ref{App_UVfield}) coming from one or both
sides of the cloud (the two intensities can be different). It solves,
in an iterative way, at each point in the cloud, the radiative transfer
in the UV taking into account the absorption in the continuum by dust
and in discrete transitions of ${\rm H}$ and $\rm{H}_{2}$.
Explicit treatment is performed for ${\rm C}$ and ${\rm S}$ photoionisation,
${\rm H}_{2}$ and ${\rm HD}$ photodissociation and ${\rm CO}$ (and
its isotopomeres) predissociating lines. The model also computes the
thermal balance taking into account heating processes such as the
photoelectric effect on dust, chemistry, cosmic rays, etc. and cooling
resulting from infrared and millimeter emission of the abundant ions,
atoms and/or molecules. Chemistry is solved, and abundances of each
species are computed at each point. The excitation state of a few important
species is then computed. The program is then able to calculate column
densities and emissivities/intensities.

One major restriction is the strict steady-state approximation, so
that model results cannot be compared directly to observations of
rapidly evolving regions. However, the time scales of photoprocesses
are modest at lower extinction and/or high radiation fields and shorter
than the typical two body chemical reaction time scale in diffuse
environments. The time scale given by the ${\rm H}{}_{2}$ photodissociation
is typically $1000/\chi$ years at the edge of a cloud (with $\chi$
the FUV radiation strength -- see Sect~\ref{sub_UV-Continuum}).
The steady state approximation is then satisfactory.

\subsection{Variables and parameters}

As modellers, we can define the \textit{parameters} which describe
the system and which can be tuned at our will. As a consequence, some
\textit{variables} will adjust themselves under those choices and
the constraints of physical laws.

Our first hypothesis is that each cell of gas is small enough for
all physical quantities to be constant in it, but large enough for
statistical mean to be meaningful. We can thus speak of quantities
such as {}``kinetic temperature'' ($T_{\rm K}$) as a function
of position. Note that this single very general hypothesis rules out
some interesting problems, such as the presence of shocks. 

The two most important physical quantities considered are density
and temperature. Here density is the total number of hydrogen nuclei
in all forms $n_{\rm H}=n({\rm H})+2\, n({\rm H}_{2})+n({\rm H}^{+})$
in ${\rm cm}^{-3}$. To keep enough flexibility, they may be either
parameters or variables depending on the user's choice. In the first
case, a complete cloud structure can be specified (by analytic expressions,
or in a {}``profile'' input file). Temperature becomes a variable
if thermal balance equations are solved, and density itself becomes
a variable if some kind of state equation is used. The most usual
cases are to solve for thermal balance and use either a constant density
or a constant pressure case. Other variables have a better defined
status and are described in Table~\ref{Tab_var}.

\placetable{Tab_var}

Model parameters are much more numerous. They fall into two main categories: 

Astrophysical parameters define the cloud characteristics and environment
and are displayed in Table~\ref{Tab_model}. They are tunable over
several orders of magnitude as will be discussed below.

\placetable{Tab_model}

(Micro)physical properties, on the other hand, should have well defined
values (e.g. chemical reaction rate coefficients) within the range
of considered variables but are in fact also parameters and displayed
in Table~\ref{Tab_cte}. Indeed, some properties suffer from different
uncertainties and offer an opportunity of choice to the modelist.
These uncertainties have to be remembered and taken into account to
assess the sensitivity of results.

\placetable{Tab_cte}

\subsection{Structure of the code}

Running a model consists of three distinct steps. First, choices have
to be decided on a set of parameters appropriate to the goal of
the computation. This is done by filling the input data files, which
allows handling very different conditions. It is generally meaningless
to try and use some published graph in the hope of interpreting some
specific observational data. A much better way consists in running
and adapting a model to its particular needs. Second, the physical structure
of the cloud is computed from the input file parameters and the result
is saved as a data base in a binary file. This is the computationally
intensive part. Finally a post-processor code allows the user to dig
into this binary file to extract quantities of interest and perform
the physical and chemical interpretation. It is only during this final
step that integration of local properties along the line of sight
is achieved and {}``observational'' quantities (such as, e.g., column
densities, line intensities) are computed.

During the second stage, and to maintain reasonable computation times,
many aspects of {}``real'' clouds have to be idealised. While making
changes to the code over the years, some approximations have been refined
and more accurate processes included, but some very basic assumptions
can not be overcome without writing a new code from scratch. The most
fundamental of these are that this a one-dimensional, plane parallel,
steady-state code:

\begin{itemize}
\item All quantities, either input parameters or computed variables depend
at most on one spatial coordinate (some are even constant but are probably varying in a true cloud,
as e.g. grain characteristics).
\item The spatial coordinate is taken as the dust optical depth in the visible
$\tau_{\rm{V}}$ computed perpendicular to a semi-infinite plane. 
Code design makes it nearly impossible to turn it e.g. to a spherical geometry.
\item In all physical equations, terms involving $\frac{\partial}{\partial t}$
are set to $0$. This implies that we compute the state of the cloud
after infinite time has passed. In practical, infinity means longer
than any characteristic time of the problem. This holds for all physical
processes (e.g. formation and destruction times of molecules), and
for the external environment of the cloud.
\end{itemize}
These conditions place limits on the kinds of objects that can be modelised.
For example, it is not possible to compute intensities from a strictly
edge on PDR since the line of sight in that direction is infinite.
It is also not very wise to compare emissivities to those from a very
young pre-stellar object, since time dependent effects are bound to
dominate.

The precise structure of the modelled cloud is described in Appendix~\ref{App_cloud_struct}.

Coupling of various physical processes (such as radiation transfer
and chemistry, thermal and chemical balance) requires using various
numerical methods during the computations. As the radiation field
decreases as the optical depth $\tau_{\rm{V}}$ increases, the
energy density inside the cloud is reduced accordingly. Therefore,
physical conditions at a given point depend more on parts of the cloud
closer to the nearest edge than on more shielded parts. This property
is used to compute physical quantities from one edge up to deeper
parts. However, it is not possible to reach a complete self consistent
solution in one step. First, if radiation is allowed to come from
both sides, the computation eventually reaches regions where the influence
of not yet computed parts of the cloud is far from negligible. Second,
when the gas opacity is taken into account in the UV radiative transfer,
it is mandatory to know in advance the level populations, which depend
on the radiation field to be computed. So the solution is reached
through an iterative process, where an approximate structure of the
whole cloud is saved at step $i$ and used to compute a refined one
at step $i+1$. Convergence properties are shown in Appendix~\ref{App_convergence}.

At each iteration, physical processes are treated in turn in the following
order:

\begin{enumerate}
\item Radiative transfer in the UV
\item Chemistry
\item Level populations and thermal balance
\end{enumerate}
As radiative transfer depends on the populations of the various species
and chemistry is sensitive to temperature, a second level of iteration
is needed at each point into the cloud. Chemical equations themselves
are solved using a Newton-Raphson scheme which is itself an iterative
method. Thus, there is a hierarchy of embedded iterative processes,
each of which has to proceed until a convergence criterium is reached. 

Finally, physical quantities are supposed to vary (almost) continuously
with $\tau_{\rm{V}}$, the visual optical depth. However, these
variations may be very steep (e.g. at the ${\rm {H}/{\rm H_{2}}}$
transition), and the optical depths where gradients of the abundances
are high, are not known in advance. Also, the intrinsically exponential
character of radiation attenuation with depth implies that very different
scales must be taken into account. We deal with those constrains by
using an adaptative spatial grid with logarithmic variations. The
first point is at $\tau_{\rm{V}}=0$ (with perfect vacuum behind
in place of ionised gas), and the second at $\tau_{\rm{V}}=10^{-8}$.
$\tau_{\rm{V}}$ grows from that value by factors of $10$ until
some variations of the main physical quantities are obtained. From
this point, the spatial step is kept short enough to ensure that relative
variations of all important quantities are kept below a predefined
threshold.

Although unphysical, discontinuities may appear at some particular
value of the optical depth $\tau_{\rm{V}}$. This comes from
the possible occurrence of multiple solutions to the steady state
equations for some set of parameters \citep{LPRS93}. As $\tau_{\rm{V}}$
increases, physical conditions vary smoothly until the point where
one of the solutions vanishes.
If the model result follows this branch of solution, it will {}``jump''
to the other branch with a discontinuity. In a {}``real'' cloud,
diffusion or turbulent mixing would smooth those discontinuities.
This point is further discussed in Sect.~\ref{Sec_chemistry}.

\section{Grain properties\label{Sec_grains}}

Grain properties are involved in, at least,  three important physical
aspects :

\begin{itemize}
\item They determine the cloud extinction curve, which is needed for UV
radiative transfer.
\item They may catalyse some chemical reactions, and are accountable for
all $\rm{H}_{2}$ formation in standard galactic conditions.
\item They take part in thermal balance through photo-electric heating,
and weak collisional coupling with the gas.
\end{itemize}
In an ideal model, these three contributions should derive from a
uniquely defined distribution. However, this is
impossible to achieve within the present knowledge of grain physical
properties and more empirical approaches are required. We have extracted
three sets of parameters which are mainly involved in the radiative
UV transfer and the thermal balance :

- the grain size distribution, which is a power law function, following
the analysis of \cite{MRN77} and referred as the MRN law: 

\[
dn(a)\propto a^{\alpha}da\]

where $dn(a)$ the number of spherical grains per unit of volume to
have a radius between $a$ and $a+da$ and $\alpha = -3.5$.

- the UV extinction curve as a function of the wavelength. A standard
galactic expression can be used as well as specific functions depending
on the environment. We use the parametrization of \cite{FM90}:\[
\frac{E_{\lambda-\rm{V}}}{E_{\rm{B-V}}}=c_{1}+c_{2}\, x+c_{3}\frac{x^{2}}{\left(x^{2}-y_{0}^{2}\right)^{2}+x^{2}\gamma^{2}}+c_{4}\, F(x)\]
\[
\rm{with}\qquad F(x)=0.5392\,(x-5.9)^{2}+0.05644(x-5.9)^{3}\]
if $x>5.9\,\mu{\rm m}^{-1}$ and $0$ elsewhere, with $x=1/\lambda$
in $\mu{\rm m^{-1}}$.

The six parameters $c_1$,$ c_2$, $c_3$, $c_4$, $\gamma$ and $y_0$ are taken to be mainly a function of $R_{\rm{V}} = A_{\rm{V}}/E(\rm{B}-\rm{V})$.

- grain absorption cross sections taken from \cite{DL84} and \cite{LD93}.
These values are determined from experimental analysis of extinction
properties of spherical particles of graphite and silicates for a sample
of radii between $1\,{\rm nm}$ and $10\,\mu{\rm m}$. Values
corresponding to radii smaller than 50 nm are considered to mimic
PAH (Poly Aromatic Hydrocarbons) properties.

Currently, graphite and silicates are mixed in constant proportion
within the code. Additional grain properties such as the dust to gas
mass ratio, the volumic mass and the mean distance between adsorption sites
on a grain etc. which affect the reactivity on dust, are displayed in
Table~\ref{Tab_grains}.

\placetable{Tab_grains}

The physical effects resulting from the gas-grain interaction have
been described in \cite{LPRF95} and the model follows the corresponding
prescription.

The present choice of parameters differs somewhat from other conventions
found in other PDR models. It is indeed often assumed that a single
parameter, the effective UV continuum absorption cross section
(\textbf{$\sigma=\pi a^{2}\,Q_{\rm ext}(100nm)\, n_{\rm g}/n_{\rm H}$})
at 100 nm allows to take care of the continuum UV flux attenuation.
\cite{SD89} introduced an effective grain absorption cross section of $1.9\,\times10^{-21}\; \rm{cm}^{2}$,
supposed to hold in the far-ultraviolet spectrum where ${\rm H}_2$ absorption occurs. It is interesting to note
that, with our choice of parameters, the mean geometrical cross section
of grains is equal to $1.1\,\times\,10^{-21}\; \rm{cm}^{2}$ with
the Mathis size distribution of grains and $5.8\,\times\,10^{-22}\; \rm{cm}^{2}$
with a unique value of the radius of $0.1\; \mu{\rm m}$.

\section{U.V. Radiative transfer\label{Sec_transfer}}

\subsection{Strategy}

The resolution of the complete radiative transfer equation is an exceedingly
time consuming computation because :

\begin{itemize}
\item The whole spectral range from the far ultraviolet (FUV) to millimetre
wavelength needs to be considered, implying a variety of processes
to consider
\item The spectral resolution in discrete lines requires a high number of
grid points (even using a variable grid size)
\item radiative processes are coupled to chemical and thermal processes
in a non-linear way.
\end{itemize}
Therefore the problem has to be split, and various approximations
introduced to render it tractable. Our main assumptions are the following:

\begin{itemize}
\item Incoming UV radiation is decoupled from outgoing IR and millimetre
radiation.
\item There is no internal UV source function, so that transfer at UV wavelengths
is scattering or pure absorption followed by emission. Photodissociation
and photoionisation resulting from secondary UV photons generated
by cosmic rays on the molecular gas are not explicitly calculated.
Scaling factors to the cosmic ray ionisation are introduced for each
species with the appropriate dissociative or ionised channel as computed
by \cite{GLDH89}.
\item Redistribution of radiation effects are neglected except for the anisotropic
UV scattering by grains. In particular, cooling lines are treated
within the {}``on-the-spot'' approximation: Photons are either re-absorbed
where they are emitted or escape from the cloud. Therefore, level
populations can be computed from detailed balance using only local
quantities.
\item During the first stage of the computation (\emph{building the data
base}) the source function in various millimeter or submillimeter
cooling transitions is obtained with an escape probability approximation
\citep{dJBD80} for computing the line opacity; however, no approximation 
is done in the integration over optical depth at line center for each line.
\item During the post-processing phase, when temperature, abundances and
populations are determined at each point inside the cloud, the full
transfer equation can be solved along a line of sight to obtain line
profiles and integrated intensities (see Fig.~(\ref{Fig_struct})).
\item We do not solve the thermal balance of the grains resulting from the
IR continuum emission of the dust, nor the emission from PAH.
\end{itemize}

\subsection{UV Continuum\label{sub_UV-Continuum}}

The external UV radiation field is described in Appendix~\ref{App_UVfield}.
By default, the  \cite{Draine78} radiation field is used and is scaled by
a multiplicative factor $\chi$ given in the entry file. Its attenuation
into the cloud is computed by solving the radiative transfer equation
on a variable grid of wavelengths using the spherical harmonics method
\citep{FRR80,Roberge83}.

This equation reads:\[
\mu\,\frac{\partial I(r,\mu)}{\partial s}=-(\kappa_{\lambda}+\sigma_{\lambda})\, I(r,\mu)+\frac{\sigma_{\lambda}}{2}\int_{-1}^{+1}p(\mu,\mu')I(r,\mu')\, d\mu'\]
where $\mu = cos \theta$ with $\theta$ the angle
between the direction perpendicular to the slab of gas and the direction
of propagation of the beam of light (see Fig.~(\ref{Fig_struct})),
$s$ is the curvilinear abscissa in the direction perpendicular to
the slab of gas. $I(r,\mu)$ is the specific intensity in ${\rm erg}\,{\rm s}^{-1}\,{\rm cm}^{-2}\,{\rm sr}^{-1}\,{\rm \mathring{A}^{-1}}$
 at the position $r$ in the direction given by
$\mu$ where $\kappa_{\lambda}$ is the total absorption coefficient
(gas plus dust) in $\rm{cm}^{-1}$, and $\sigma_{\lambda}$
the dust scattering coefficient ($\rm{cm}^{-1}$). $p(\mu,\mu')$
is the angular redistribution function by dust of the radiation field.
Gas extinction in selected lines of ${\rm {H}}$ and ${\rm H_{2}}$
may be included by using as an independent variable the true opacity:\[
d\tau_{\lambda}=-ds\,(\kappa_{\lambda}^{G}+\kappa_{\lambda}^{D}+\sigma_{\lambda})\]
where $\kappa_{\lambda}^{D}$ is the dust extinction and $\kappa_{\lambda}^{G}$
is the gas absorption coefficient, all in ${\rm cm^{-1}}$ and at
the wavelength $\lambda$.

The contribution from dust is easily computed from (see also Appendix~\ref{App_cloud_struct})\[
d\tau_{\lambda}^{D}=d\tau_{\rm V}^{D}\,\left(1+\frac{1}{R_{\rm V}}\,\frac{E_{\lambda-{\rm V}}}{E_{\rm B-V}}\right)\]
and%
\[ds=1.086\,\frac{C_{\rm D}}{R_{\rm V}}\,\frac{1}{n_{\rm H}}\, d\tau_{\rm V}\]

Extinction curves $\left(\frac{E_{\lambda-\rm{V}}}{E_{\rm{B}-\rm{V}}}\right)$
can be selected on a per object basis from an easily expandable data
base, using the analytical fits of \cite{FM86,FM88,FM90}. Many authors
have given the extinction curves using this same set of 6 parameters.
The $R_{\rm{V}}$ factor can be determined observationally \citep{Cardelli94,PMPB01,PMPB03}
and has a mean value close to 3.1. If unknown, in diffuse gas, the standard mean Galactic
parameters reported in \cite{FM90} are used.

\subsection{UV discrete absorption}

Discrete absorption in the UV also occurs and is mainly due to
atomic ${\rm {H}}$ Lyman lines and molecular ${\rm H_{2}}$ electronic
transitions within the Lyman and Werner system bands. Photodissociation
of ${\rm H_{2}}$ takes place via these discrete transitions coming
from specific rotational levels (J=0, 1, etc.). These lines may become
very wide due to saturation effects. The set of transitions to introduce
in the radiative transfer can be chosen from none to any at the user's
choice and is determined from the value of the rotational quantum
number of the lower level of the UV transition. A huge computation
time can result if many lines are included. Adding more and more lines 
does not lead to a simple scaling between the number of lines and the computing time.
CPU time is proportional to the  number of points in the wavelength grid 
and line overlap tends to  limit its growth as the number of lines increases.

For the lines taken into account in the radiative transfer the method
is the following. For a given absorption line from lower level $l$
to upper level $u$, the contribution to opacity at wavelength $\lambda$
is computed from\[
\kappa_{l\rightarrow u}^{G}=\frac{\sqrt{\pi}e^{2}}{mc\Delta\nu_{\rm D}}\, n_{l}\, f_{lu}\, H(a_{lu},\xi)\]
where $\Delta\nu_{\rm D}=\frac{\nu_{0}}{c}\sqrt{\frac{2kT_{\rm K}}{m}+v_{\rm turb}^{2}}$
is the line Doppler width, $T_{\rm K}$ the gas kinetic temperature,
$v_{\rm turb}$ the turbulent velocity, $n_{l}$ the density of
the lower level population, $f_{lu}$ the oscillator strength of the
transition $l\rightarrow u$.

The Voigt profile is defined by \[
H(a,\xi)=\frac{a}{\pi}\int_{-\infty}^{+\infty}\frac{e^{-y^{2}}dy}{a^{2}+(\xi-y)^{2}}\]
where $a=\frac{\Gamma}{4\pi\Delta\nu_{\rm D}}$, $\Gamma$ is the
upper level inverse life time and $\xi=\frac{\nu-\nu_{0}}{\Delta\nu_{\rm D}}$.
The Voigt function is computed using \cite{Wells99}%
\footnote{Source code and corrections available at \url{http://www.atm.ox.ac.uk/user/wells/voigt.html}%
}. Various tests have shown that it is faster to compute the contribution
of all lines at all wavelengths than to try and devise
position and wavelength dependent tests to include
only significant contributions.

The transfer equation then becomes:\begin{equation}
\mu\frac{\partial I(\tau,\mu)}{\partial\tau}=I(\tau,\mu)-S(\tau,\mu)\label{mathed:Eq_transfer}\end{equation}
with:\begin{equation}
S(\tau,\mu)=\frac{\omega(\tau)}{2}\int_{-1}^{+1}p(\mu,\mu')I(\tau,\mu')\, d\mu\label{mathed:Eq_redistrib}\end{equation}

Differences with the method of \cite{Roberge83} come from the dependence
of the effective albedo $\omega$ with $\tau$. This dependence results
from the contribution of the gas to the UV absorption. Details are
given in Appendix~\ref{App_sph_harm}.

We use a variable grid in wavelengths to maintain accuracy while limiting
memory and CPU requirements. The grid is built during the initialisation
phase in two stages:

\begin{itemize}
\item first, all {}``required'' wavelengths are listed. They include various
values, either physical (ionisation thresholds, line positions, etc.)
or imposed by input files (grids from external files, etc...).
\item Then the grid is completed by choosing points in the profiles of each
line (if any) using increasing steps from each centre of line. A test
integral is evaluated, and the grid refined until adequate convergence
is reached.
\end{itemize}
Various numerical tools have been designed to manipulate the radiation
field, allowing dynamical memory allocation. All parameters are tunable,
but not included in the user input file: this part is considered too technical for
promotion to general {}``user space''.

\begin{figure}
\epsscale{0.65}
\includegraphics[angle=270,scale=0.65]{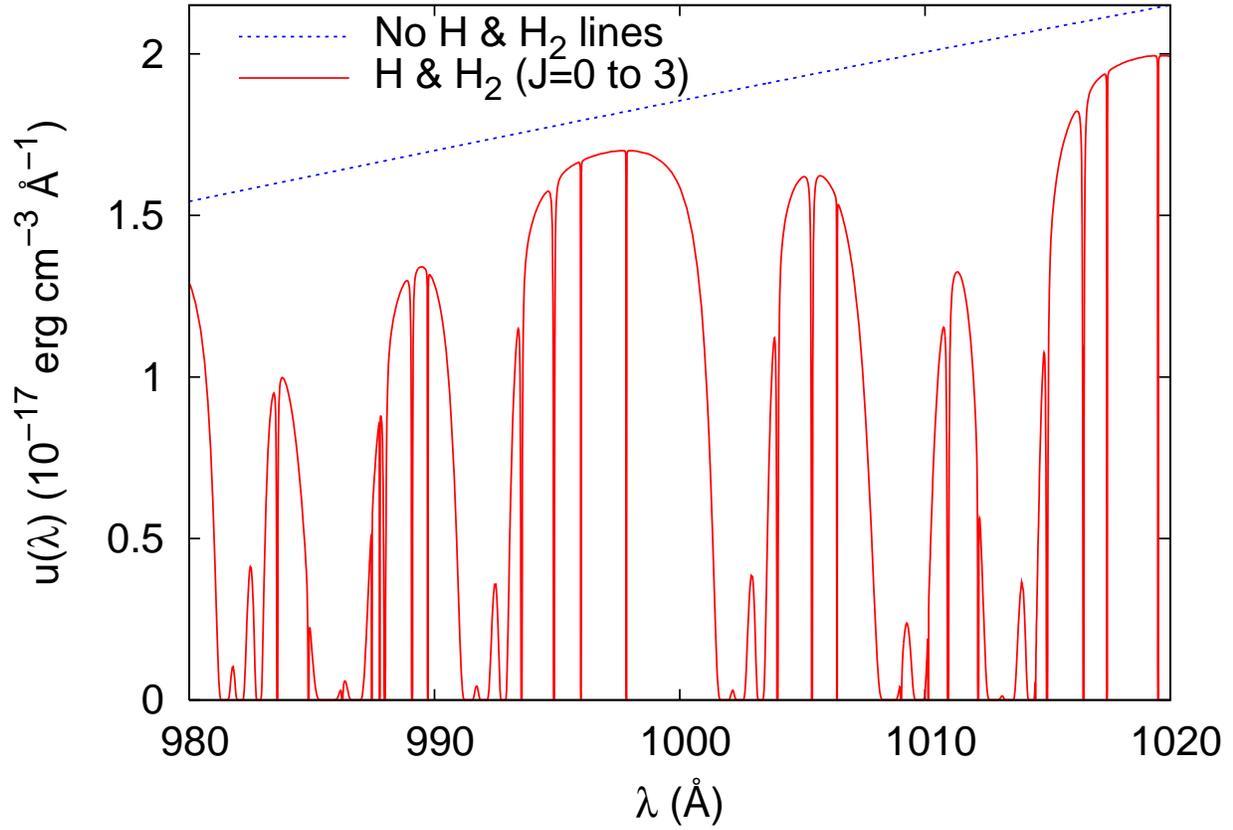}
\caption{Radiative energy density inside a $0.5\,{\rm {mag}}$ clump, $n_{\rm{H}}=100\,{\rm cm^{-3}}$,
irradiated by the standard ISRF on both sides. \label{Fig_RadUV}}
\end{figure}

As an example, a portion of the radiation field inside a small clump
($A_{\rm{V}}=0.5$ mag) of density 100 $\rm{cm}^{-3}$ and
irradiated by the standard ISRF on both faces is presented in Figure (\ref{Fig_RadUV}).
Note that line transfer is a CPU time consuming process. Including
no lines in the radiative transfer, only ${\rm H_{2}}$ lines from
J=0 and 1, or ${\rm H_{2}}$ lines from J=0 to J=3, CPU times are
in a ratio $1:10:50$.

\placefigure{Fig_RadUV}

\subsection{Other lines\label{sec_FGK}}

The ${\rm H_{2}}$ lines above the threshold mentioned in the previous
paragraph and the ${\rm {CO}}$ predissociating lines (as well as
the ones of their isotopes since ${\rm {D}}$, $^{13}{\rm {C}}$ or
$^{18}{\rm {O}}$ are introduced in the model) are not included in
the complete radiative transfer described above. We compute self-shielding
effects by using the approximation of \cite{FGK79} (hereafter FGK
approximation). It gives a fair approximation of the optical depth
at the centre of the line and of the resulting self-shielding from
a knowledge of the abundance and temperature along the line of sight,
known from a previous iteration. Note that absorption is computed
using the local ($\tau_{\rm{V}}$ dependent) radiation field,
which includes absorption by ${\rm H_{2}}$ strong lines. Protection
of ${\rm {CO}}$ and ${\rm {HD}}$ by ${\rm H_{2}}$ is then taken
into account.

\subsection{UV escape probability}

From \cite{FRR80} we have  an elegant mean to compute the escape
probability of UV photons produced within the cloud, e.g. within ${\rm H_{2}}$
cascades. Those photons are emitted isotropically so that the fraction
which reaches one side of the cloud is a transfer problem
similar to the one described in the previous section. The cloud is split into
two parts at the position $\bar{\tau}$ of the emitting molecule.
Boundary conditions (i.e. impinging specific intensities) are set
to $1$ at $\bar{\tau}$ and to $0$ at $\tau=0$ and at $\tau=\tau_{max}$
(where $\tau_{max}$ is the maximal optical depth of the cloud), and
Eq~(\ref{mathed:boundary_cond}) is solved for the two half-clouds.
The probability for a photon emitted at $\bar{\tau}$ to reach $0$
(respectively $\tau_{max}$) is then the ratio $J(0)/J(\bar{\tau})$
(respectively $J(\tau_{max})/J(\bar{\tau})$) from that sub-system.
Those escape probabilities are computed once per global iteration,
thus reducing  the CPU requirements.

The main drawback of this method is the {}``on the spot''
approximation: photons that do not escape the cloud are assumed to
be absorbed where they were emitted. This underestimates the energy
loss from that point and changes slightly the level populations. However,
lifting that approximation would require a full treatment of radiative
transfer (including the line source function) which is out of the
scope of our model.

\section{Excitation\label{Sec_excitation}}

Level populations are computed for a number of species (see Table~\ref{Tab_Coll}).
Detailed balance is solved explicitly including collisional transitions,
radiative decay, interaction with the cosmological
background, and state to state formation and destruction processes.

\placetable{Tab_Coll}

We follow \cite{dJBD80} to evaluate a resonant photon escape probability.
Solving the non-linear coupled steady state equations provides both
level populations and cooling rates through radiative decay. Dust
extinction is neglected for infrared and millimeter lines.

Note that for one-sided models, the escape probability towards the
interior of the cloud is strictly zero. As we use the {}``on the
spot'' approximation, this results in a modification of the local
source function that may modify significantly the populations. Thus,
it is always better to compute two-sided models if an estimation of
the real size is known, even if the radiation field is very different
on both sides. See Sect.~\ref{sub_Fine-structure-transitions} for
a quantitative example.

\begin{figure}
\epsscale{0.65}
\includegraphics[angle=270,scale=0.65]{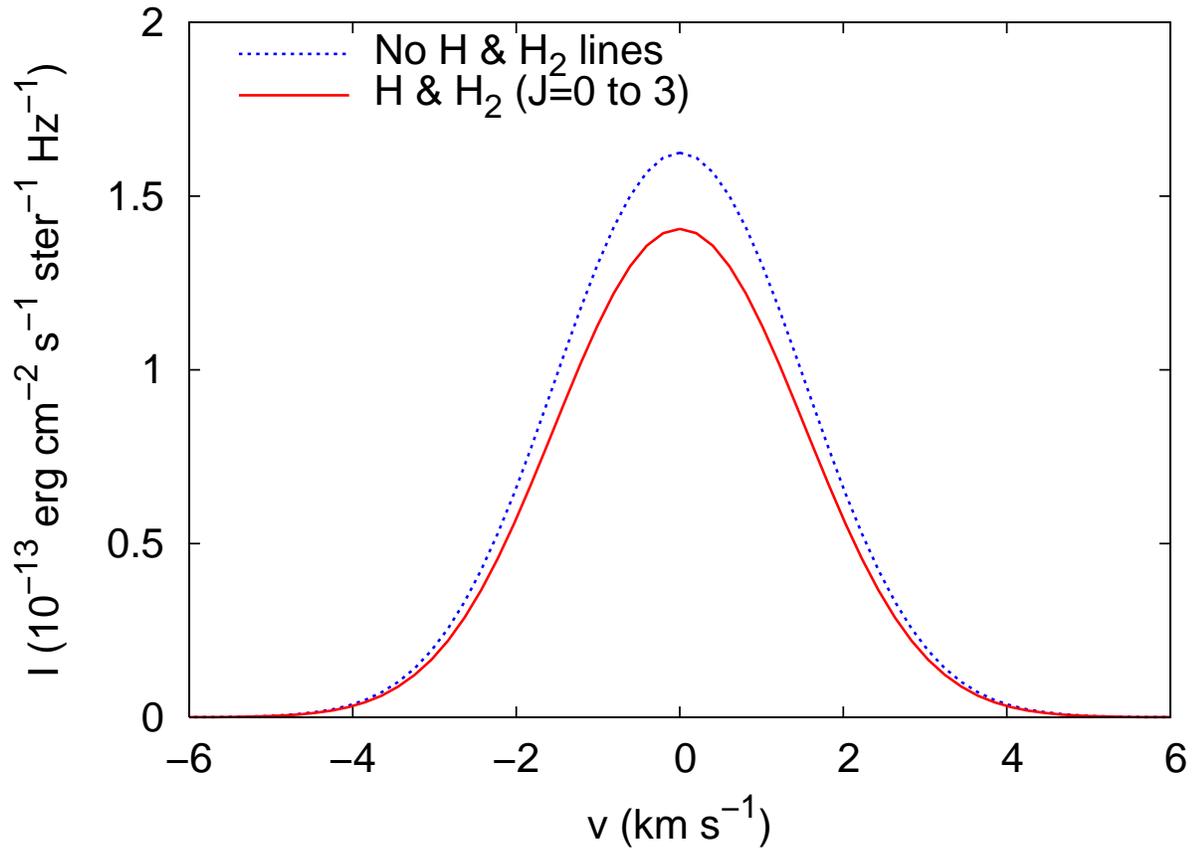}
\caption{${\rm C^{+}}$ emission line, for the two cases of Fig~(\ref{Fig_RadUV}).\label{Fig_C+line}}
\end{figure}

After all iterations have converged, the whole cloud state is known,
and the radiative transfer equation can be solved to compute a line
emissivity in a given direction and the line profile. Fig.~(\ref{Fig_C+line})
shows the profile of ${\rm C^{+}}$ fine structure line at $158\,{\rm {\mu m}}$
for the same cloud as in Fig.~(\ref{Fig_RadUV}). The profile is essentially
Gaussian, dominated by turbulent velocity (here $2\,{\rm {km}\,{\rm s^{-1}}}$),
and with very little saturation (optical depth at line centre: $0.24$).
The effect of radiation attenuation by ${\rm H_{2}}$ lines is barely
detectable here: the integrated line intensity decreases from $3.8\,10^{-6}$
to $3.3\,10^{-6}$ ${\rm {erg}\,{\rm cm^{-2}\,{\rm s^{-1}\,{\rm sr^{-1}}}}}$.
This results from the reduced ionisation of ${\rm {C}}$ inside the
cloud due to shielding by ${\rm H_{2}}$ lines (${\rm {C}}$ column
density varies from $2.1\,10^{14}$ to $2.9\,10^{14}$ ${\rm cm^{-2}}$).

\placefigure{Fig_C+line}

\section{Thermal Balance\label{Sec_Ther_Bal}}

\subsection{Heating}

Heating results from impinging UV photons and cosmic rays, but several
different processes are involved to convert either directly or indirectly
the energy input into kinetic energy.

\subsubsection{Photo-electric effect on grains}

We follow the development of \cite{BT94}, but add some significant
upgrades. Grain absorption coefficients are taken from \cite{LD93}.
Integration over radiation field is performed using the local field
as computed from the radiative transfer (see Sect.~\ref{Sec_transfer}). The charge
distribution is computed for each grain size. Finally, integration
over size is performed.

\subsubsection{${\rm H_{2}}$ formation on grains}

${\rm H_{2}}$ formation releases about $4.5$ ${\rm {eV}}$. However
that energy is unevenly distributed between grain excitation, internal
energy and kinetic energy of released molecules. The branching ratios
are not known and probably depend on conditions in the cloud and the
nature of the grain itself. Our treatment is based on an equipartition
argument, but with technicalities,
and ${\rm H}_{2}$ formation as discussed further in \citet{LBPRDG95}.

\begin{itemize}
\item First, the internal energy distribution of newly formed ${\rm H_{2}}$
on grain can be chosen among several options. This results in some
internal energy $E_{\rm int}$ which is computed before each run.
Note that some hypotheses may lead to most available energy to be
in $E_{\rm int}$, precluding any heating.
\item Secondly, the fraction of energy remaining in the grain is set to
one third of ${\rm H_{2}}$ formation energy at most, provided there
is enough available.
\item The remaining fraction is set to kinetic energy.
\end{itemize}
Using standard options, this results in effectively  $1.5$ ${\rm {eV}}$
released in the gas by molecule formation. But the code adjusts itself
automatically to other prescriptions, providing one is available.

The initial population of ${\rm H}_2$ upon formation is the same
throughout the modeled cloud.
Our standard option is to distribute an energy of $E_{\rm int}= 1.478\; {\rm eV}$
with a Boltzmann distribution. Note that this is \emph{not} equivalent to
using a Boltzmann factor of $T_B=E_{\rm int}/k$. Given $n_{\rm max}$, the
 highest included level of ${\rm H}_2$, we have:

\[
E_{\rm int} = \frac{\sum_{n=1}^{n_{\rm max}}\, E_n\; g_n\, \exp \left( -\, \frac{E_n}{kT_B} \right)}
               {\sum_{n=1}^{n_{\rm max}} g_n\, \exp\left(-\, \frac{E_n}{kT_B} \right)}
\]
where $E_n$ is the energy of level $n$ and $g_n$ its statistical weight. Including
all levels of ${\rm H}_2$, we find that $T_B=8734\; {\rm K}$ 
is required to recover $E_{\rm int}/k=17322\; {\rm K}$ ($1.478\; {\rm eV}$),
while using $T_B=17322\; {\rm K}$ leads to $E_{\rm int}/k = 28237\; {\rm K}$.
This may induce very significant differences in the computed gas temperature
in regions where active formation/destruction of ${\rm H}_2$ has a
significant contribution to the thermal balance.

The initial ortho to para ratio upon formation is set to $3$. That value
may be changed if needed, which has a large influence on the final 
ortho to para ratio in regions where photodestruction occurs on time scales
smaller than the conversion ones. However, there is no firm experimental
indication yet that this occurs on interstellar grains.

\subsubsection{Photodissociation and Photoionisation}

The kinetic energy of the atomic and/or molecular fragments released
in photodissociation and photoionization may also contribute to the
heating of the gas. As these processes may also lead to internal excitation
which is rarely known, we have introduced different assumptions, depending
on the species involved. The kinetic energy released in the photodissociation
of ${\rm H}_{2}$ has been computed by \citet{ARD00} for each discrete
level involved. The corresponding mean kinetic energy is about 1 eV,
in agreement with the previous estimation of \citet{SD73}. We have
assumed the same energy release for ${\rm HD}$, ${\rm CO}$ and its
isotopes. The energy input (${\rm erg}\,{\rm cm}^{-3}\,{\rm s}^{-1}$)
is then obtained by multiplying this energy by the photodissociation
probability (${\rm s}^{-1}$) and the abundances of the species involved.
For other species, we have assumed a mean photon energy of 13 eV.
In atomic photoionization (${\rm C}$, ${\rm S}$, ${\rm Fe}$, ..)
we assume that all the available energy is transferred in kinetic
form. When molecules are produced, we assume that one third of the
available energy is in kinetic form and participates to the heating.
Abundant species such as ${\rm H}_{2}$, ${\rm C}$, ${\rm CO}$ give
the main contribution but this term is never dominant. Similar assumtions
are used for secondary UV photons.

\subsubsection{Cosmic rays destruction}

The heating rate due to cosmic ray ionisation is
also poorly known. \citet{Black87} estimated that this contribution
should be between 4 and 6 eV. We assume $4$ ${\rm {eV}}$ per ionization
by cosmic rays.

\subsubsection{Chemical equilibrium}

We calculate the energy balance of chemical reactions from the variation
in formation enthalpies between the products and the reactants. This
term contributes to the heating of the gas and becomes important in
dark and cold regions when UV photons are almost completely absorbed.
We note that the estimated contribution is probably an upper limit
: indeed, the main fraction is due to dissociative recombination reactions
which involve large exothermicities and exothermic ion-neutral reactions.
We assume that molecular species are produced at thermal equilibrium
and do not consider branching ratios in excited states. The actual
source of this heating is cosmic rays which produce the ions.

\subsubsection{Other heating sources}

Additional heating terms coming from macroscopic motions, such as
turbulence or shock waves may be introduced via empirical formulae
such as those derived by \citet{Black87}. Although we have done various
tests, these are not included in the standard version of the code
as most of the physics is still poorly known or can not be expressed
via simple analytical formulae.

\subsection{Cooling}

Cooling comes mainly from discrete radiative transitions in lines
of various species following collisional excitation.

\subsubsection{Fine-structure transitions\label{sub_Fine-structure-transitions}}

These are described in Sect.~\ref{Sec_excitation} and Table~\ref{Tab_Coll}. Note that cooling
is underestimated for one-sided models since the {}``on the spot''
approximation forbids photons from escaping towards the {}``rear''
side of the cloud.

\begin{figure}
\includegraphics[angle=270,scale=0.65]{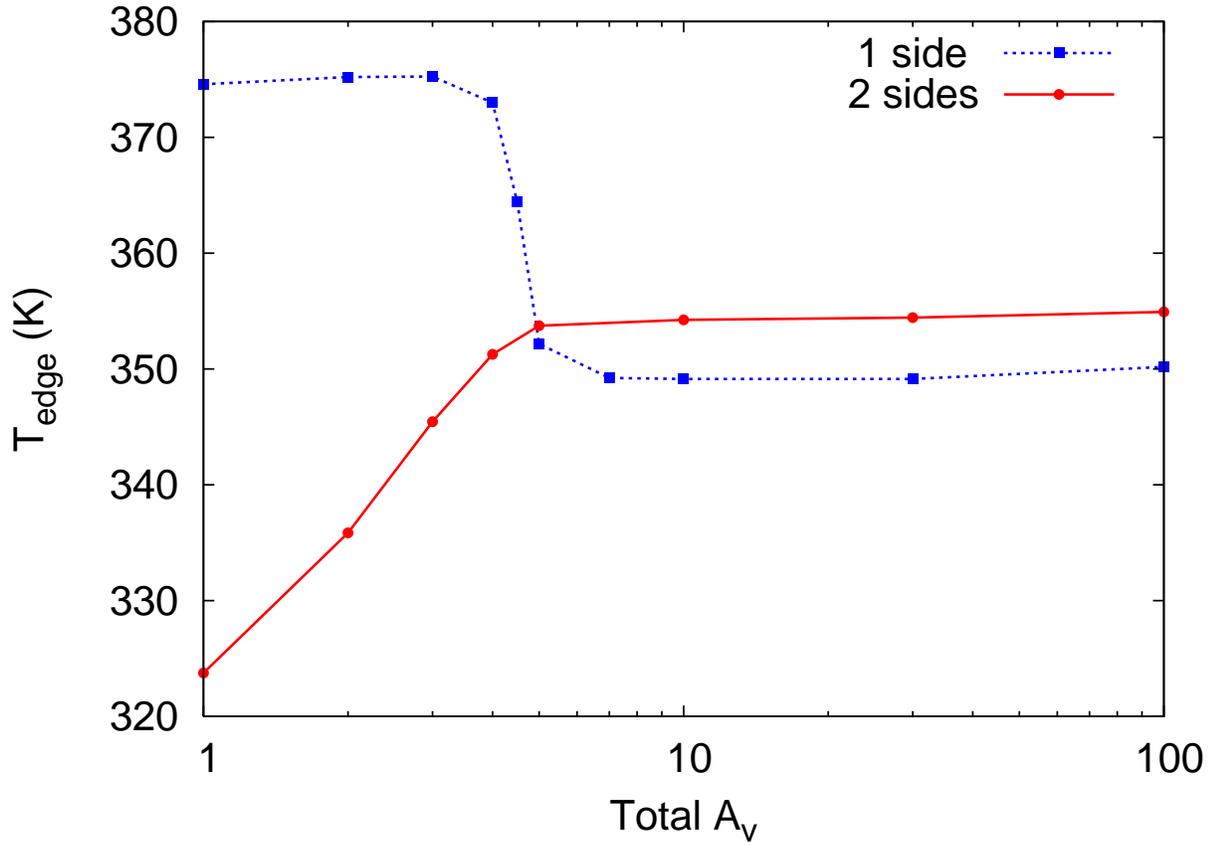}
\caption{ Effect of cloud geometry on edge temperature, $T_{K}^{\rm edge}$\label{Fig_T_edge}}
\end{figure}

\placefigure{Fig_T_edge}

As an example, we show in Figure~\ref{Fig_T_edge} the temperature
computed at the edge of a cloud of density $n_{\rm{H}}=10^{3}\,{\rm {cm}^{-3}}$
illuminated by a radiation field of $\chi=10^{3}$. The total depth
of the cloud varies from $A_{\rm{V}}=1$ to $A_{\rm{V}}=100\,{\rm {mag}}$,
and we use either radiation coming from one side only or from both
sides (with $\chi=1$ on the far side in this case). Three features
are worth noticing:

\begin{enumerate}
\item For $A_{\rm V}>5\,{\rm {mag}}$, $T_{\rm K}^{\rm edge}$ is
constant, and 1 side and 2 sides models agree. The small difference
(a few degrees) can be ascribed to numerical convergence effects.
\item For $A_{\rm V}<5\,{\rm {mag}}$, in the 2 sides models $T_{\rm K}^{\rm edge}$
rises with $A_{\rm V}$. This is physically consistent with the
fact that cooling photons have a greater and greater path to cross
before escaping from the \emph{far} side of the cloud.
\item For $A_{\rm V}<5\,{\rm {mag}}$, in the 1 side models $T_{\rm K}^{\rm edge}$
is significantly higher at low $A_{\rm V}$ ($50\,{\rm {K}}$, about
15\%), then drops brutally. This is a numerical effect. One side models
make the assumption that the cloud extension is large compared to
the size of the slab computed. However, one still needs to compute
\emph{some} escape probability for cooling lines towards the far side.
This probability is estimated from level populations in the deepest
part of the cloud; we use the last few points computed and extrapolate
to a large (rather arbitrary) depth. If the total depth computed is
too small, then the optical depth may be overestimated leading to
an underestimated cooling.
\end{enumerate}
Here, ${\rm {C}^{+}}$ contributes significantly to the cooling function
at the edge, but the size of the ${\rm {C}^{+}}$ slab is about $3\,{\rm {mag}}$.
Photons crossing those $3\,{\rm {mag}}$ are essentially free to escape
(and thus contribute to cooling the gas \emph{at} the edge), but this
(relatively) large escape probability can only be computed in models
with a total $A_{\rm{V}}$ larger than the ${\rm {C}^{+}/{\rm {C}}}$
transition zone, so that the drop in line opacity may be detected.

As a consequence, the user should remember that it is unwise to compute
1 sided models with a too small total $A_{\rm{V}}$, even if the
radiation field is large on one side.

\subsubsection{${\rm H_{2}}$ transitions}

Detailed balance of ${\rm H_{2}}$ is quite complex due to the variety
of processes involved. Besides transitions between ro-vibrational
levels, one must include level specific formation and destruction
processes which have a strong influence on populations. We include:

\begin{itemize}
\item UV pumping to Lyman and Werner bands followed by de-excitation towards
either the continuum (leading to dissociation) or a ro-vibrational
bound state of the ground electronic state.
\item Electric quadrupole radiative cascades within the ro-vibrational levels
of the ground state.
\item Collisional transitions induced by ${\rm {H}}$, ${\rm {H}_{2}}$,
${\rm {He}}$, ${\rm {H}^{+}}$, ${\rm {H}_{3}^{+}}$, including reactive
processes leading to ortho/para transfer.
\item Level specific formation on grains.
\end{itemize}
The ${\rm {H}_{2}}$ photodissociation rate (used in the chemistry) is
a by product of this computation. Given the steady state populations
of ${\rm {H}_{2}}$, it is possible to compute the net energy balance
between processes leading to energy loss (typically quadrupole transitions),
and energy gain (pumping by external UV radiation followed by collisional deexcitation).
Thus, radiative transitions lead either to heating (mainly close to
a brightly illuminated edge) or cooling (in deeper parts of the cloud).
Note that radiative transitions may arise from ${\rm {H}_{2}}$ newly
formed on grain, so that the heating rate from ${\rm {H}_{2}}$ formation
above must include both internal and translational energy to keep
a correct balance.

\subsubsection{Thermal coupling to grains}

We do not compute a detailed energy balance for grains. Dust temperature
is evaluated using the empirical expression of \cite{BHT90}. Thus,
we are not able to compute an IR continuum spectrum. However, that
temperature is used to compute an energy transfer between gas and
grains in a manner similar to \cite{BH83}.

\section{Chemistry\label{Sec_chemistry}}

We consider mainly elementary two body processes with the exception
of possible 3 body reactions involving two atomic hydrogen atoms.
Reactions are divided into several different classes, and read from
an input file described in Appendix~\ref{App_chemistry}. Chemistry
for isotopes of ${\rm C}$ and ${\rm O}$ is automatically generated
if $^{13}{\rm {C}}$ or $^{18}{\rm {O}}$ are included in the species
list. However selective reactions involving the main and rare isotopes
(leading to isotopic fractionation) need to be retrieved and specified.

From chemical reactions, for each species ${\rm {X}}$, chemical balance
leads to:\begin{equation}
\frac{dn({\rm {X})}}{dt}=F_{\rm X}-D_{\rm X}\equiv0\label{mathed:chem1}\end{equation}
with $F_{\rm X}$ the formation rate, and $D_{\rm X}$ the destruction
rate.

With $N_{\rm s}$ species included, built from $N_{\rm a}$ atoms,
Eq~(\ref{mathed:chem1}) leads to a non-linear system of $N_{\rm s}$
equations $\mathbf{f}(n({\rm {X}))=0}$ to solve. However, the system
is under-determined, and only $N_{\rm s}-N_{\rm a}$ equations
are independent. The $N_{\rm a}$ missing conditions come from conservation
equations of each atom. Numerical stability is ensured by excluding
at each point the evolution equation of the most abundant species
for each atom. A Newton-Raphson scheme is used to compute the solution.

This is a dynamical system, whose steady states are solutions of our
stationary problem. Depending on physical
conditions, there may be one or more steady state at each point into
the cloud as described in \cite{LPR95}. Computation starts from the
edge of the cloud, so that the solution at $\tau=0$ is the {}``High
ionisation Phase'' (HIP). As $\tau$ is increased, that solution
is followed by continuation even if a new {}``Low ionisation Phase''
(LIP) exists. However, deeper into the cloud, there may be a point
where the HIP solution ceases to exist and only the LIP one is possible.
At this critical value of $\tau=\tau_{\rm c}$, the bifurcation
from one type of solution to the other is characterised by a discontinuity
of many variables. Although not physical, this is not a bug, nor a
weakness of the model. It is a natural consequence of the lack of
coupling between adjacent regions: it could be solved with the introduction
of a small diffusive term that would smooth the discontinuity (see
\citet{DLB02} for an example of the effect of turbulence on chemistry).
However, we feel that the computational burden is not worth the trouble.
In real clouds both solutions could coexist if, e.g., different clumps
of gas with different histories were mixed. In that case, reaction-diffusion
fronts are bound to arise whose dynamics is, to our knowledge, still
unknown.

Once convergence has been reached for radiative transfer, chemical
equilibrium and thermal balance, the chemical network can be analysed
at each point of the cloud using the post-processor code.

The different classes of chemical reactions are described in Appendix~\ref{App_chemistry}.
We focus here on two important reactions: formation of ${\rm H}_{2}$
and photodestruction processes.

\subsection{Formation of ${\rm H_{2}}$\label{Sec_H2_form}}

It is customary to write the ${\rm H_{2}}$ formation rate (in ${\rm cm}^{-3}\,\rm{s}^{-1}$)
as $R_{f}\, n_{\rm H}\, n({\rm H})$. The commonly adopted value
of $R_{f}$, derived from Copernicus observations, is $3\,10^{-17}\,{\rm cm^{3}\,{\rm s^{-1}}}$\citep{Jura75}.
Recent works from FUSE observations \citep{GBNPHF02} have confirmed
that order of magnitude. However, this value results from observational
data integrated over whole lines of sight. It includes in one single
term different physical processes (adsorption of an ${\rm {H}}$ onto
a grain, migration on the surface, reaction with another particle
which is not necessarily an hydrogen atom, and desorption of the resulting
${\rm H_{2}}$). All these processes may vary with depth into the
cloud as the grain surface certainly does, so the local ${\rm H_{2}}$
formation rate depends on $\tau$ and also on various adopted parameters
of the model, particularly the grain model.

We have chosen not to impose a single constant value of $R_{f}$.
${\rm H_{2}}$ production results from independent processes:\[
{\rm {H}+dust\,\rightarrow\,{\rm {H_{ad}}}}\qquad(k_{ad})\]

\[
{\rm {H_{ad}}+{\rm {H_{ad}}\,\rightarrow\,{\rm H_{2}}}}\qquad(k_{s})\]

where ``$\rm{H_{ad}}$'' means an atom of hydrogen adsorbed
on dust, $k_{s}$ is the reaction rate (in $\rm{cm}^{3}\,\rm{s}^{-1}$)
at the surface of the grains and $k_{\rm ad}$ the adsorption rate
(in $\rm{s}^{-1}$) of H atoms on grains. The $\rm{H_{2}}$ abundance may depend also on other processes (various
desorption processes, or other reactions). If only these two processes
are included and steady state applies, then the production rate of
${\rm H_{2}}$ may be computed by elimination of ${\rm {H}}$ adsorbed:
\[
\left.\frac{d{\rm H_{2}}}{dt}\right|_{form}=k_{s}\, n(\rm{H_{ad}})^{2}=\frac{1}{2}\, k_{\rm ad}\, n(\rm{H})\]

The rate is independent of how adsorbed ${\rm {H}}$ atoms eventually manage to reach one another.
This gives:\[
\left.\frac{d{\rm H_{2}}}{dt}\right|_{form}=\frac{1}{2}\, s\,
\left\langle \sigma n_{g}\right\rangle \,\overline{v}_{\rm{H}}\, n({\rm {H})}\]

where $s$ is the sticking coefficient of ${\rm {H}}$ upon collision
with grains, $\overline{v}_{\rm{H}}$ its mean velocity, and
$\left\langle \sigma n_{g}\right\rangle $ the mean cross section
of grains per unit volume. Results from \cite{LPRF95} show the cross section is given by:\[
\left\langle \sigma n_{g}\right\rangle =\frac{3}{4}\,\frac{1.4\, m_{\rm{H}}\, G}
{\rho_{\rm g}}\,\frac{1}{\sqrt{a_{\rm min}\times a_{\rm max}}}\, n_{\rm{H}}\]
where $G$ is the dust to gas mass ratio, $\rho_{\rm g}$ the
density of grain material and the size distribution law of grains
from \citet{MRN77} has been used with $\alpha=3.5$. Using $G=10^{-2}$,
$\rho_{\rm g}=3\,{\rm {g}\,{\rm cm^{-3}}}$, $a_{\rm min}=3\,10^{-7}\,{\rm {cm}}$,
$a_{\rm max}=3\,10^{-5}\,{\rm {cm}}$ one finally obtains:\[
\left.\frac{dn({\rm H_{2}})}{dt}\right|_{form}=s\,1.4\,10^{-17}\,\sqrt{T_{\rm K}}\, n_{\rm{H}}\, n({\rm {H})}\]

This gives about $4\,10^{-17}\,{\rm cm}^{3}\,{\rm s}^{-1}$ for $R_{f}$
at $10\,{\rm K}$ where $s$ should be close to 1. However, it does
not include other reaction paths for {}`` ${\rm {H_{ad}}}$ '' (mainly
direct evaporation from warm grains) which may lower that rate. Furthermore,
at higher temperature the sticking coefficient is expected to decrease.
The local value of $R_{f}$ is thus one of the results of the model,
and depends on the physics and reactions included.

An effective constant value of $R_{f}$ may be enforced by using only
the two reactions above and adjusting the sticking coefficient. For
a given grain model, the value of $<\sigma\, n_{g}>$ is easily computed
and $R_{f}$ adjusted from the parameter $\gamma$ of the adsorption
reaction (see \ref{sub_chim_ads}). However, this procedure should
be avoided in normal usage since it is unphysical and introduces inconsistencies.

\begin{figure}
\includegraphics[angle=270,scale=0.65]{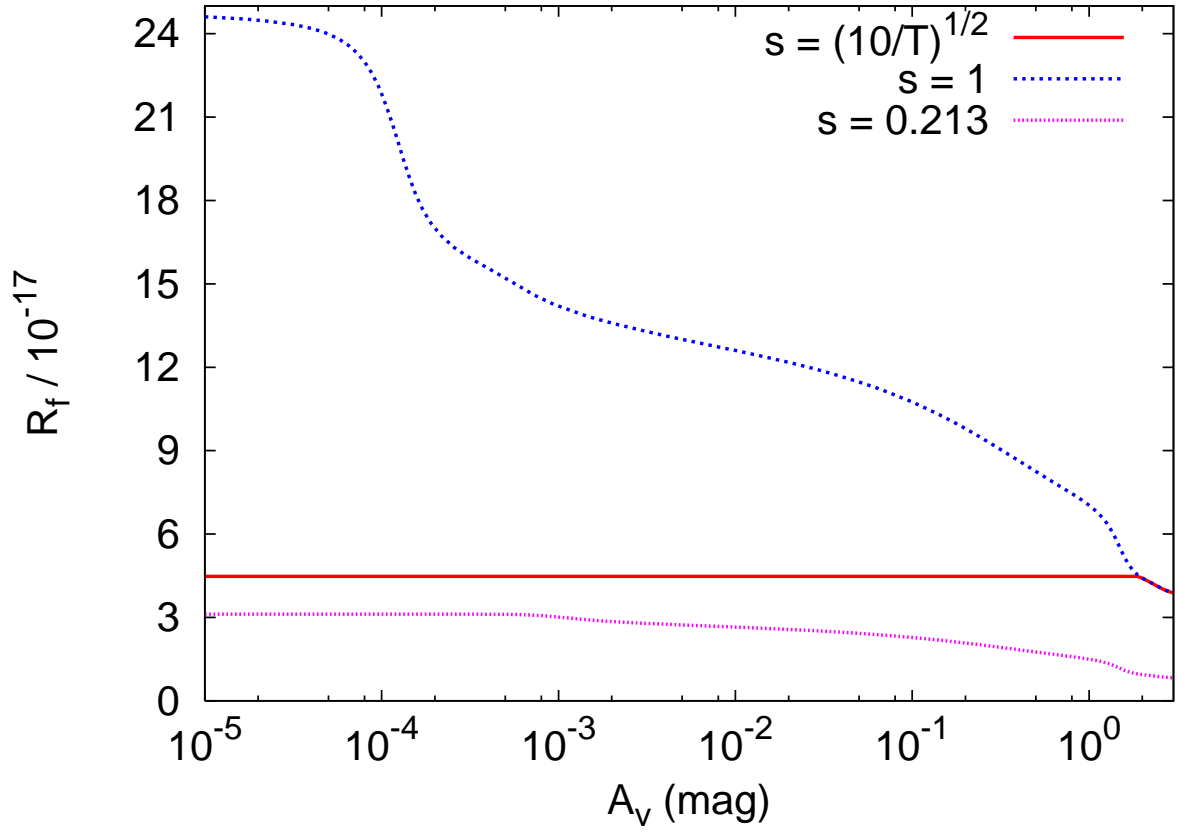}

\caption{ ${\rm \rm{H}_{2}}$ formation rate $R_{f}$ ($10^{-17}\,{\rm cm}^{3}\,{\rm s}^{-1}$). 
Parameters of the model : $n_{\rm{H}}=10^{4}\,{\rm cm}^{-3},\,\chi=10$.\label{Fig_fh2_a}}
\end{figure}

\begin{figure}
\includegraphics[angle=270,scale=0.65]{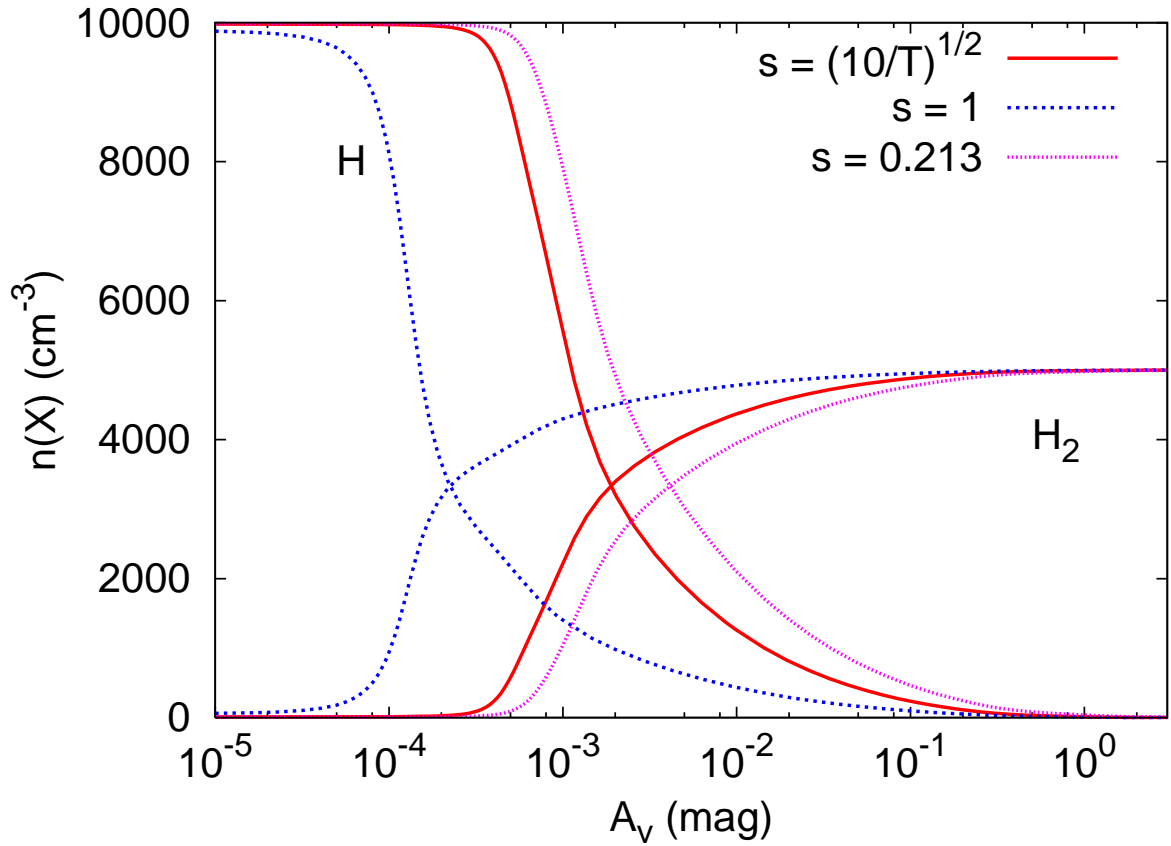}

\caption{Resulting ${\rm {H}/{\rm \rm{H}_{2}}}$ transition for
various sticking coefficient s. Same parameters as Figure \ref{Fig_fh2_a}.
Note the large shift in the ${\rm H}/{\rm H}_{2}$ transition.\label{Fig_fh2_b}}
\end{figure}

As an example, Fig.~(\ref{Fig_fh2_a}) shows the resulting $R_{f}$
with various sticking coefficients and (\ref{Fig_fh2_b}) shows the effect of varying $s$ on the $\rm{H}/\rm{H_2}$ transition. The test model has $n_{\rm H}=10^{4}\,{\rm cm^{-3}}$
and $\chi=10$. $s=1$ leads to a high formation rate, with a ${\rm {H}/{\rm H_{2}}}$
transition very close to the cloud edge. The {}``canonical value''
of $3\,10^{-17}$ is recovered at the edge by adjusting the sticking
coefficient to $s=0.213$, but this is rather artificial. Our empirical
prescription, $s=\sqrt{\frac{10}{\max(10,T_{\rm K})}}$ gives a
constant formation rate close to $3\,10^{-17}$ in the transition
region, where most of ${\rm H_{2}}$ emission comes from. Various
other laws have been proposed for $s$, however that parameter is
still very uncertain (see \citet{HT99} and references therein).

\placefigure{Fig_fh2_a}

\placefigure{Fig_fh2_b}

\subsection{${\rm H}/{\rm H_{2}}$ transition\label{Sec_H_H2_trans}}

It is well known that the position of the ${\rm H}/{\rm H_{2}}$ transition
depends on the ratio $n_{\rm H}/\chi$ (\cite{HT99}). The higher this ratio, the
closer to the edge of the cloud is the point where
$f = \frac{2\; N({\rm H}_2)}{N({\rm H})+2\;N({\rm H}_2)} = 0.5$.
However, in a photon dominated region, the same abundance does not
translate to the same excitation and in a denser cloud submited to
a higher radiation field, UV pumping is more efficient \emph{for
the same column density} than in a less excited cloud.

\begin{figure}
\includegraphics[angle=270,scale=0.65]{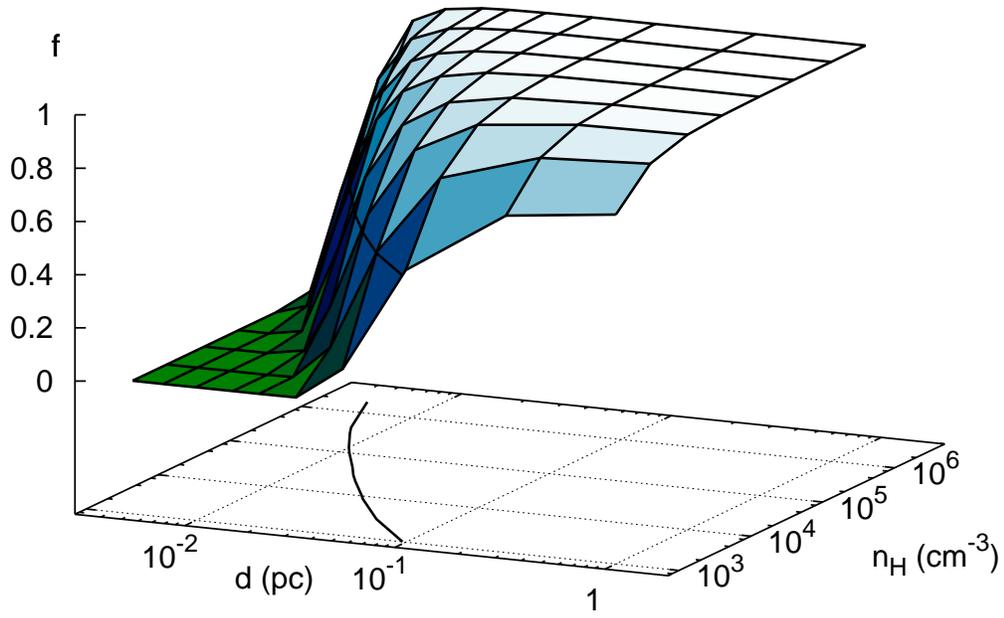}

\caption{${\rm H}/{\rm H}_2$ transition for a $A_{\rm v} = 10^{-2}$ slab of
gas as a function of density $n_{\rm H}$ and distance $d\; ({\rm pc})$ to
the star HD~102065. The solid line in the plane $d - n_{\rm H}$ corresponds 
to $f_{\rm H_2} = 0.5$.
\label{Fig_f_H2b}}
\end{figure}

\placefigure{Fig_f_H2b}

\begin{figure}
\includegraphics[angle=270,scale=0.65]{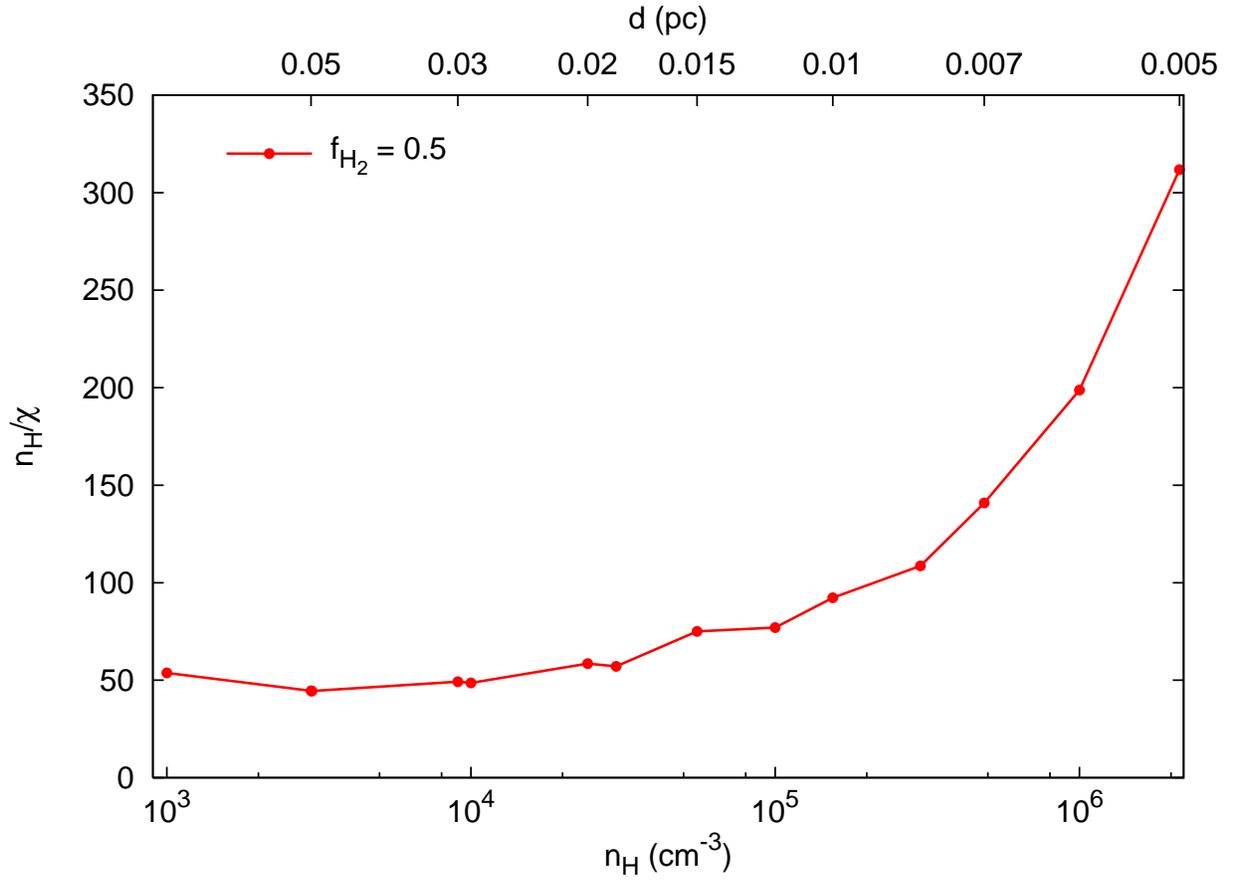}

\caption{$n_{\rm H}/\chi$ ratio for the iso-$f_{\rm H_2} = 0.5$ curve. Fig. \ref{Fig_f_H2b} shows that for each value of n$_{\rm H}$ (lower x axis) a single radiation field leads to $f_{\rm H_2} = 0.5$. The associated distance to the star is shown on the upper x axis. 
\label{Fig_nsurc}}
\end{figure}

This may be illustrated by computing the ${\rm H}_2$ excitation in a
small slab of gas of constant thickness (say $A_{\rm v} = 10^{-2}$)
placed at various distances from a star. To be specific, we used a late
B star ressembling HD~102065 (see also \citealp{NLGBP05}).
$n_{\rm H}$ runs from $10^3$ to $3\; 10^6\; {\rm cm}^{-3}$, and
$d$ from $5\; 10^{-3}$ to $1\; {\rm pc}$, which gives an 
equivalent UV field in Draine's unit varying from $\sim 7000$ to $1$.
Fig.~(\ref{Fig_f_H2b}) shows the abrupt transition from atomic to
molecular hydrogen. The $f_{\rm H_2} = 0.5$ isocontour shows that
the transition is well defined, although not a straight line.
Fig.~(\ref{Fig_nsurc}) shows that to reach the same amount of ${\rm H}_2$
on the line of sight, density must rise faster than the radiation field.

\placefigure{Fig_nsurc}

\begin{figure}
\includegraphics[angle=270,scale=0.65]{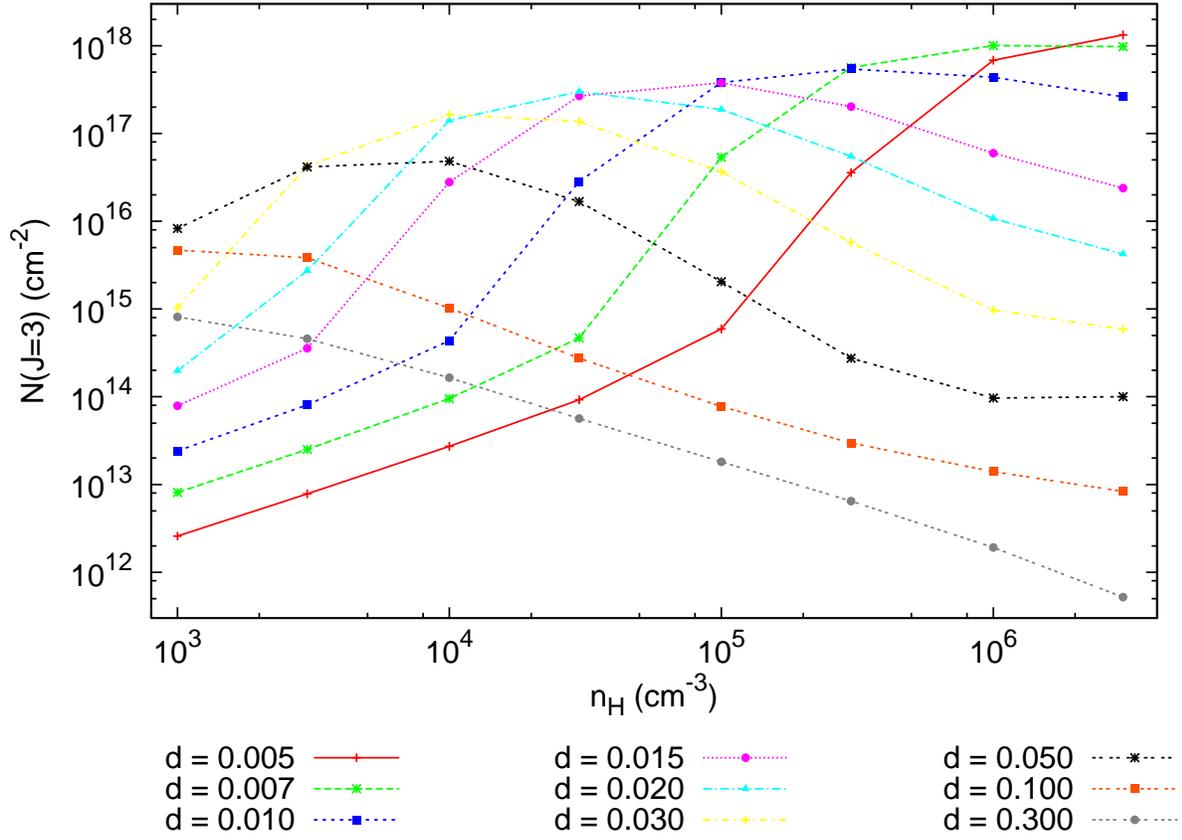}

\caption{Column density of ${\rm H}_2\; (J=3)$ as a function of $n_{\rm H}$
for various distances to HD~102065.
\label{Fig_nh2b_3}}
\end{figure}

\begin{figure}
\includegraphics[angle=270,scale=0.65]{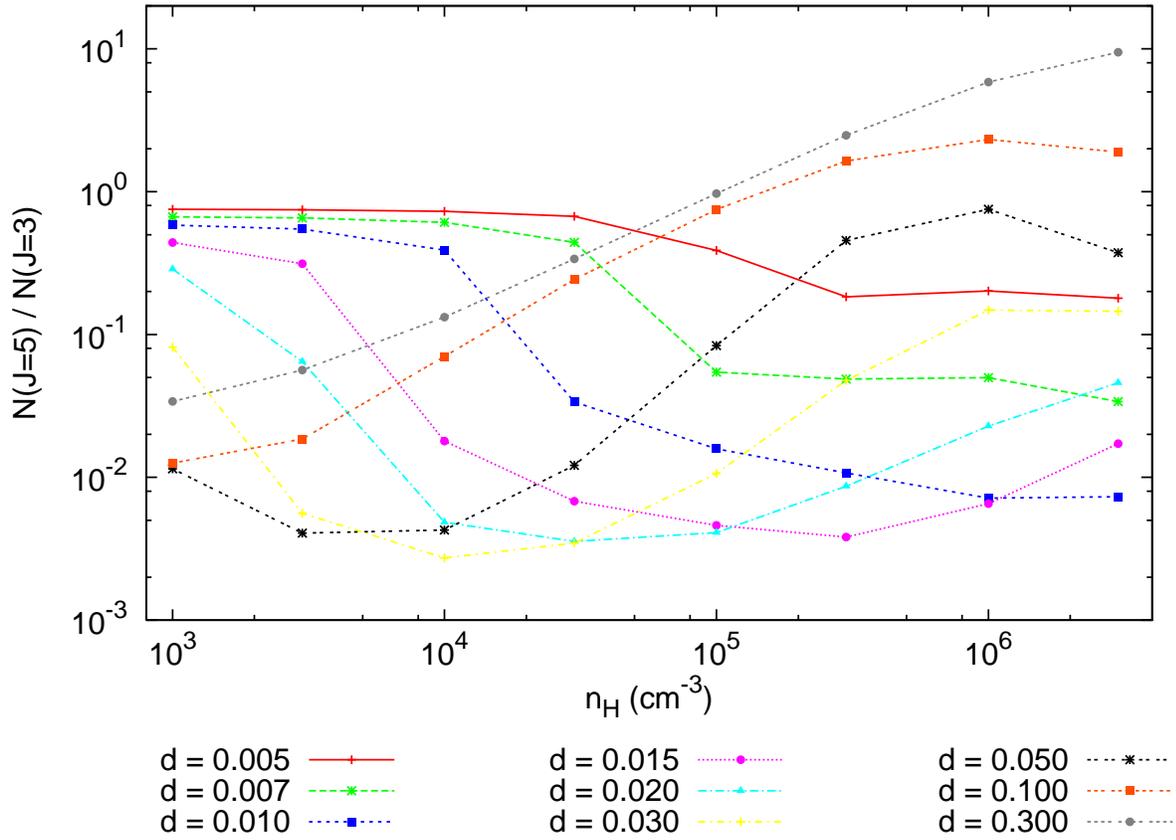}

\caption{$N({\rm H}_2\; (J=3))/N({\rm H}_2\; (J=5))$
as a function of $n_{\rm H}$
for various distances to HD~102065.
\label{Fig_nh2b_r}}
\end{figure}

Increasing the radiation field also increases radiative pumping
of excited rotational levels of ${\rm H}_2$ through cascades. It is thus
possible to reach higher excitation temperature for a given molecular
fraction. Fig.~(\ref{Fig_nh2b_3}) shows $N({\rm H}_2 (J=3))$ rising first then 
decreasing as excitation grows. This intermediate level is first populated
from lower lying levels, then gets depopulated in favour of higher
lying ones. This is clearly illustrated in Figure~(\ref{Fig_nh2b_r})
which shows the ratio of column densities $(J=5)/(J=3)$.
\cite{NLGBP05} use these results to try and understand observations
of rotationally excited ${\rm H}_2$ towards HD~102065. They show that the
observational constraints are met by their models. However, it is unlikely that the
required high pressure in a slab of gas close to the star will last long enough.
Furthermore, the observed ortho to para ratio requires that, upon 
formation on grains, ${\rm H}_2$ be released with a ratio of $1$
instead of $3$, which is not supported on physical grounds.

\placefigure{Fig_nh2b_3}

\placefigure{Fig_nh2b_r}

\subsection{Photodestruction processes\label{Sec_photo_D}}

\subsubsection{Dissociation in lines}

For ${\rm H}_{2}$, ${\rm CO}$ and their isotopes, photo dissociation
proceeds in lines by absorption of an ultraviolet photon from the
ground state to an electronicaly excited state, followed by fluorescence
either to vibrationally excited bound states or non radiative transitions
in a dissociative state via non-adiabatic couplings as described in
\citet{evD88}. These processes are described in details in \citet{Abgrall92}
for ${\rm H}_{2}$, \citet{LHPRL96} for ${\rm CO}$ and its isotopes
and \citet{LRL02} for ${\rm HD}$. The numbers of discrete levels
included in a computation for each molecule are parameters of the
model.

\subsubsection{Continuous processes}

\begin{figure}
\includegraphics[angle=270,scale=0.65]{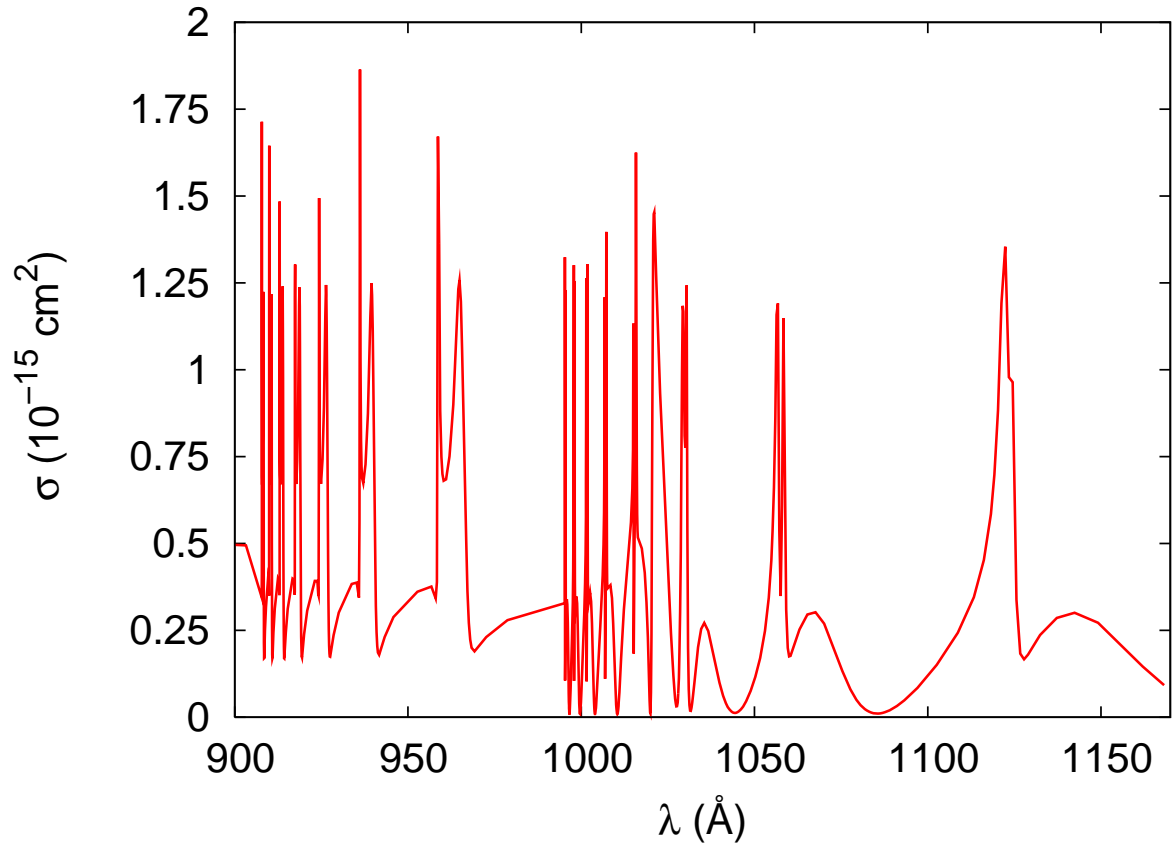}
\caption{Sulfur ionisation cross section from TopBase\label{Fig_sigma_S}.}
\end{figure}

\begin{figure}
\includegraphics[angle=270,scale=0.65]{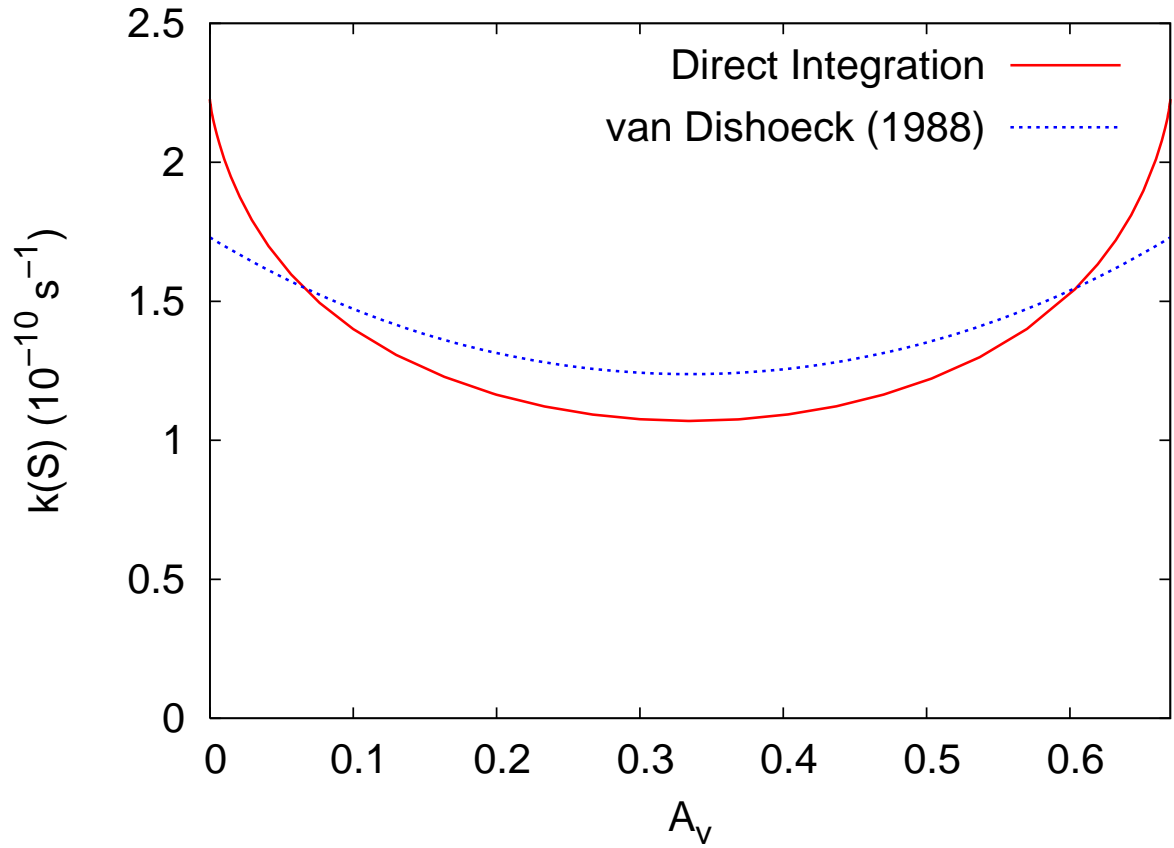}
\caption{Comparison of ${\rm S}$ photoionisation rate by direct integration
and from \citet{evD88}.\label{Fig_k_ionis_S}}
\end{figure}

If ionisation or dissociation cross sections are known, the corresponding
photodestruction rate is obtained from direct integration over the radiation
field by:

\[
k=\frac{1}{h}\,\int_{912\,\rm{Å}}^{\lambda_{\rm t}}\sigma_{\lambda}\, u_{\lambda}\,\lambda\, d\lambda\]
where $k$ is the destruction rate (${\rm s}^{-1}$), $\sigma_{\lambda}$
the photo-destruction cross section (${\rm cm}^{2}$), $u_{\lambda}$
the radiative energy density (${\rm erg}\,{\rm cm}^{-3}\,{\rm \mathring{A}^{-1}}$)
and $\lambda_{\rm t}$ the destruction threshold (${\rm \mathring{A}}$),
which should be larger than the Lyman cut-off. This is done in our
code for ${\rm CI}$ and ${\rm SI}$ with cross sections taken from
TopBase (\url{http://vizier.u-strasbg.fr/topbase/xsections.html}).
The ${\rm CI}$ ionisation cross section happens to be constant ($\sigma_{\rm C}=1.6\,10^{-17}\,{\rm cm}^{2}$)
from the Lyman limit to the ionisation threshold at $\lambda_{\rm t}=1101\,\rm{Å}$.
For ${\rm SI}$, the section is shown on Fig.~(\ref{Fig_sigma_S})
($\lambda_{\rm t}=1168.2\,\rm{Å}$). The resulting ionisation
rate for a test model is shown on Fig.~(\ref{Fig_k_ionis_S}).

\placefigure{Fig_sigma_S}

\placefigure{Fig_k_ionis_S}

If the full cross section is not included into the code, then we use
the results of \citet{evD88}. Fig.~(\ref{Fig_k_ionis_S}) shows a
comparison of the two computations for ${\rm S}$ for a $A_{\rm V}=1$
slab of gas illuminated by a standard radiation field. It is seen
that the effect is small here. However, the full computation includes
the effect of protection by ${\rm H}_{2}$ in regions of a cloud where
that molecule is already abundant and radiation still penetrates between
the saturated lines. Then it may lead to significant differences in
some line intensities as seen on Fig.~\ref{Fig_C+line}.

\subsection{Uncertainties in the chemical rates}

Chemical networks are presently available on different websites. The
latest UMIST database for Astrochemistry is available at \url{http://www.rate99.co.uk/}
and described in \citet{LTMM00}. The chemical reaction rate coefficients
are displayed as analytical formulae of the temperature as $k\,\left(\frac{T_{\rm K}}{300}\right)^{\alpha}\,\exp\left(-\beta T_{\rm K}\right)$
and are supposed to be relevant in a specific temperature range. Chemical
networks are also downloadable from the web site of prof. E. Herbst
from Ohio state University (OSU) at \url{http://www.physics.ohio-state.edu/~eric/research.html},
with different versions and some comments included. However, these
data have been initially derived for dark cloud conditions for which
very few measurements are available. Some additions have been implemented
regarding neutral-neutral reactions involving activation barriers
and/or endothermicity (mainly in the UMIST data base) and photodissociation
probabilities, which are of special interest for PDR models. The corresponding
photodissociation rates are expressed as $k\,\exp\left(-\beta A_{\rm V}\right)$
where $k$ is in ${\rm s}^{-1}$ and the exponential factor depends
on the grain properties. \citet{evD88} describes in detail the derivation
of the photodissociation and photoionization probabilities and derives
expressions for different species. Such expressions are very convenient
but may be very crude as in the case of ${\rm H}_{2}$ and ${\rm CO}$
where the formulae displayed in OSU chemistry are not adequate. Note
that UMIST chemistry does not contain any photodissociation rate for
${\rm H}_{2}$. In our previous studies and the present paper, we
compute explicitly the photodissociation probabilities of ${\rm H}_{2}$,
${\rm HD}$, ${\rm CO}$ and the photoionisation probabilities of
atomic carbon and sulfur as described in Sect.~\ref{Sec_photo_D}.

The chemical network used in the present paper is also downloadable
from our website and maintained by us. However it is clear that the
corresponding data may suffer from many uncertainties due mainly to
the large temperature range involved in PDRs. The sensitivity of the
model results to such uncertainties is seldom considered. \citet{RLBP96}
have studied the fluctuations in steady state computed abundances
resulting from the uncertainties in the chemical reaction rate coefficients
in the case of dark cold cloud conditions. In such cases, the variation
of a reaction rate coefficient by a factor less than 2 may even change
the chemical phase describing steady state, leading to huge differences
in the model results. A specific example is the dissociative recombination
rate coefficient of the ${\rm H}_{3}^{+}$ molecular ion which has
been the subject of many theoretical and experimental studies (see
for example \citealp{LPR01}). In PDRs conditions, we have shown in
the case of the horsehead PDR \citep{Teyssier04}, that the results
obtained by using the UMIST and OSU 2003 chemistries are consistent
with each other when considering carbon chain abundances. The crucial
step in PDR is to accurately describe the ${\rm H}$, ${\rm H}_{2}$,
${\rm C}^{+}$, ${\rm C}$, ${\rm CO}$ variations which require specific
treatment as discussed in the present paper.

\section{Diffuse clouds\label{Sec_diffuse}}

Diffuse and translucent clouds are permeated by the UV interstellar
radiation field, and are defined as moderately dense media ($n_{\rm{H}}\simeq10-100\,\,\rm{cm}^{-3}$)
with $A_{\rm{V}}$ of a few 1-3. Visible and UV observations
in absorption, have probed their molecular content: $\rm{H}_{2}$,
${\rm HD}$, ${\rm CO}$ (Copernicus, IUE, FUSE, HST), ${\rm CH}$,
$\rm{CH}^{+}$, ${\rm CN}$, $\rm{C}_{2}$, $\rm{C}_{3}$,
${\rm OH}$, $\rm{H}_{2}\rm{O}$, $\rm{H}_{3}^{+}$,
etc. Compared to dark clouds, the chemistry of these regions is relatively
simple and so they are good places to test the physics of the numerical
models.

In order to discuss the trends which can be eventually compared to
observations, we present a grid of 56 models corresponding to a total
visual extinction of 1.0 with different densities, $n_{\rm{H}}$,
and radiation scaling factors, $\chi$. These two parameters take
respectively the values : 20, 50, 100, 200, 500, 1000, 1500 and 2000
$\rm{cm}^{-3}$ and $\chi=$0.1, 0.2, 0.5, 1, 2, 5 and 10. Thus
the ratio $n_{\rm{H}}/\chi$ ranges from 2 to $20000.$ In these
low visual extinction conditions, the incident radiation field has
to be taken impinging on both sides of the cloud. The FGK approximation
is used (cf Sect. \ref{sec_FGK}). This allows us to derive the general
trends and to save computing time. The elemental abundances and the
dust properties are fixed and given in Table~\ref{Tab_paramdiff}.

\placetable{Tab_paramdiff}

\subsection{$T_{01}$ vs $T_{\rm K}$}

$T_{01}$, which is often used as a measure of the
kinetic temperature (cf. \citealp{Rachford02}), is defined as:

\[
T_{01}=-170.5/ln\left(\frac{N({\rm H}_{2},J=1)}{9\times N({\rm H}_{2},J=0)}\right)\]

\begin{figure}
\includegraphics[angle=270,scale=0.65]{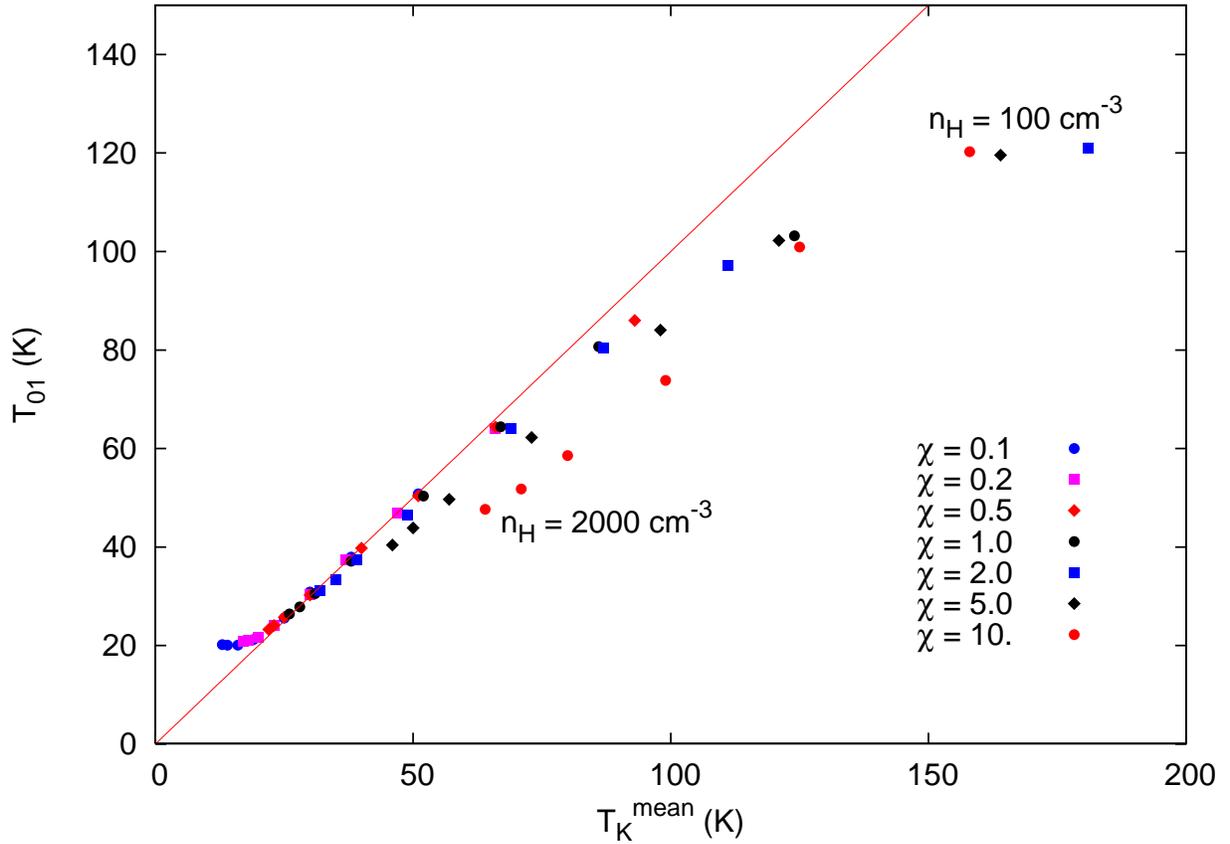}
\caption{Excitation temperature of levels J = 0, 1 of $\rm{H}_{2}$
versus the mean kinetic temperature. Each point corresponds to a model
($A_{\rm{V}}=1$). Densities are 100, 200, 500, 1000, 1500 and 2000 $cm^{-3}$. The straight line displays $T_{01}=T_{\rm K}^{\rm mean}$.\label{Fig_t01_tmoy}}
\end{figure}

We check this assumption by plotting on Fig.~\ref{Fig_t01_tmoy}
the calculated $T_{01}$ as a function of the mean kinetic temperature
of the gas, $T_{\rm K}^{\rm mean}$, derived from thermal balance:

\[
T_{\rm K}^{\rm mean}=\frac{1}{A_{\rm V}^{\rm tot}}\int_{0}^{A_{\rm{V}}^{\rm tot}}T_{K}(A_{\rm{V}})\,\, dA_{\rm{V}}\]
$T_{01}$ is a good approximation to $T_{\rm K}^{\rm mean}$ for $T_{\rm K}^{\rm mean}$
below $100\,{\rm K}$. In the other cases, $T_{01}$ is a lower limit
as the $J=1$ population is not thermalized. They correspond
to  low ratios $n_{\rm{H}}/\chi$ and  processes such
as radiative pumping in Lyman and Werner bands followed by cascades
compete with pure collisional excitation. The mean temperature, $T_{01}$, observed by Copernicus is $55\pm8\,{\rm K}$ and observed by FUSE is $68\pm15\,{\rm K}$ \citep{Rachford02} and thus, assuming $T_{01}$ is close to the kinetic temperature is justified. 

\placefigure{Fig_t01_tmoy}

\subsection{$\rm{H}_{2}$ excitation}

\begin{figure}
\includegraphics[angle=270,scale=0.65]{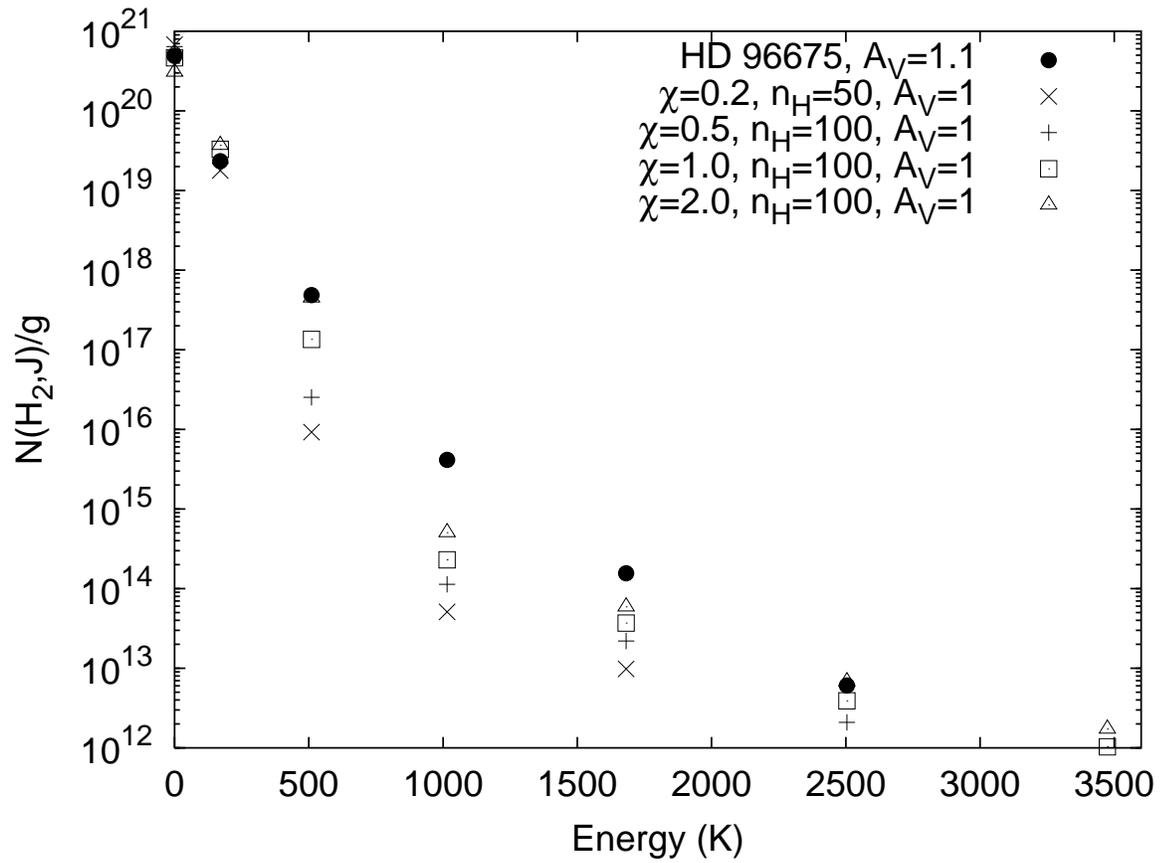}
\caption{Comparison of the excitation diagram towards HD~96675 \citep{GBNPHF02}
with some models from our grid. \label{Fig_diagex}}
\end{figure}

\placefigure{Fig_diagex}

Fig.~\ref{Fig_diagex} compares the excitation diagram towards HD~96675
\citep{GBNPHF02} with some diagrams obtained with our grid of models
with $A_{\rm{V}}=1$ for $\chi$ between $0.2$ and $2.0$.
HD~96675 has been chosen because on this line of sight the probed
gas is unperturbed by any special radiation field. Among all our models,
we select those with densities giving the closest values to the observed
$J=0$ and $J=1$ levels. The purpose of this figure is to show that,
in classical diffuse clouds, radiative pumping is unable to explain
the observed excitation of $\rm{H}_{2}$ in its levels
$J>3$. Other mechanical processes such as C-shock \citep{FPdF98}
or turbulence \citep{JFPdF98,Casu03} are required to explain it.

\subsection{Atomic to molecular fraction}

The atomic and molecular abundances of hydrogen
are the result of the balance between the formation of $\rm{H}_{2}$
on dust and photodissociation. As a consequence we expect that their
column densities are a function of the ratio of the density to the
scaling factor of the incident radiation field: $n_{\rm{H}}/\chi$.
Observers define the molecular fraction as :\[
f=\left(\frac{2\times N(\rm{H}_{2})}{N(\rm{H})+2\times N(\rm{H}_{2})}\right)\]

\begin{figure}
\includegraphics[angle=270,scale=0.65]{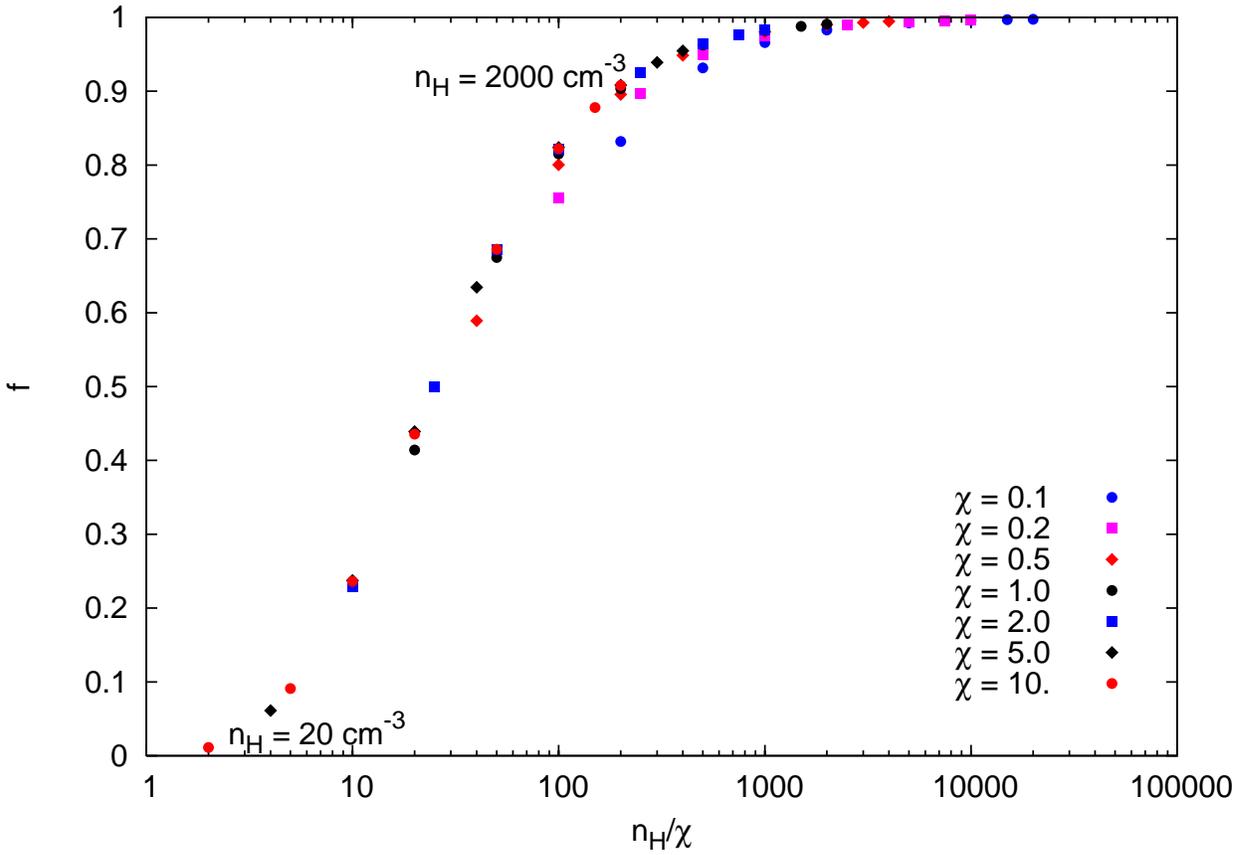}
\caption{Molecular fraction as a function of $n_{\rm{H}}/\chi$. Each
point corresponds to a model ($A_{\rm{V}}=1$). Densities corresponding to these models are 20, 50, 100, 200, 500, 1000, 1500 and 2000 $cm^{-3}$.\label{Fig_Hvsnhc}}
\end{figure}

Fig.~\ref{Fig_Hvsnhc} displays the variation of
the molecular fraction as a function of $n_{\rm{H}}/\chi$.
First, as expected, $f$ reaches a value of $1$ for high $n_{\rm{H}}/\chi$
ratio. From FUSE and Copernicus observations of diffuse and translucent clouds,
\citet{Rachford02} brought to the fore a group a 10 lines of sight
\footnote{These lines of sight are $\zeta$ Oph, $o$ Per, $\zeta$ Per, HD~24534,
HD~27778, HD~62542, HD~73882, HD~96675, HD~154368 and HD~210121.
} with $f\simeq0.7$. According to Fig.~\ref{Fig_Hvsnhc}
this corresponds to $n_{\rm{H}}/\chi~\simeq~60$. So if we assume
that $\chi\simeq1$, this gives $n_{\rm{H}}$ around $60\,\rm{cm}^{-3}$
which is the order of magnitude of the expected value of the density
in diffuse/translucent clouds. Secondly, we see that models with different
parameters but the same ratio $n_{\rm{H}}/\chi$ give the same
results: for a given visual extinction, the molecular fraction of
a cloud is controlled by its ratio $n_{\rm{H}}/\chi$.

\placefigure{Fig_Hvsnhc}

\subsection{Chemical results}

\begin{figure}
\includegraphics[angle=270,scale=0.65]{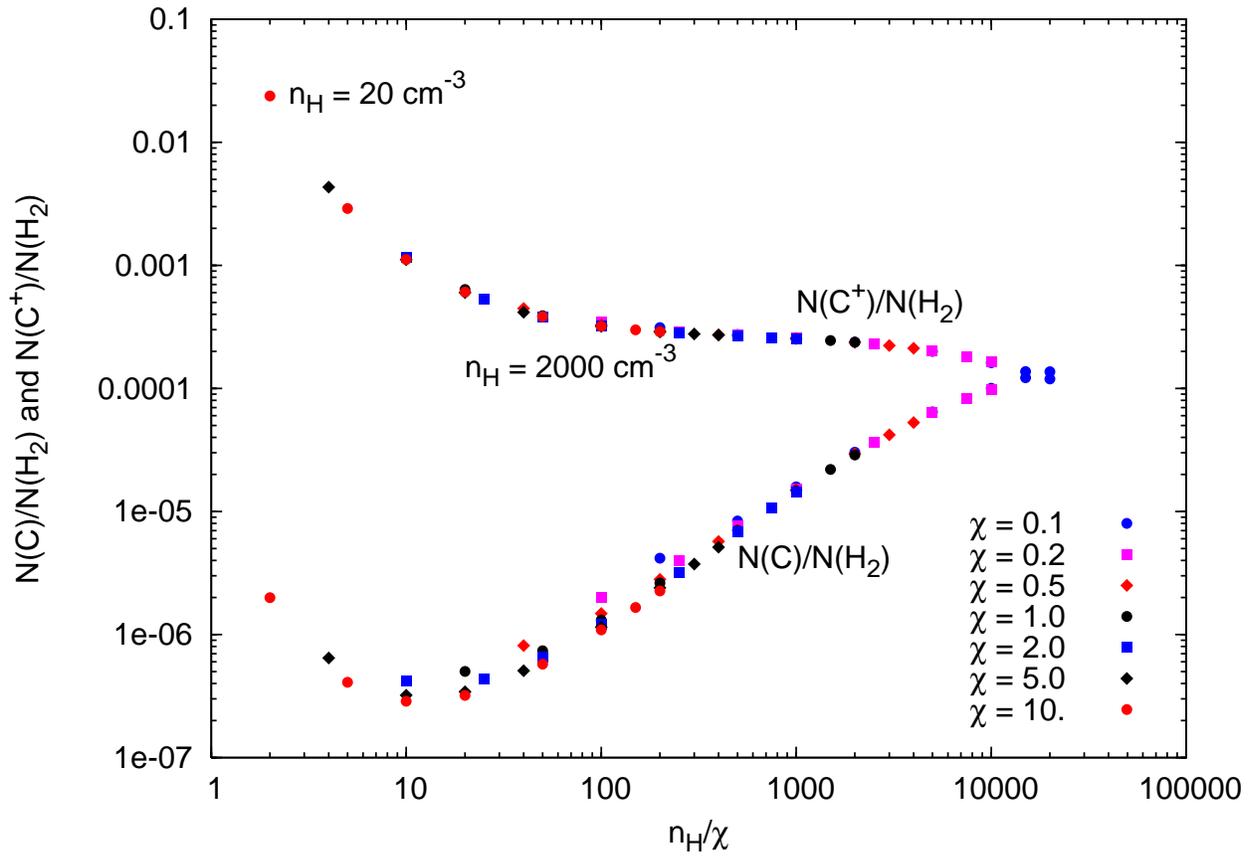}

\caption{Column densities of $\rm{C}^{+}$ and ${\rm C}$
as a function of $n_{\rm{H}}/\chi$.\label{Fig_CHvsnhc_a}}
\end{figure}

\begin{figure}
\includegraphics[angle=270,scale=0.65]{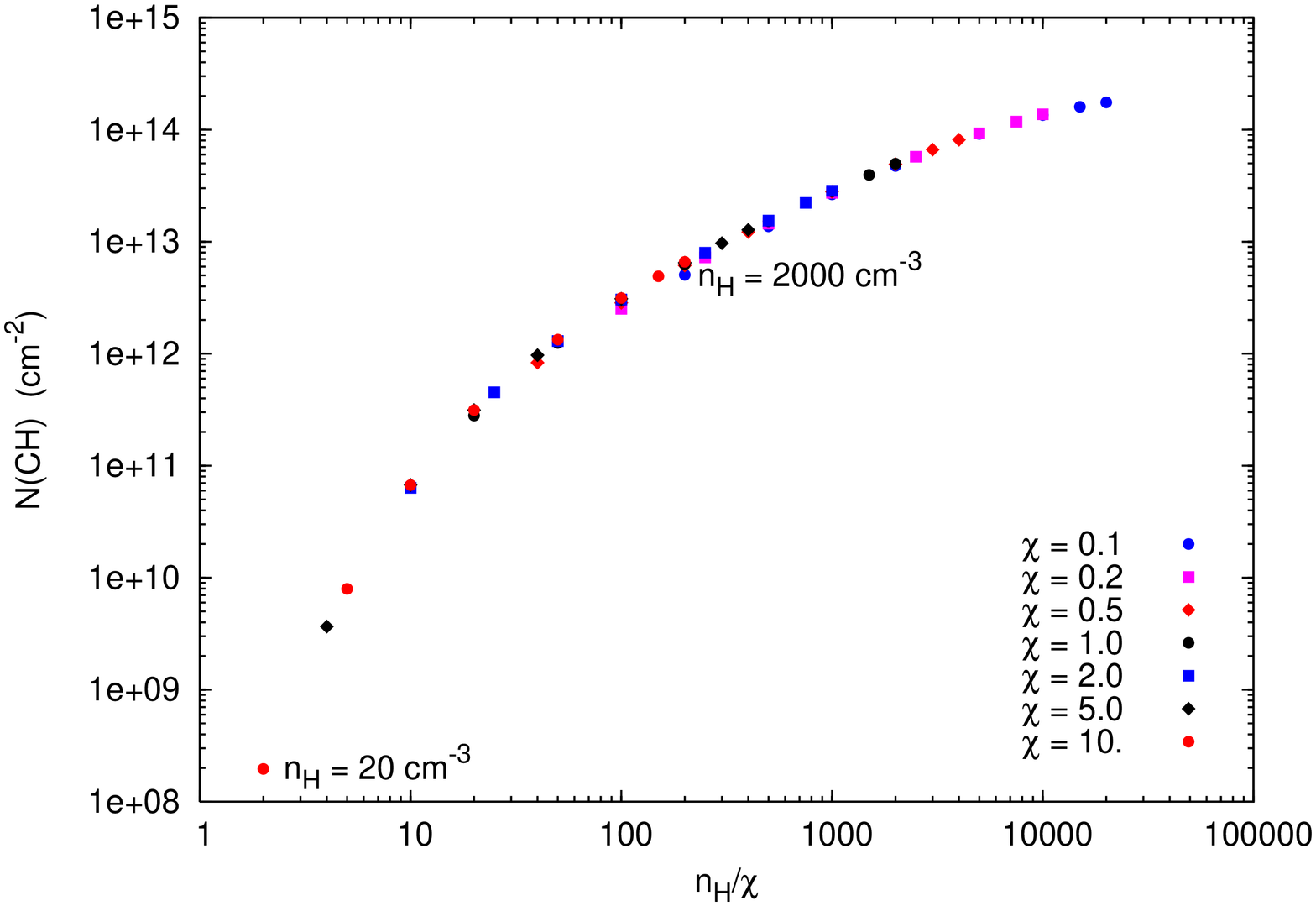}

\caption{Column densities of ${\rm CH}$
as a function of $n_{\rm{H}}/\chi$.\label{Fig_CHvsnhc_b}}
\end{figure}

\begin{figure}
\includegraphics[angle=270,scale=0.65]{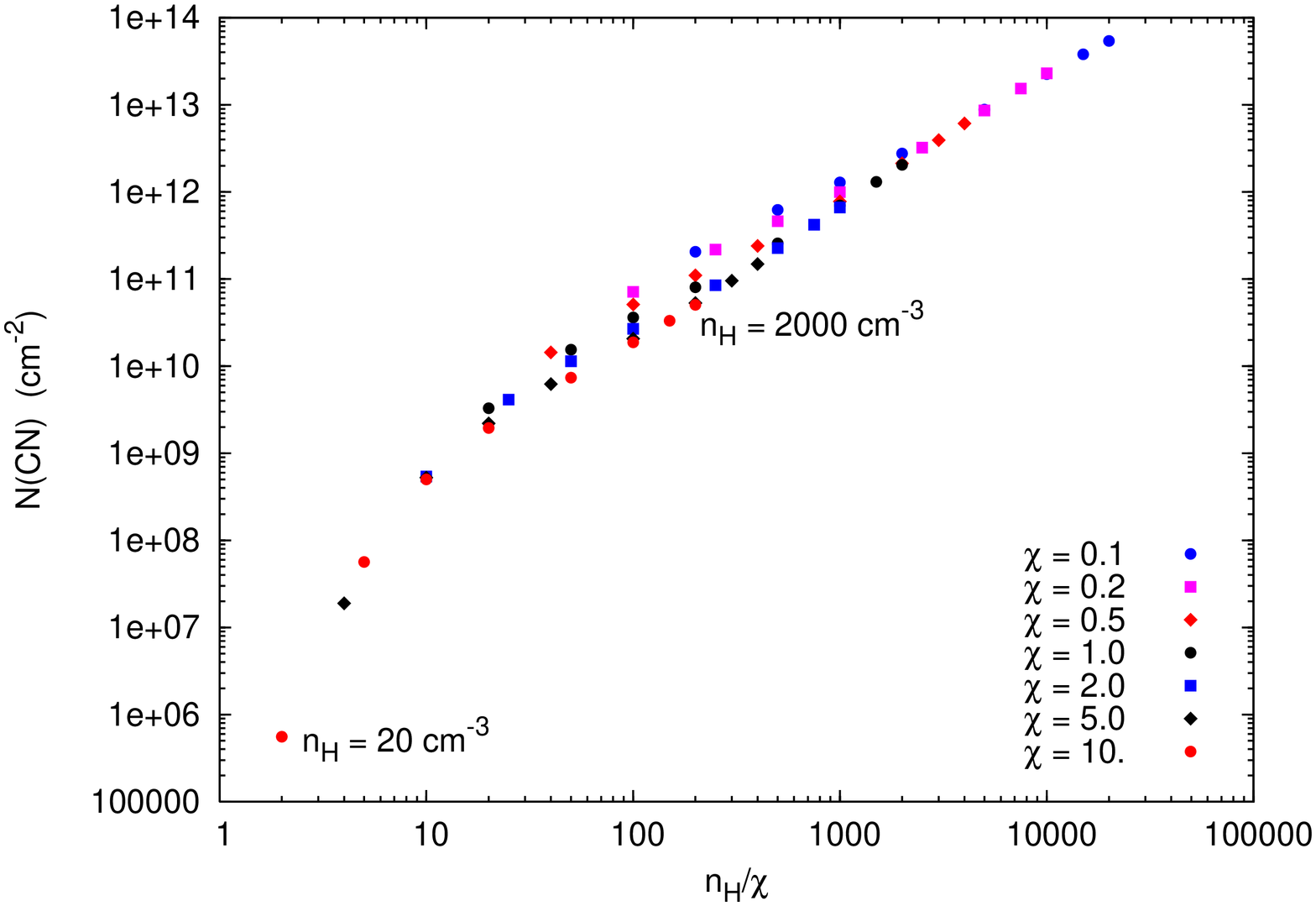}

\caption{Column densities of 
${\rm CN}$ as a function of $n_{\rm{H}}/\chi$.\label{Fig_CHvsnhc_c}}
\end{figure}

\begin{figure}
\includegraphics[angle=270,scale=0.65]{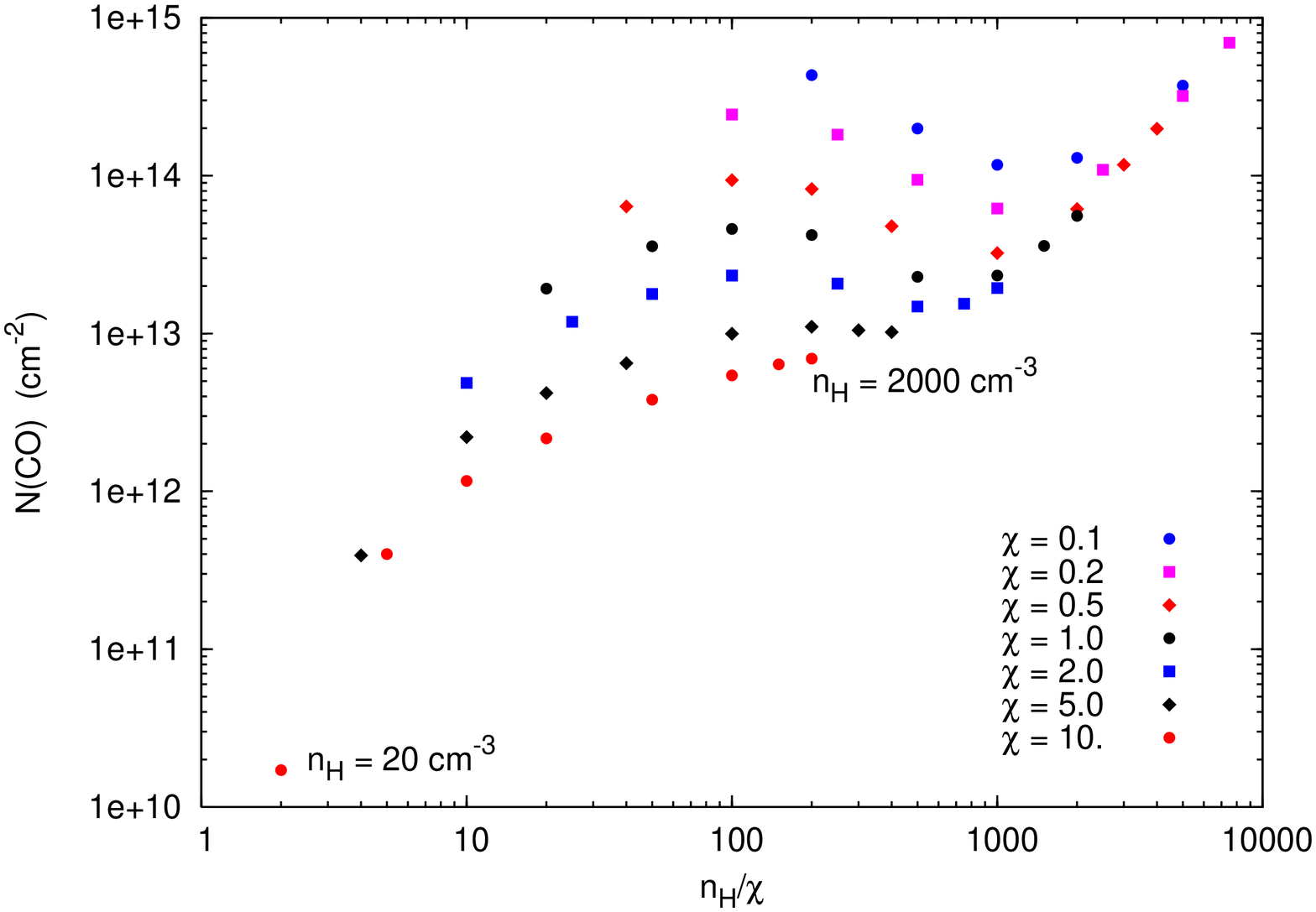}

\caption{Column densities of ${\rm CO}$ as a function of
$n_{\rm{H}}/\chi$.
\label{Fig_COvsnhc_a}}
\end{figure}

\begin{figure}
\includegraphics[angle=270,scale=0.65]{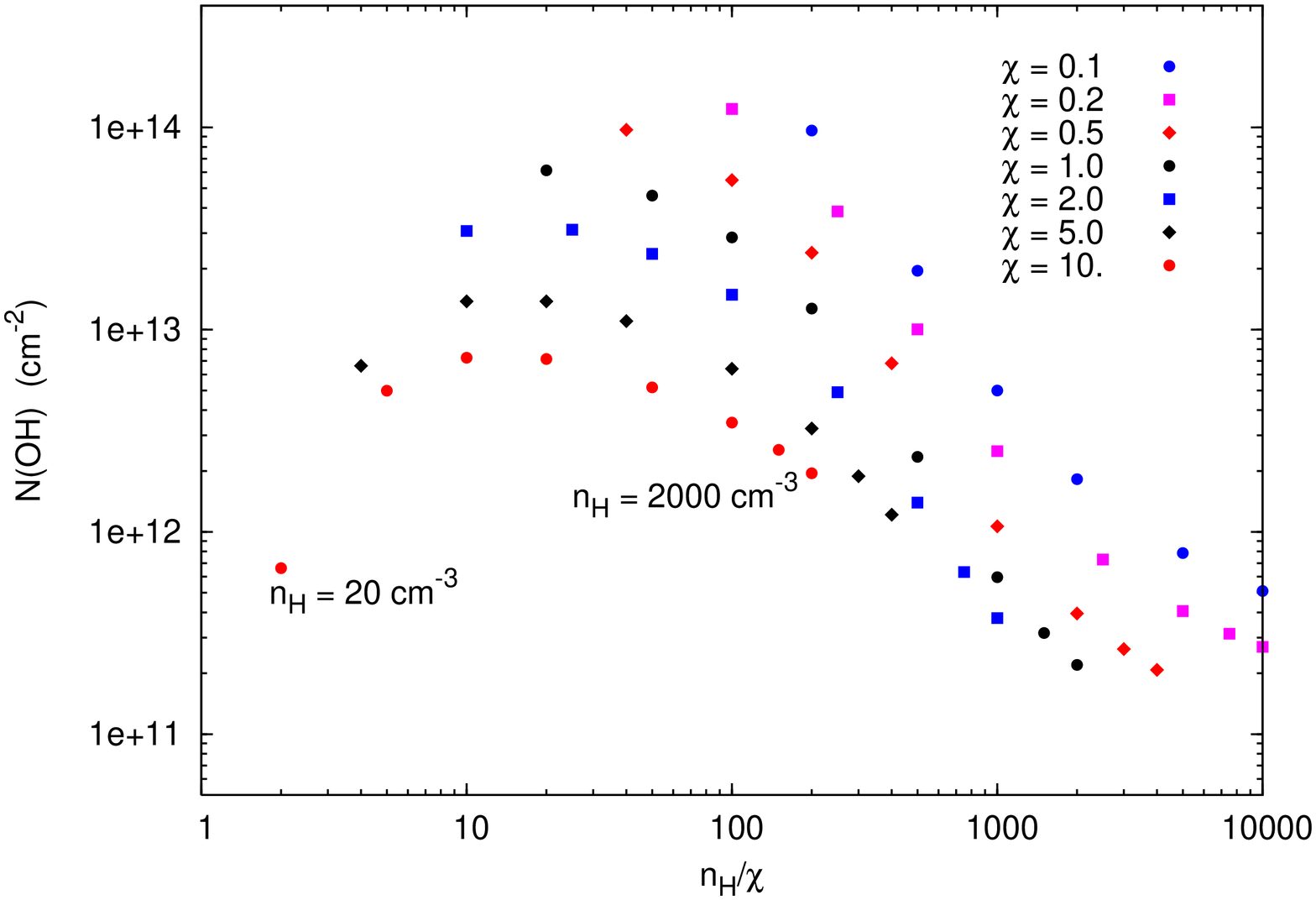}

\caption{Column densities of ${\rm OH}$ as a function of
$n_{\rm{H}}/\chi$.
\label{Fig_COvsnhc_b}}
\end{figure}

Fig.~\ref{Fig_CHvsnhc_a}, \ref{Fig_CHvsnhc_b}, \ref{Fig_CHvsnhc_c},
\ref{Fig_COvsnhc_a} and \ref{Fig_COvsnhc_b}
present the column densities of some observed species in diffuse clouds
as a function of $n_{\rm{H}}/\chi$. The column densities of
${\rm CH}$, ${\rm CN}$, C and $\rm{C}^{+}$ are directly controlled
by the ratio $n_{\rm{H}}/\chi$. This behaviour appears when
species are formed by two body reactions and are destroyed by photo-processes
(or the opposite as for $\rm{C}^{+}$). For species such as
${\rm CO}$, ${\rm OH}$ and ${\rm NH}$ there is no simple relationship
since other mechanisms, which do not involve two body reactions but
specific reactions such as cosmic ray ionisation (for ${\rm OH}$
and ${\rm NH}$), come into play . The case of ${\rm CO}$ is particular
and can be explained by its forming reaction. For models with $n_{\rm{H}}/\chi~\simeq~100$,
its formation occurs via ${\rm OH}+\rm{C}^{+}$. Whereas for
higher values ($n_{\rm H}/\chi\simeq10^{4}$) the reaction ${\rm O}+\rm{C}_{2}$
is the main route of formation of ${\rm CO}$. Thus the behaviour
of $N({\rm CO})$ vs $n_{\rm{H}}/\chi$ follows the one of ${\rm OH}$
or $\rm{C}_{2}$ depending on the value of $n_{\rm{H}}/\chi$.

\placefigure{Fig_CHvsnhc_a}

\placefigure{Fig_CHvsnhc_b}

\placefigure{Fig_CHvsnhc_c}

\placefigure{Fig_COvsnhc_a}

\placefigure{Fig_COvsnhc_b}

\subsection{UV radiative transfer in $\rm{H}_{2}$ lines}

\begin{figure}
\includegraphics[angle=270,scale=0.65]{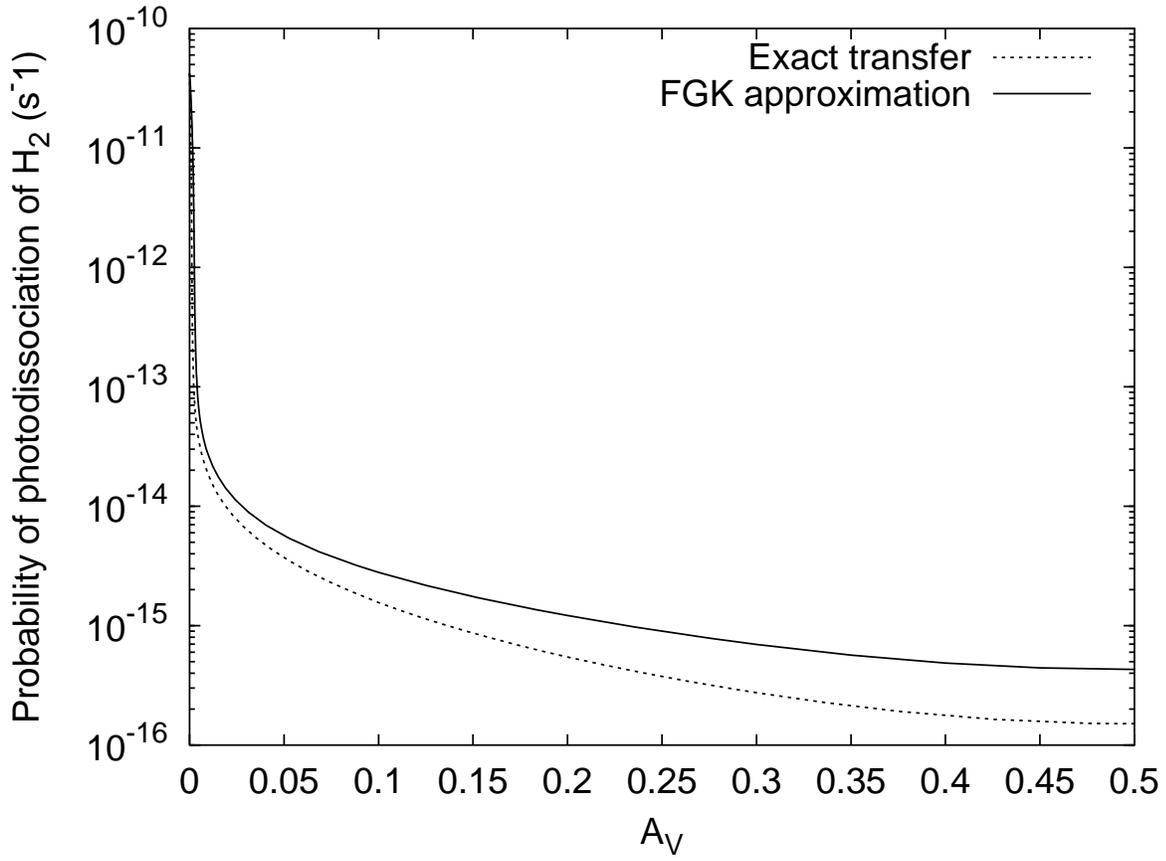}
\caption{Photodissociation probability of $\rm{H}_{2}$ for a model
($n_{\rm{H}}=100\,\,\rm{cm}^{-3}$, $\chi=1$, $A_{\rm{V}}=1$)
computed within the FGK approximation and the exact radiative transfer
in the $\rm{H}_{2}$ lines up to $J=5$.\label{Fig_phodih2}}
\end{figure}

\begin{figure}
\includegraphics[angle=270,scale=0.65]{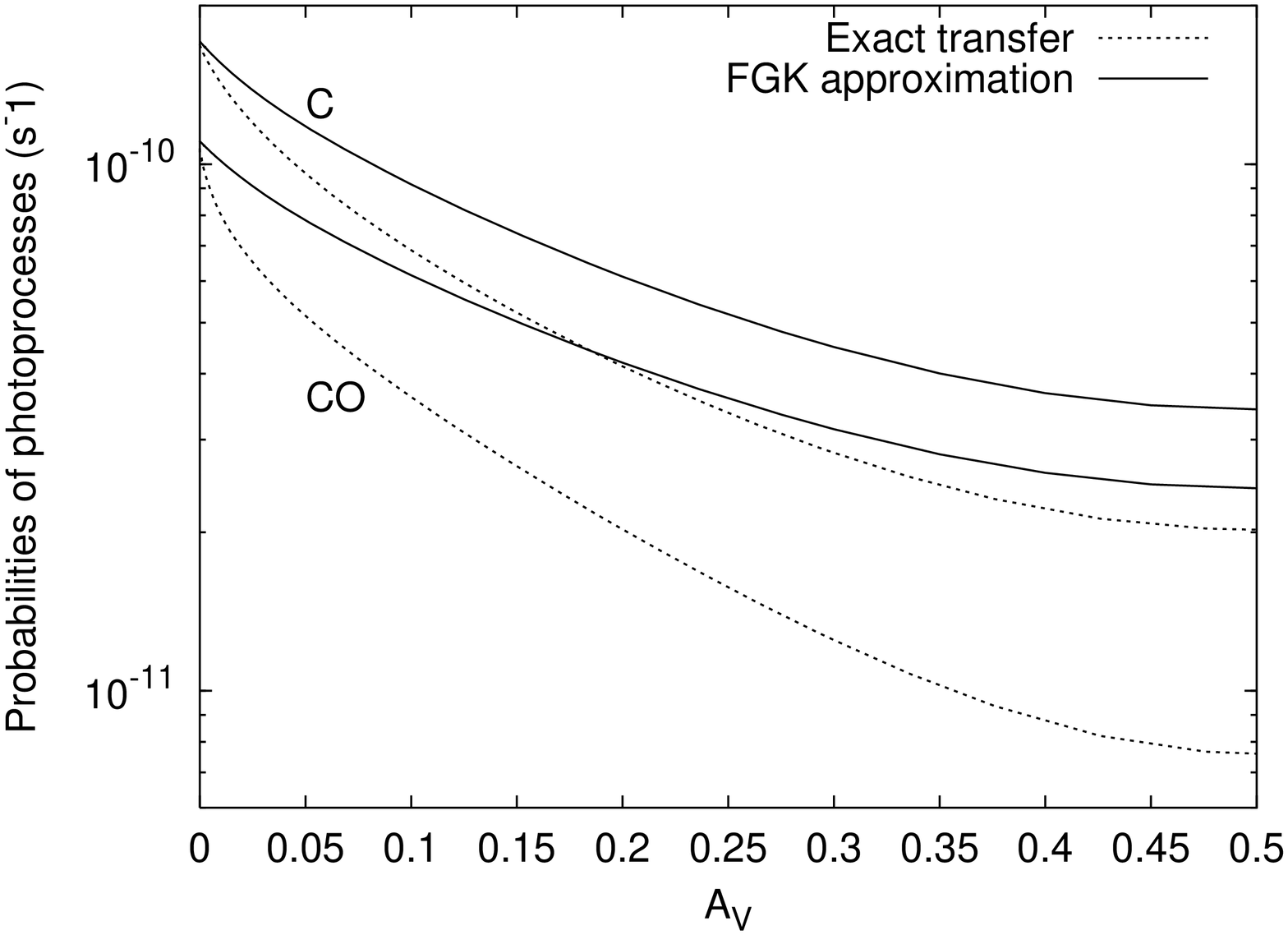}
\caption{Photodissociation and photoionisation probability of CO and C for
a model ($n_{\rm{H}}=100\,\rm{cm}^{-3}$, $\chi=1$, $A_{\rm{V}}=1$)
within the FGK approximation and the exact radiative transfer in the
$\rm{H}_{2}$ lines up to $J=5$.\label{Fig_phodico}}
\end{figure}

We checked the validity of the FGK approximation for the model $n_{\rm{H}}=100\,\rm{cm}^{-3}$
and $\chi=1,$ $A_{\rm V}=1$, by solving the full radiative transfer
as described in Sect.~\ref{Sec_transfer} up to $J=5$. The photodissociation
probabilities of $\rm{H}_{2}$, ${\rm HD}$, ${\rm CO}$ and
photoionisation probabilities of ${\rm C}$ and ${\rm S}$ are the
same at the edge of the cloud (Tab. \ref{Tab_proba_edge}) in the
two treatments. However, the values differ significantly at $A_{\rm V}$
greater than $0.05$ as displayed on Figs.~\ref{Fig_phodih2} and
\ref{Fig_phodico}.

Table \ref{Tab_transf} presents the column densities obtained in
the two cases. The column density of CO is increased by a factor of $2$
when proper account of the radiative transfer is performed. This produces
also a decrease of $N({\rm H})$ and an increase of $N(\rm{H}_{2})$
and $N({\rm C})$. For molecules (${\rm CH}$, ${\rm OH}$, etc.) whose
photodissociation rate is given by the formula $\gamma\,{\rm e}^{-\beta A_{\rm{V}}}$
(\citep{evD88}) column densities are obviously similar with the two
treatments.

Extending our photodestruction cross-sections database to compute
photorates by direct integration over the true radiation field is
under way.

\placetable{Tab_proba_edge}

\placetable{Tab_transf}

\placefigure{Fig_phodih2}

\placefigure{Fig_phodico}

\subsection{One side and Two sides models}

The numerical model allows the incident radiation field to come from
one side or from two sides of the plane parallel cloud. In this
latter case, the computing time is increased significantly in order
to converge. A two sides model is required when the radiation field
coming from the other side is still significant, i.e. for low value
of $A_{\rm{V}}$. We study this effect by comparing two models
($n_{\rm{H}}=100\,\rm{cm}^{-3}$, and $\chi=1$):
model 1 where the radiation field comes from one side and model 2
with the radiation field coming from both sides of a cloud with a
visual extinction $A_{\rm{V}}^{tot}$. In order to compare column
densities between the two models, in model 1, they are computed integrating
abundances up to $A_{\rm{V}}^{tot}/2$ and are then multiplied
by 2. 

Results (Table~\ref{Tab_deuxf}) show that differences are important
for $N({\rm H}_{2})$ only for low visual extinction clouds ($<0.5$).
However, for other molecules such as ${\rm CO}$, the differences
are significant at much higher $A_{\rm V}$ (we have a factor of
2 differencies in column densities at $A_{\rm V}=1$). For diffuse lines of sight,
using a 2 sided model is mandatory.

\placetable{Tab_deuxf}

\section{Dense PDRs\label{Sec_Dense}}

In this section, we consider dense clouds in a young stellar environment.
We present a grid of models with radiation field coming from one side
to model a dark cloud with high extinction illuminated by a young
star. The total visual extinction is chosen as 20. Adopted densities
are $n{}_{\rm H}=$ $10^{3}$, $3\times10^{3}$, $7\times10^{3}$,
$10^{4}$, $3\times10^{4}$, $7\times10^{4}$ , $10^{5}$, $3\times10^{5}$,
$7\times10^{5}$, $10^{6}$, $3\times10^{6}$, $7\times10^{6}$ and
$10^{7}$ $\rm{cm}^{-3}$. The adopted multiplicative factors
of the Draine radiation field are $\chi=$ $10^{3}$, $3\times10^{3}$,
$7\times10^{3}$, $10^{4}$, $3\times10^{4}$, $7\times10^{4}$ ,
$10^{5}$, $3\times10^{5}$, $7\times10^{5}$, $10^{6}$, $3\times10^{6}$,
$7\times10^{6}$ and $10^{7}$. These parameters correspond to a wide
range of physical conditions. Other parameters are given in Table~\ref{Tab_paramdiff}.

\subsection{Temperature at the edge of the clouds}

\begin{figure}
\includegraphics[angle=270,scale=1.0]{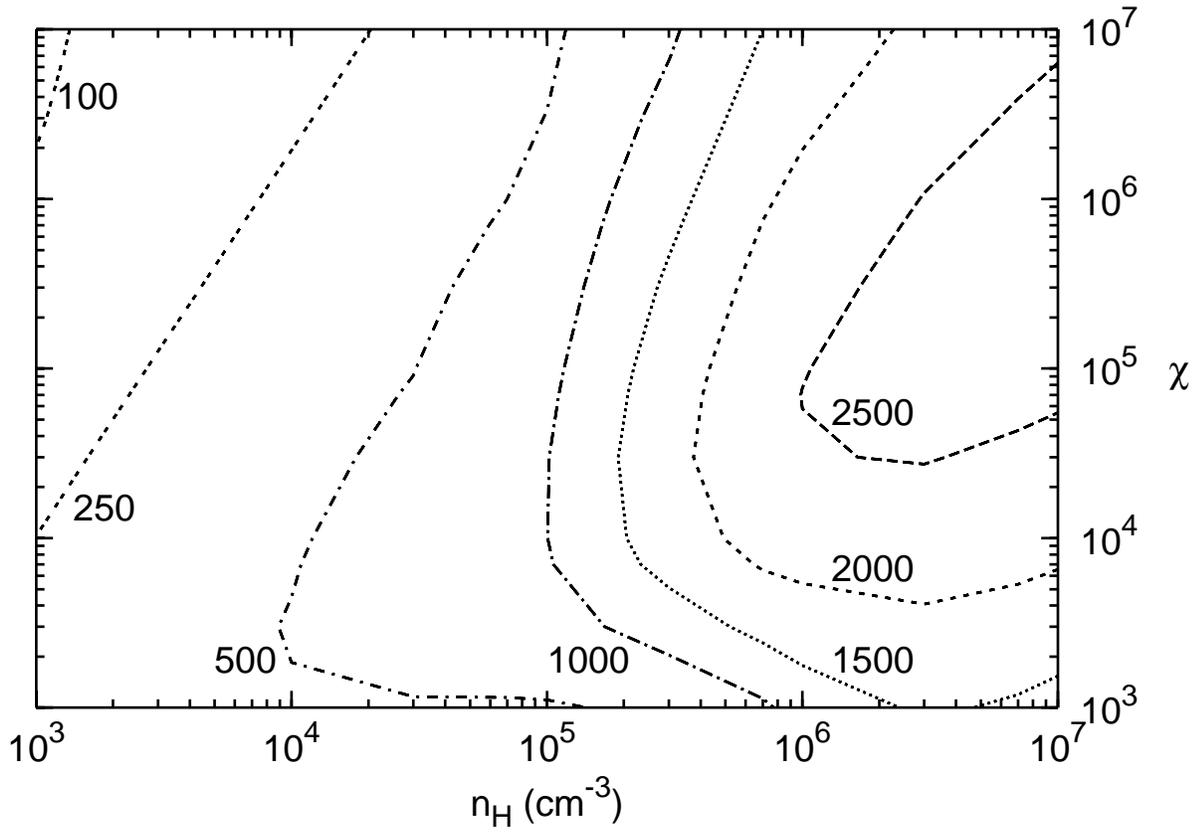}
\caption{Temperature of the gas at the edge of the PDR as a function of $n{}_{\rm H}$
and $\chi$. Values next to isocontours are temperatures in Kelvin.
\label{Fig_Tbord_gPDR}}
\end{figure}

\begin{figure}
\includegraphics[angle=270,scale=0.65]{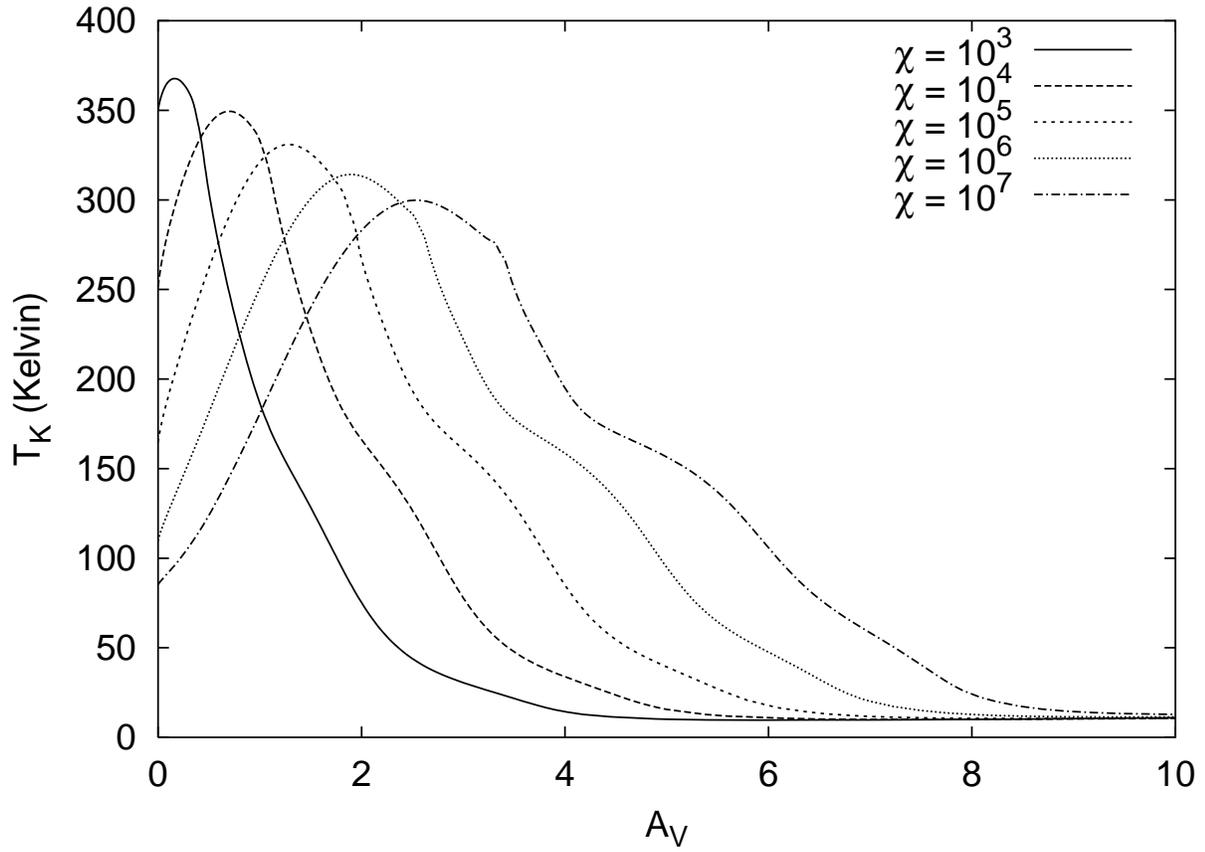}
\caption{Temperature profiles as a function of the visual extinction for models
of density $1000\,\rm{cm}^{-3}$ and different values of $\chi$.
\label{Fig_TvsAvnH3}}
\end{figure}

\begin{figure}
\includegraphics[angle=270,scale=0.65]{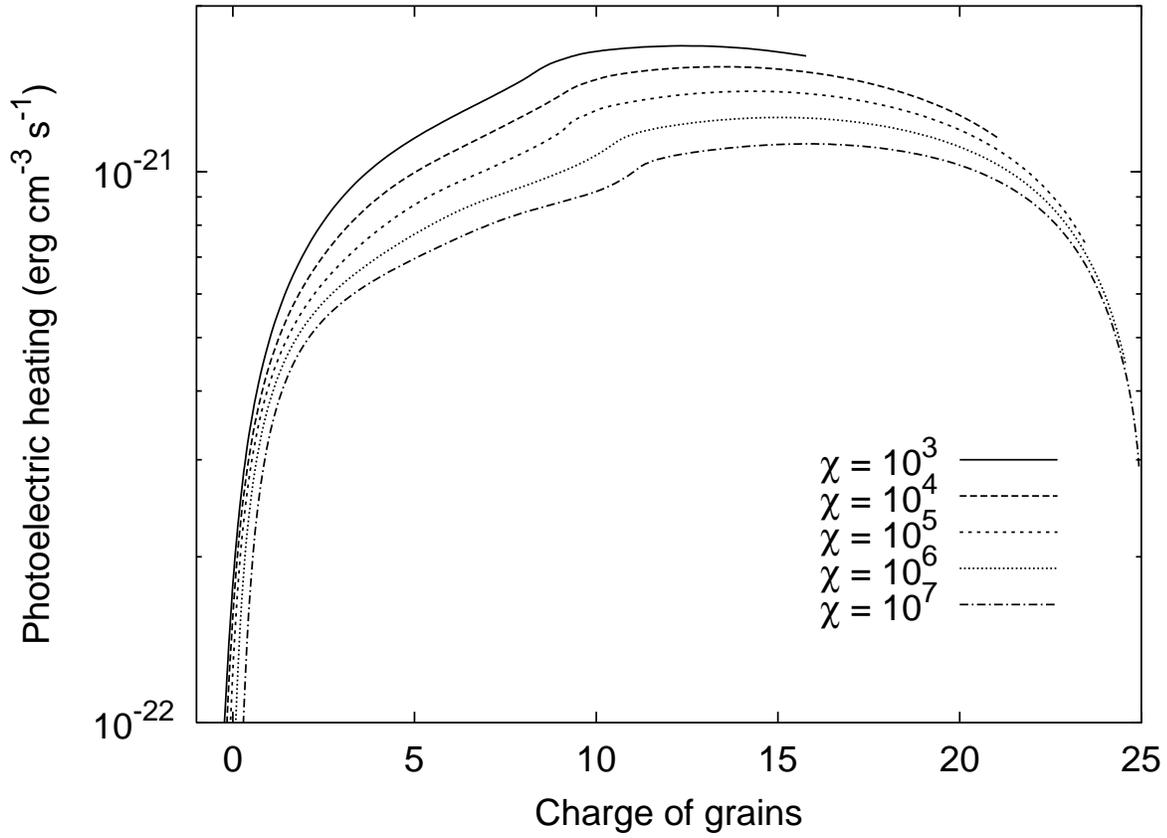}
\caption{Photoelectric heating (${\rm erg}\,{\rm cm}^{-3}\,{\rm s}^{-1}$)
as a function of the charge of grains for models with $n_{\rm{H}}=1000\,{\rm cm}^{-3}$
and various values of $\chi$. Grains considered here are the ones
with the smallest radius. The edge of the cloud is on the right of
the figure where the charge is maximal. \label{Fig_PE_charge}}
\end{figure}

At the edge of the cloud, the main heating process is the photoelectric
effect on dust. The kinetic temperature at the edge of the cloud obtained
in the grid of models is presented in Fig. \ref{Fig_Tbord_gPDR} and
varies between $100$ and $2500\,{\rm K}$. It is often used as an
indicator of the excitation of the medium. However, as shown in Fig.
\ref{Fig_TvsAvnH3}, the temperature maximum is not located at the
edge of a cloud but rather a bit deepper. For highly illuminated clouds,
grain ionization can be so large at the edge that the photoelectric
effect is less efficient than deeper where the charge of the grains decreases
as displayed in Fig. \ref{Fig_PE_charge}.

\placefigure{Fig_Tbord_gPDR}

\placefigure{Fig_TvsAvnH3}

\placefigure{Fig_PE_charge}

\subsection{Line intensities}

We compute the intensities (in ${\rm ergs}\, {\rm cm}^{-2}\, {\rm s}^{-1}\,{\rm sr}^{-1}$)
of the cooling transitions assuming that the PDR is seen ``face
on''. The results for $\rm{C}^{+}$, ${\rm C}$, ${\rm O}$,
${\rm CO}$ are in good agreement with the results of \cite{KWHL99}
and are not reported here. We focus on the molecular hydrogen infrared
transitions over a wide range of densities and UV enhancement factors\footnote{The referee mentioned a forthcoming paper by  \cite{Kau06} in which mainly rotationnal transitions of $\rm{H}_{2}$ are displayed. In the one case in which there is an overlap over a limited range of parameters, the 1-0 S(1) line intensity, the agreement is quite good.}. 

\begin{figure}
\includegraphics[angle=270,scale=1.0]{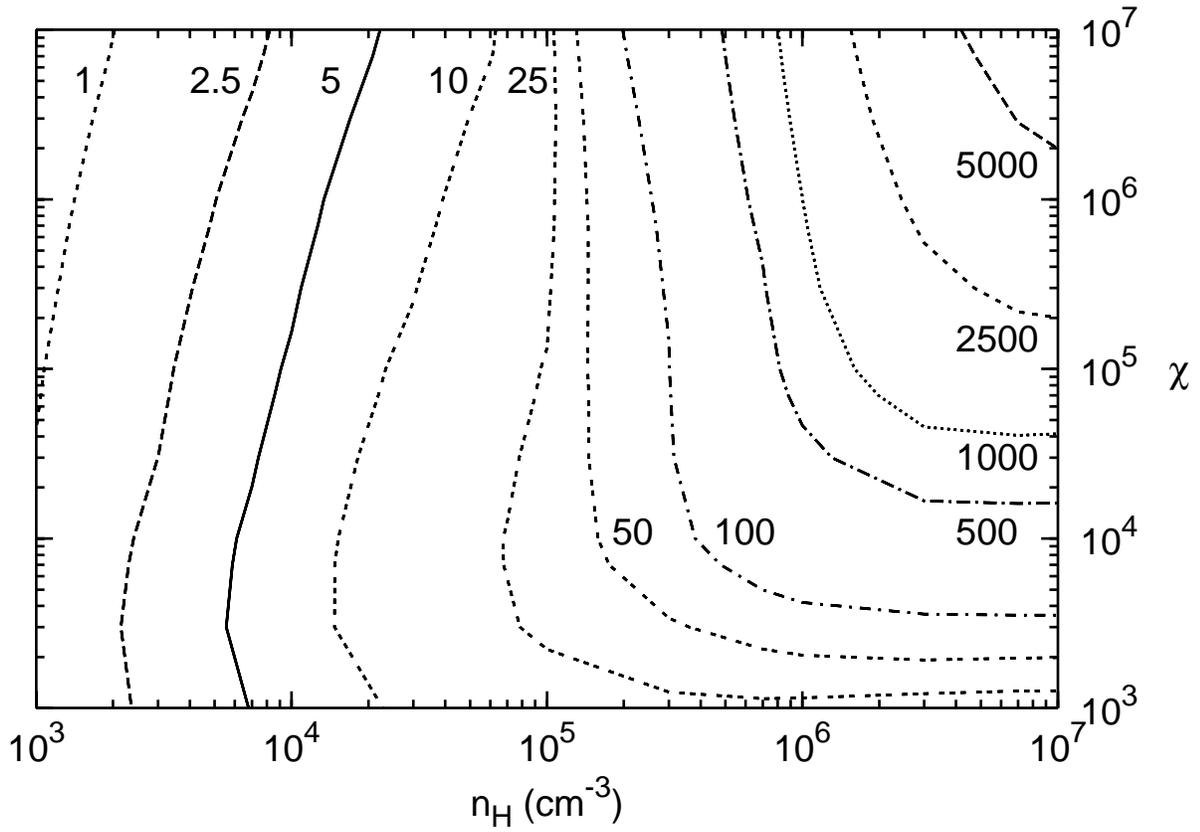}
\caption{Intensity of the 1-0 S(1) $\rm{H}_{2}$ transition in $10^{-6}\rm{erg}$
$\rm{s}^{-1}$ $\rm{cm}^{-2}$ $\rm{sr}^{-1}$ assuming
a face on geometry.\label{Fig_1-0S(1)}}
\end{figure}

The 1-0 S(1) transition at $2.12\,\mu{\rm m}$ has been widely studied
in a variety of environments and the corresponding intensity is displayed
in Fig.~\ref{Fig_1-0S(1)}.

\placefigure{Fig_1-0S(1)}

\begin{figure}
\includegraphics[angle=270,scale=1.0]{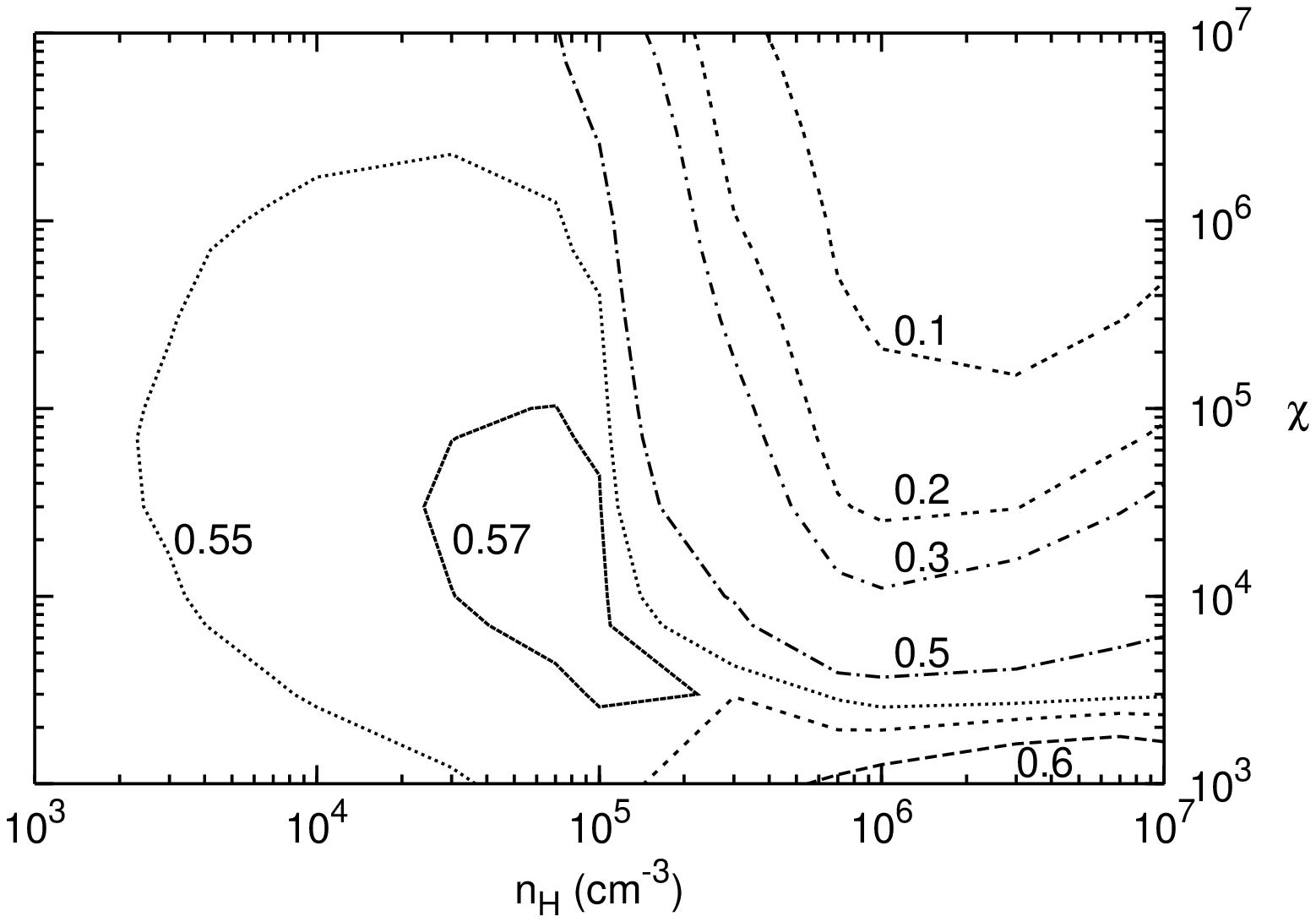}
\caption{Ratio of the intensities of the $\rm{H}_{2}$ lines 2-1 S(1)/1-0
S(1) given by the models. \label{Fig_2-1S(1)1-0S(1)}}
\end{figure}

It is often said that the ratio of the intensities of the ${\rm H}_{2}$
lines 2-1 S(1) on 1-0 S(1) is a good candidate to discriminate between
PDR and shocks: a value of $0.11$ is assumed to be characteristic
of shocks \citep{Kwan77} and a value of $0.54$ is appropriate for
radiative pumping \citep{BvD87}. Fig.~\ref{Fig_2-1S(1)1-0S(1)}
presents this ratio in the plane $n_{\rm{H}}-\chi$ obtained
with the grid. For $n_{\rm{H}}<10^{5}$ the ratio is almost
constant at a value close to $0.56$. The ratio is decreasing at higher
densities with increasing radiation fields and a ratio of 0.1 is obtained
for $n_{\rm{H}} > 10^{5}$ $\rm{cm}^{-3}$ and $\chi>10^{5}$.
So, if such parameters cannot be justified, other physical processes
such as shocks are required to explain such a ratio. 

\begin{figure}
\includegraphics[angle=270,scale=1.0]{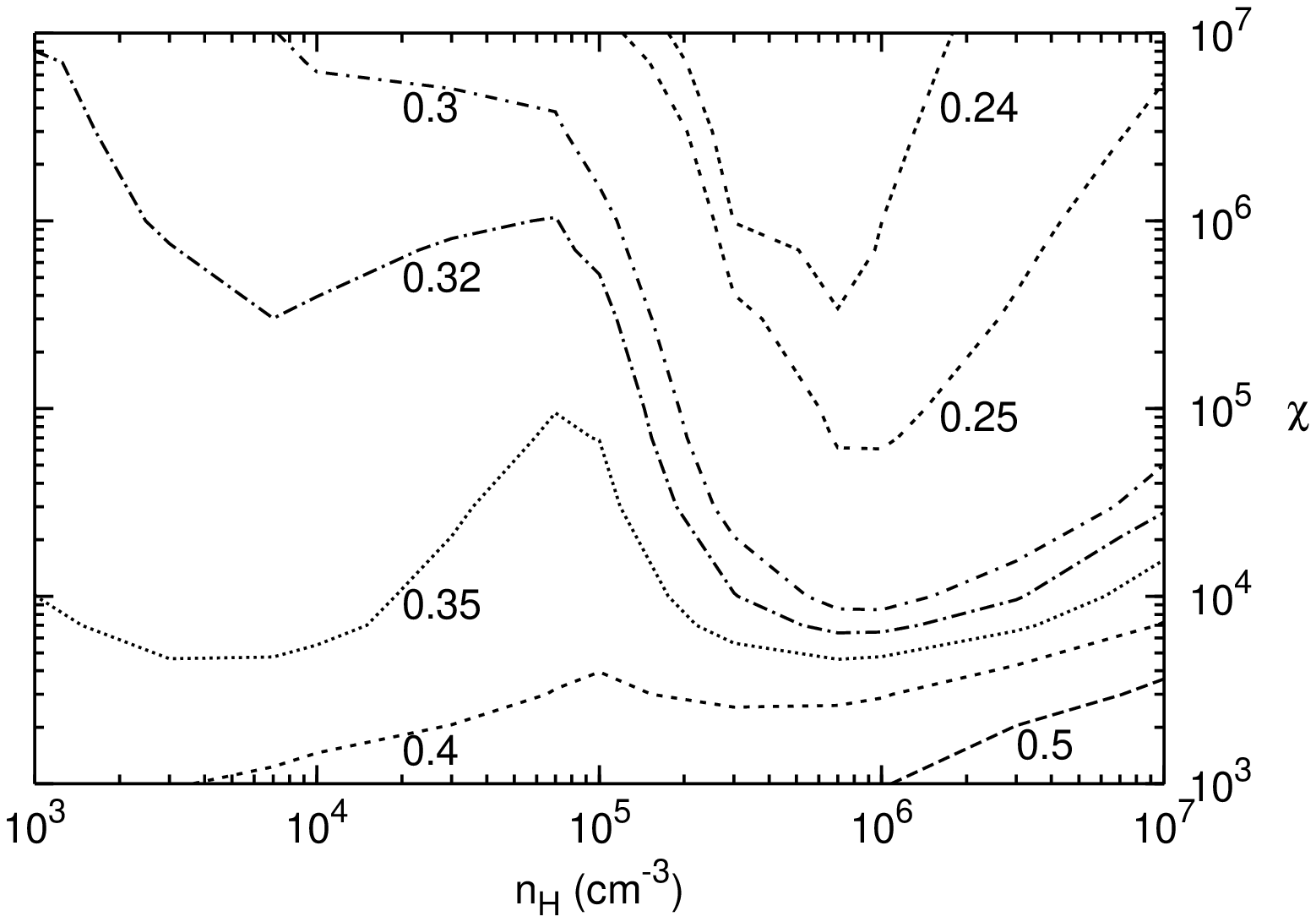}
\caption{Ratio of the intensities of the $\rm{H}_{2}$ lines 1-0 S(0)/1-0
S(1) given by the models.\label{Fig_1-0S(0)1-0S(1)}}
\end{figure}

\begin{figure}
\includegraphics[angle=270,scale=1.0]{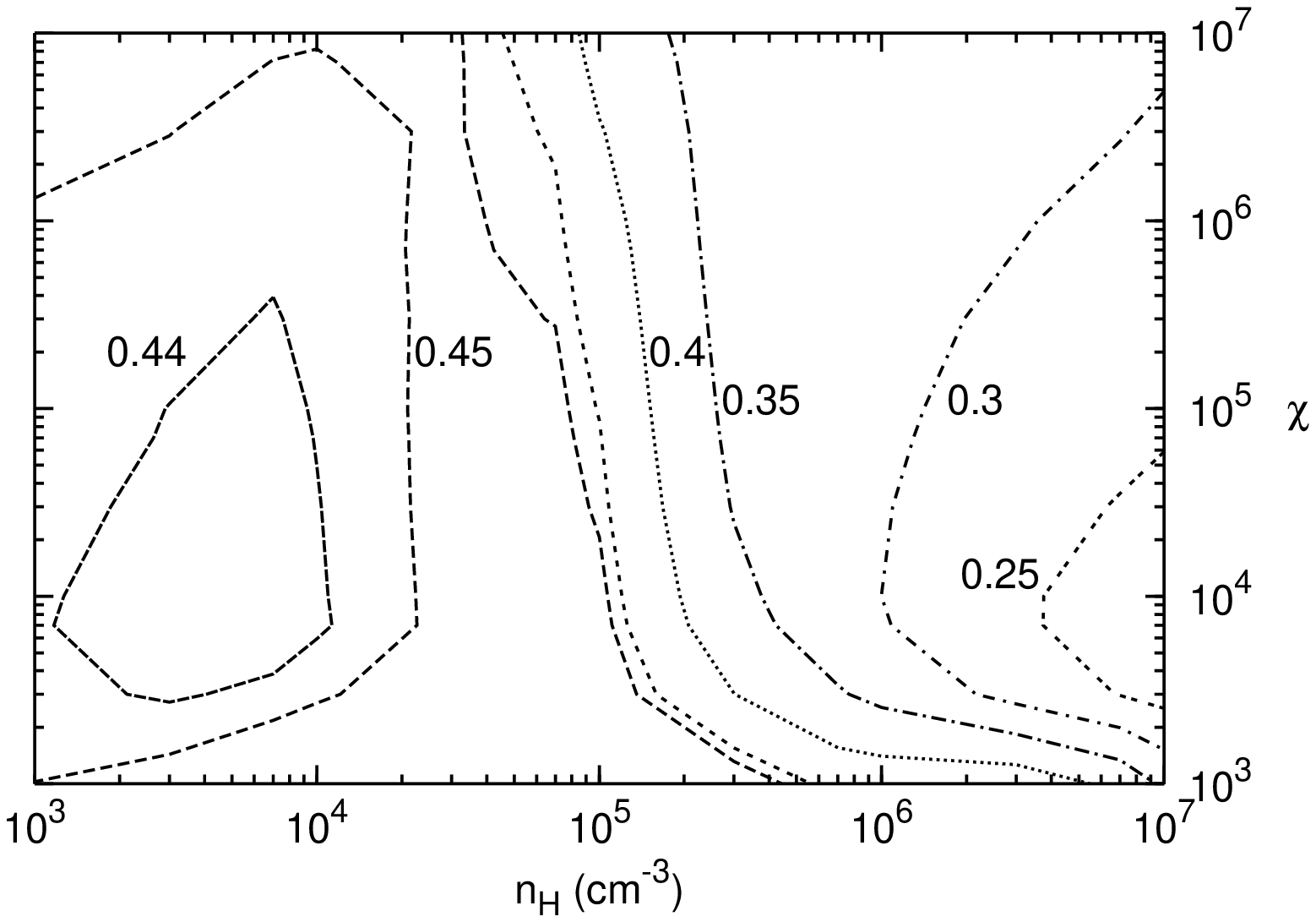}
\caption{Ratio of the intensities of the ${\rm H}_{2}$ lines 1-0 S(2)/1-0
S(1) given by the models.\label{Fig_1-0S(2)1-0S(1)}}
\end{figure}

\begin{figure}
\includegraphics[angle=270,scale=1.0]{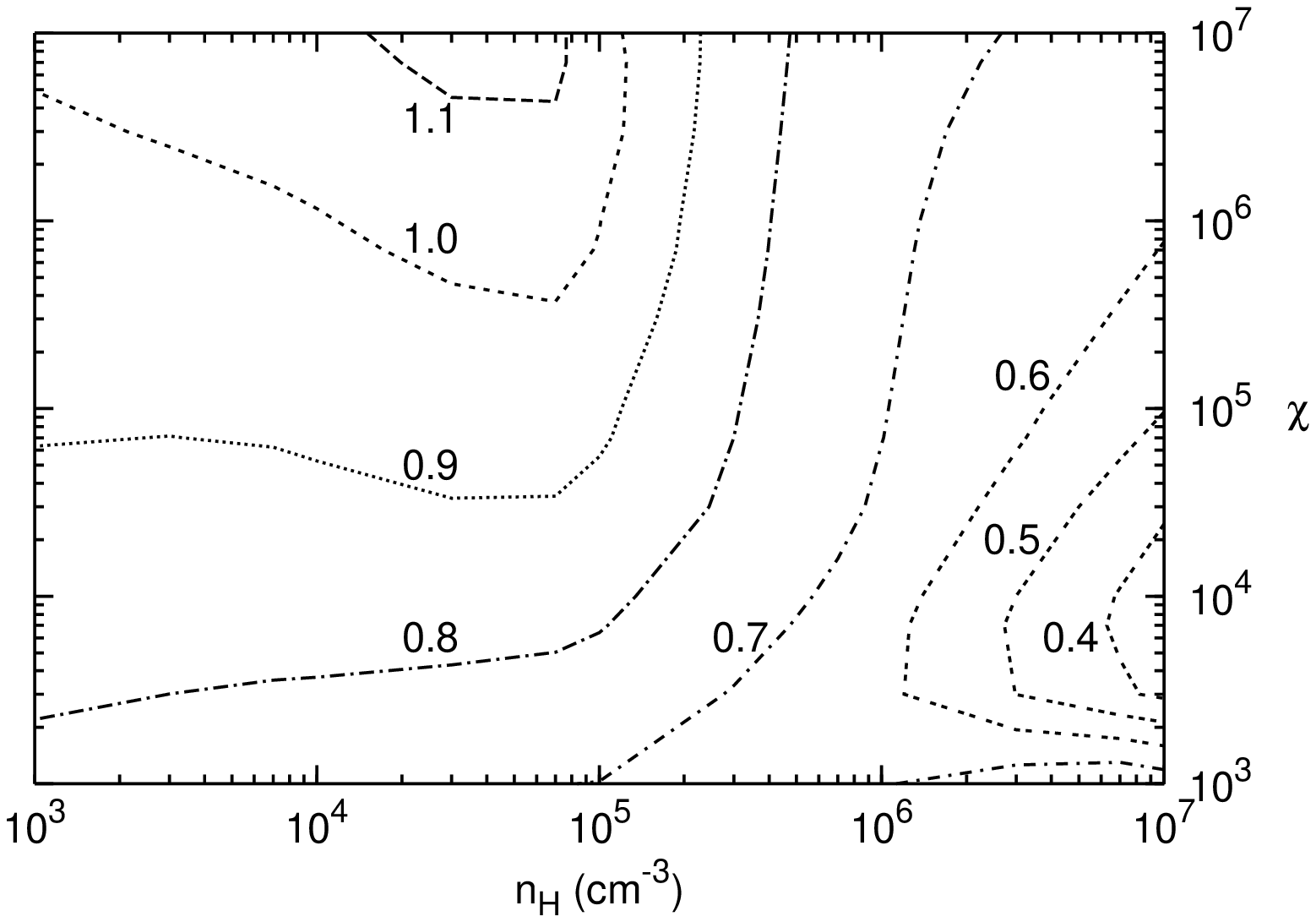}
\caption{Ratio of the intensities of the ${\rm H}_{2}$ lines 1-0 S(3)/1-0
S(1) given by the models. \label{Fig_1-0S(3)1-0S(1)}}
\end{figure}

\begin{figure}
\includegraphics[angle=270,scale=1.0]{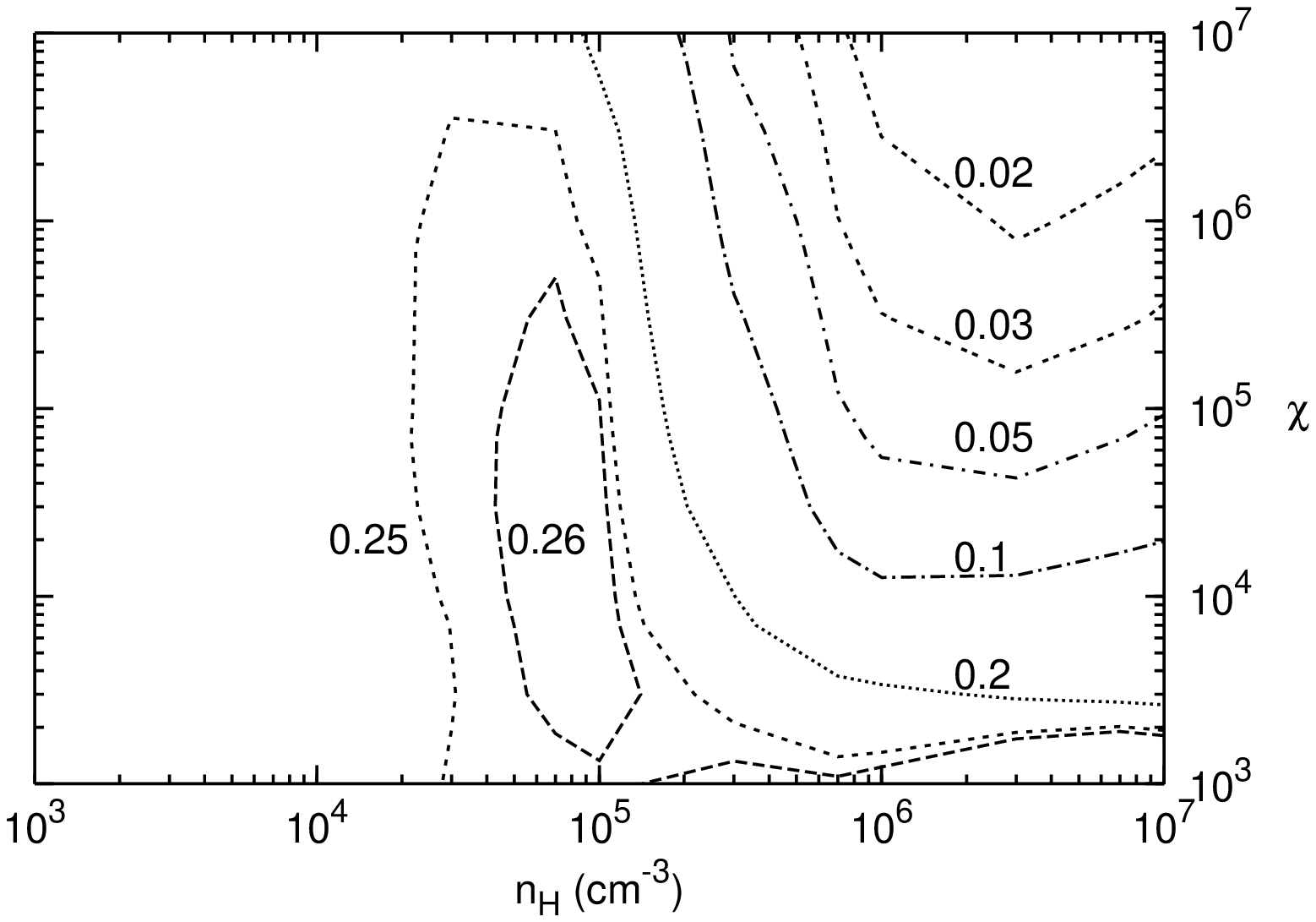}
\caption{Ratio of the intensities of the ${\rm H}_{2}$ lines 2-1 S(2)/1-0
S(1) given by the models.\label{Fig_2-1S(2)1-0S(1)}}
\end{figure}

\begin{figure}
\includegraphics[angle=270,scale=1.0]{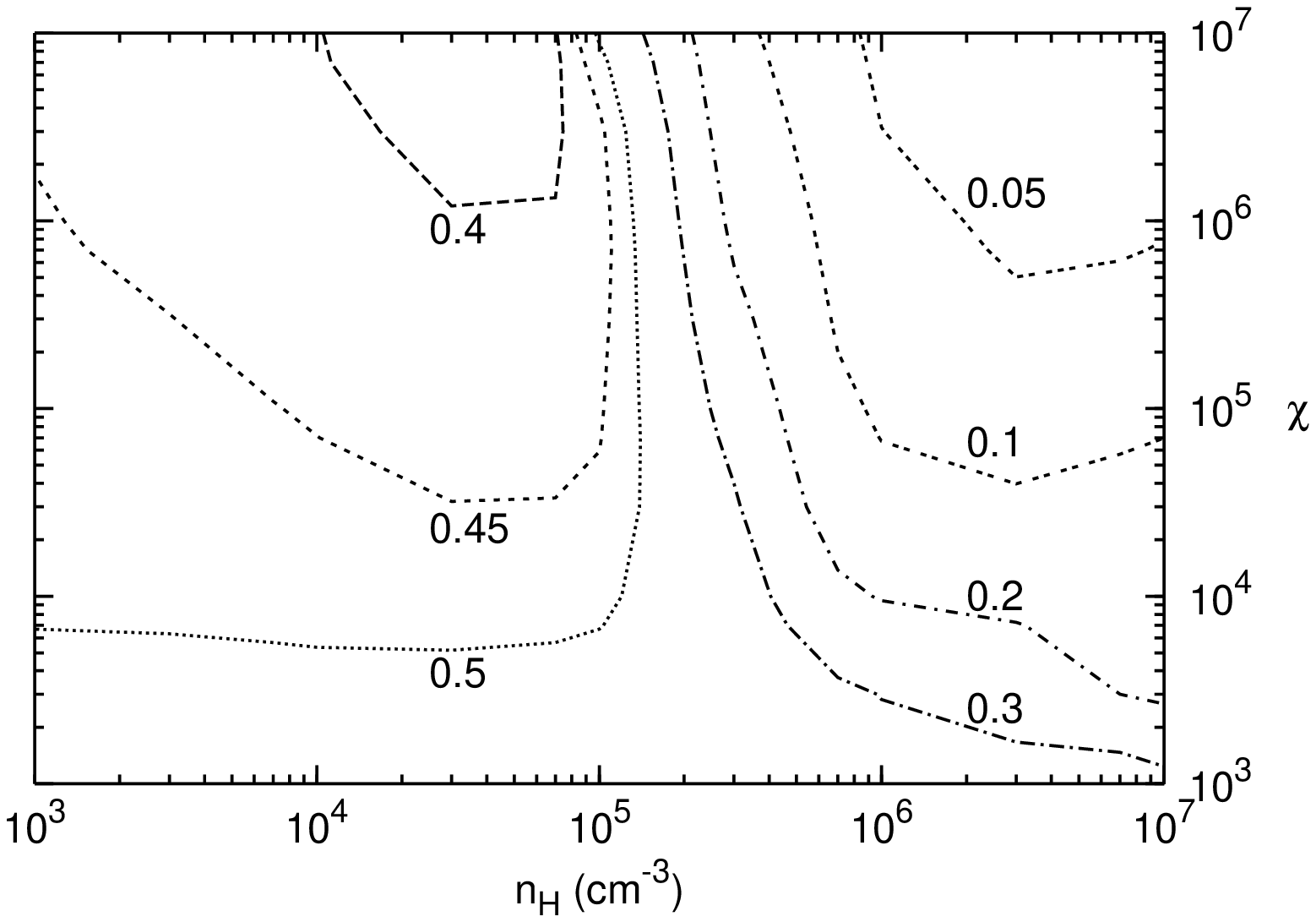}
\caption{Ratio of the intensities of the ${\rm H}_{2}$ lines 2-1 S(3)/1-0
S(1) given by the models. \label{Fig_2-1S(3)1-0S(1)}}
\end{figure}

Figs.~\ref{Fig_1-0S(0)1-0S(1)}, \ref{Fig_1-0S(2)1-0S(1)}, \ref{Fig_1-0S(3)1-0S(1)},
\ref{Fig_2-1S(2)1-0S(1)} and \ref{Fig_2-1S(3)1-0S(1)} present respectively
the ratio of the intensities of the 1-0 S(0), 1-0 S(2), 1-0 S(3),
2-1 S(2) and 2-1 S(3) $\rm{H}_{2}$ lines to the 1-0 S(1) $\rm{H}_{2}$
line which may be detected in good observing sites. These ratios are
almost always smaller than 1. The sensitivity of the ratios to density
is moderate below $10^{5}\rm{cm}^{-3}$ but more significant
at higher densities. The 1-0 S(2) / 1-0 S(1) and 1-0 S(3) / 1-0 S(1)
ratios could be used for high density determinations. On the other
hand, these ratios are highly dependent on $\chi$ and the observation
of several of these transitions can lead to a good guess of the intensity
of the incident radiation field. Ratios involving para and ortho transitions
are different for each case.

\placefigure{Fig_2-1S(1)1-0S(1)}

\placefigure{Fig_1-0S(0)1-0S(1)}

\placefigure{Fig_1-0S(2)1-0S(1)}

\placefigure{Fig_1-0S(3)1-0S(1)}

\placefigure{Fig_2-1S(2)1-0S(1)}

\placefigure{Fig_2-1S(3)1-0S(1)}

\subsection{Ratios of antenna temperatures}

\begin{figure}
\includegraphics[angle=270,scale=1.0]{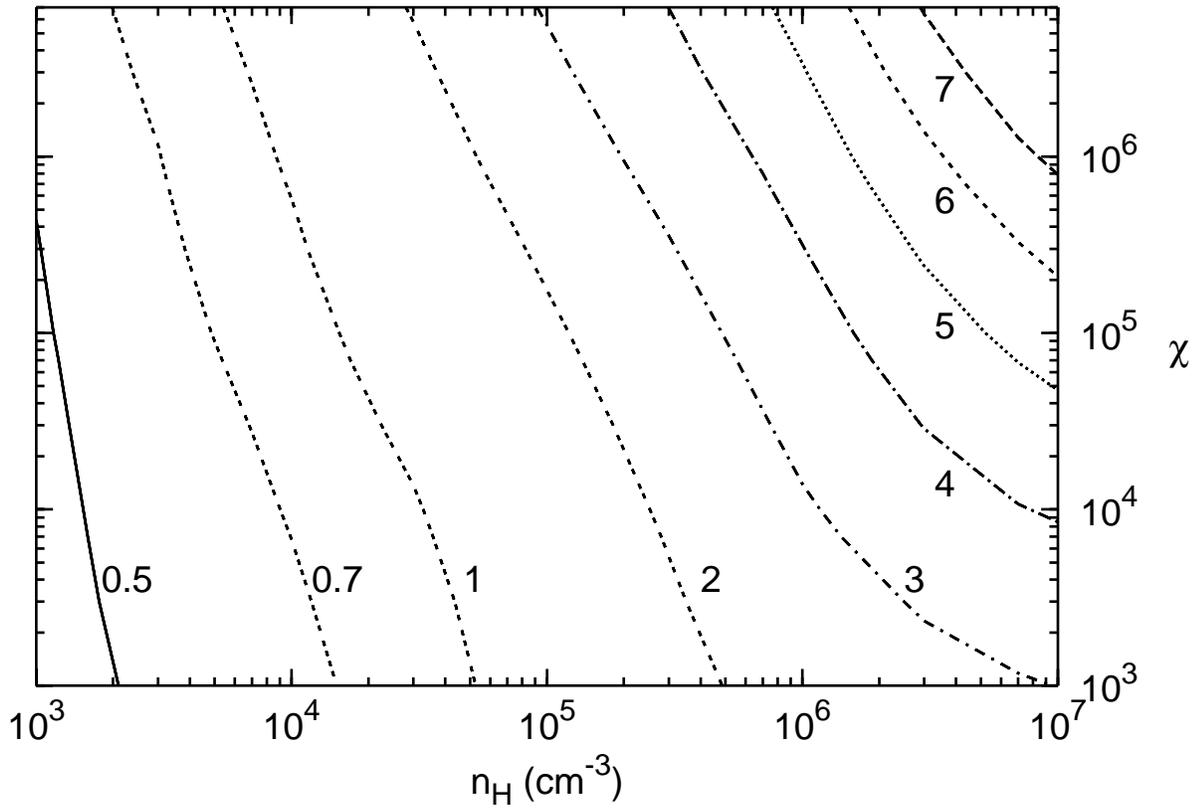}
\caption{Intensity of ${\rm CO}$ (2-1) in $10^{-6}\,{\rm erg}\,{\rm cm}^{-2}\,{\rm s}^{-1}\,{\rm sr}^{-1}$
as a function of $n_{\rm{H}}$ and $\chi$. \label{Fig_Tant_2-1}}
\end{figure}

\begin{figure}
\includegraphics[angle=270,scale=1.0]{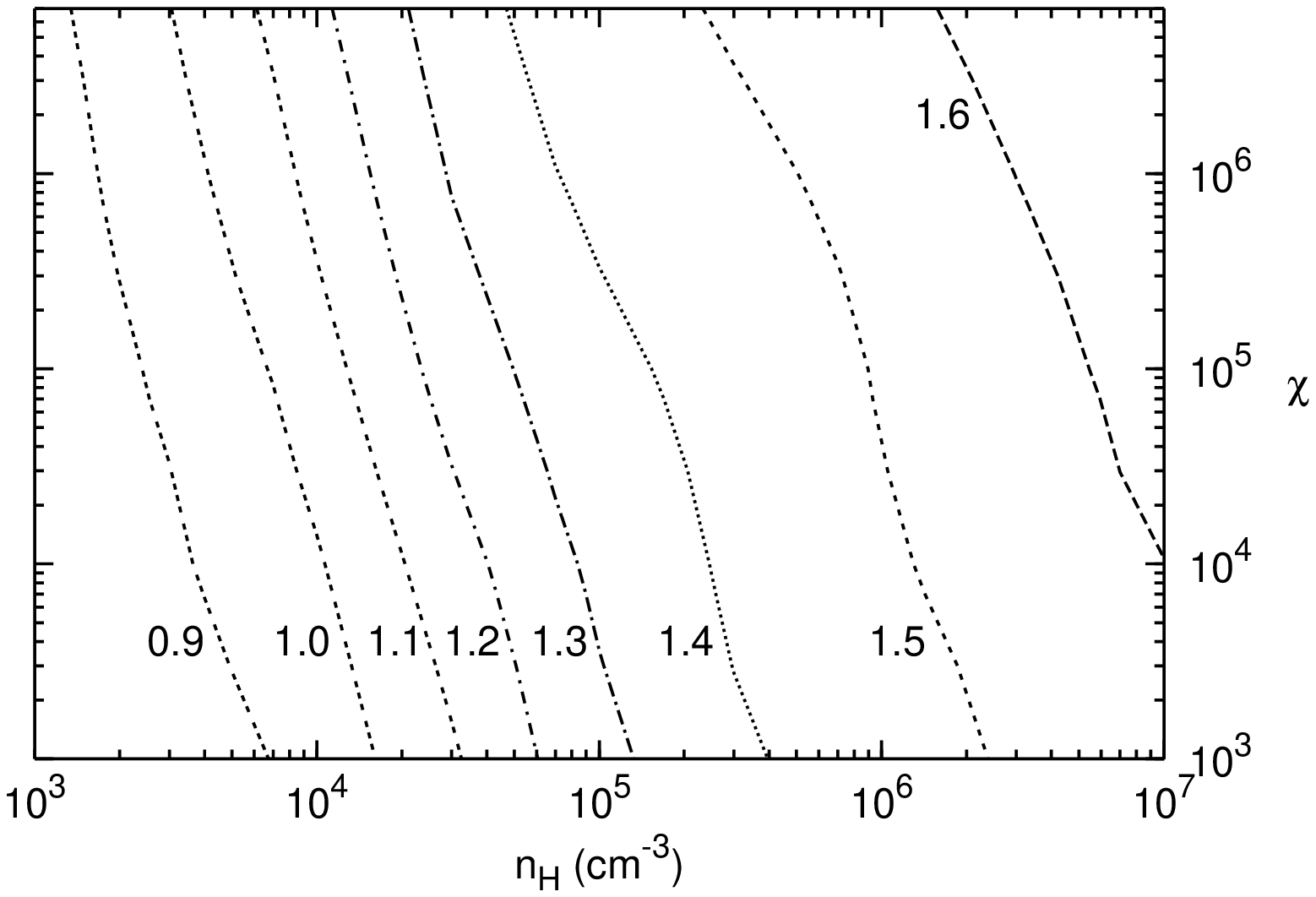}
\caption{Ratios of antenna temperatures of ${\rm CO}$ 3-2 to 2-1 transition
as a function of $n_{\rm{H}}$ and $\chi$. \label{Fig_Tant_3-2s2-1}}
\end{figure}

\begin{figure}
\includegraphics[angle=270,scale=1.0]{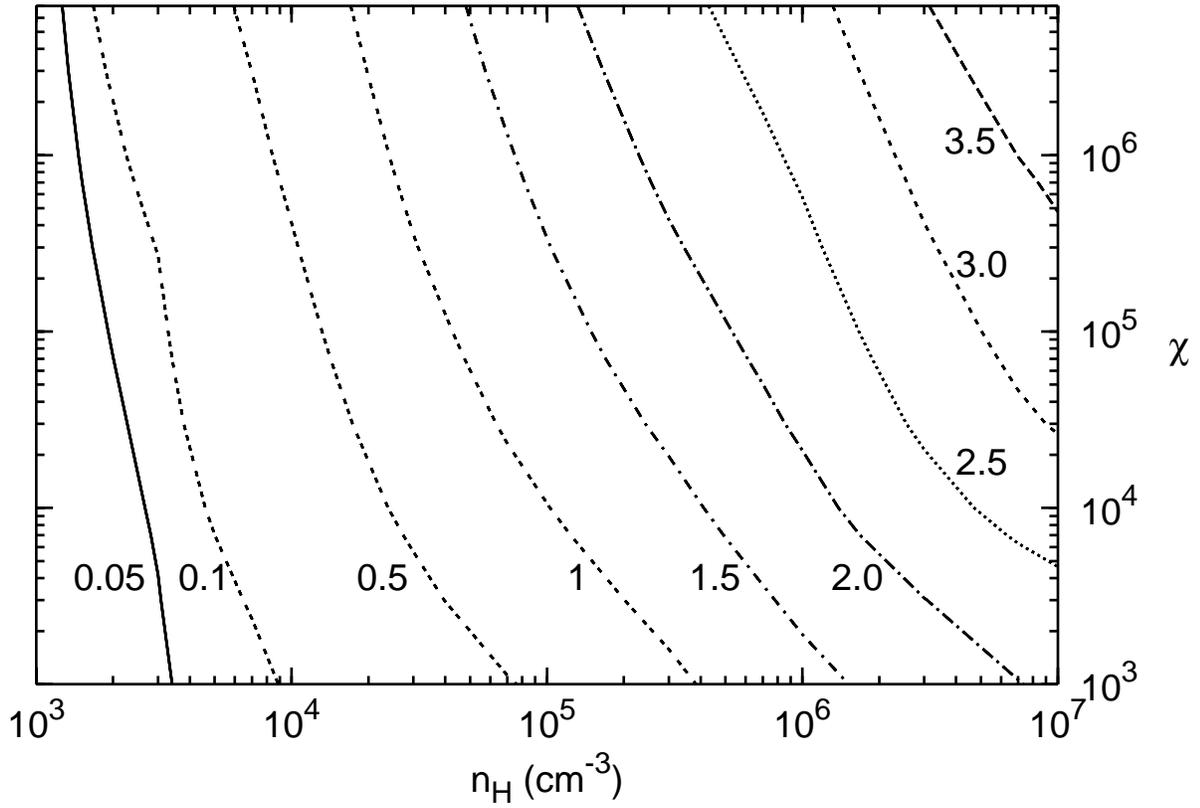}
\caption{Ratios of antenna temperatures of ${\rm CO}$ 6-5 to 2-1 transition
as a function of $n_{\rm{H}}$ and $\chi$. \label{Fig_Tant_6-5s2-1}}
\end{figure}

${\rm CO}$ is often used as a tracer of H$_{2}$ and deserves special
discussion. We display the isocontours of the intensity of the ${\rm CO}$
(2-1) transition in Fig.~\ref{Fig_Tant_2-1} in
the $n_{\rm{H}}$ -$\ \chi$ plane. We checked
that the results are in agreement with \cite{KWHL99}. We also display
the ratio of the antenna temperatures of 3-2 and 6-5 transitions to
the 2-1 transition in Fig. \ref{Fig_Tant_3-2s2-1} and \ref{Fig_Tant_6-5s2-1}.

The ratios of the antenna temperatures ($T_{\rm A}$)
are defined as (for example in the case of $T_{\rm A}^{2-1}/T_{\rm A}^{1-0}$)
, using the Rayleigh-Jeans approximation:

\[
\frac{T_{\rm A}^{2-1}}{T_{\rm A}^{1-0}}=\left(\frac{\nu_{1-0}}{\nu_{2-1}}\right)^{2}\times\frac{I_{2-1}}{I_{1-0}}\]

where $\nu_{i-j}$ is the frequency of the $i\rightarrow j$ transition
and $I_{i-j}$ is the intensity of this same transition.

In the hypothesis of an homogeneous and optically
thin medium, the ratio of the populations is given by :

\[
\frac{n_{2}}{n_{1}}\simeq\frac{T_{\rm A}^{2-1}}{T_{\rm A}^{1-0}}\quad\frac{A_{1-0}}{A_{2-1}}\quad\frac{\nu_{2-1}}{\nu_{1-0}}\]

\placefigure{Fig_Tant_2-1}

\placefigure{Fig_Tant_3-2s2-1}

\placefigure{Fig_Tant_6-5s2-1}

Whereas the $\rm{T}_{\rm A}^{3-2}$ to $\rm{T}_{\rm A}^{2-1}$
ratio is only slightly dependent on the density and illumination conditions,
much larger variations are obtained for the highly excited (6-5) to
(2-1) antenna temperature ratios. Thus highly excited ${\rm CO}$
transitions are valuable tests of dense PDR conditions.

\section{Conclusion\label{Sec_conclusion}}
We have presented and discussed the new features implemented in the Meudon PDR code first described in Le Bourlot et al. (1993).
The principal highlights are:
\begin{itemize}
\item Treatment of the UV radiative transfer : a 2 sides illumination of the gas slab is included as well as the possibility of an additional UV source impinging perpendicularly to the surface of the "cloud". The treatment of the UV radiative transfer includes the effect of discrete transitions of H and H$_2$ in the computation of the radiative energy density. This possibility, which is computing time intensive, is optional and is tuned by the number of rotational levels of H$_2$ explicitly involved. The approximate treatment following the FGK approximation allows for rapid computation with a reasonable approximation to the main features of the UV radiative transfer involving self-shielding effects but neglecting discrete line overlaps between H, H$_2$ and CO. 
Photodissociation of H$_2$, HD, CO and its isotopes by summation over all relevant transitions, as well as photoionization of C and S from the integration of photoionization cross-sections over the UV radiative field intensity are then obtained.

\item Grain properties : Scattering properties of grains are explicitly introduced for carbon and silicate particles for sizes ranging from the nanometer to the millimeter region. Corresponding data are taken from \cite{WD01}. PAH properties are mimicked by the nanometric carbon aggregates data. The photoelectric effect as well as the grain charge distribution are then computed from the actual value of the UV radiation field without analytic formula. 

\item Chemical processes :  Chemical processes involving grain surfaces are included as a function of the grain size distribution. In this respect, H$_2$ formation results from subsequent adsorption of H on the grain surface and reaction between the two adsorbed atoms. Three body collisions are implemented for dense regions where two hydrogen atoms and a third body may contribute to chemical formation .

\item Thermal balance : An effort to homogenize the various cooling processes has been achieved and includes the excitation and subsequent emission of forbidden visible transitions of the main atomic and ionic constituents. 

\end{itemize}

Examples relevant both for diffuse and dense cloud conditions are presented. In the diffuse cloud conditions, we discuss possible chemical diagnostics as a function of $n_H / \chi$. We show that whereas the molecular fraction, fractional column densities of C$^+$, C, CH, CN scale smoothly with n$_H$ / $\chi$, no specific trend is found for CO and OH. We have displayed emissivity ratios of rovibrational transitions of H$_2$ in a parameter space n$_H$, $\chi$, relevant to dense PDRs.
These plots are a guidance for the interpretation of observations. However, we recommend our readers to rather run their own model with the proper parameters of the studied line of sight. Our code is indeed downloadable at http://aristote.obspm.fr/MIS/

Several improvements are still in progress. These concern the treatment of photodissociation processes with the inclusion of appropriate cross sections when available. This allows us to include the effect of grain extinction properties without any approximation. The role of infrared pumping is also under study together with a consistent treatment of the grain properties and gas phase elemental abundances. Finally, we also plan to consider the dependence of the grain size distribution on the space position within the cloud: there is increasing observational evidence that very small grains and PAH are abundant in the illuminated parts of the cloud whereas big grains are more likely present in the shielded regions. 
We nevertheless consider that the present PDR code is a useful tool to the scientific preparation of future spatial missions such as Herschel and Spitzer and look forward to incorporate it within the Virtual Observatories.

\appendix

\section{Cloud structure\label{App_cloud_struct}}

\begin{figure}
\epsscale{0.75}
\plotone{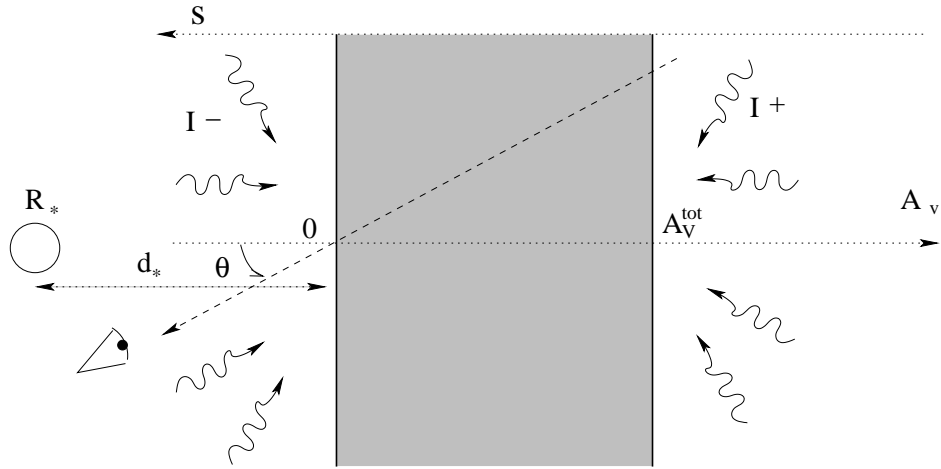}
\caption{Structure and geometry conventions. \label{Fig_struct}}
\end{figure}

The cloud is 1D, plane parallel, with sharp edges. It may be semi-infinite
or have a finite extent $A_{\rm{V}}^{tot}$ in magnitude, in
which case radiation comes from both sides. Geometry and sign convention
are displayed on Fig.~(\ref{Fig_struct}). The origin is at the left
edge of the cloud. The observer is always on the negative side. Integrated
line intensities are computed following an angle $\theta$ from the
normal of the surface of the cloud. We define $\chi^{+}$ and $\chi^{-}$
respectively as the enhancement factors of the Draine radiation field
on the positive and negative sides of the cloud. If $\chi^{+}=0$,
the cloud is semi-infinite and $A_{\rm{V}}^{\rm tot}$ is
the limit up to which the structure is computed. If $\chi^{+}\neq0$,
$A_{\rm{V}}^{\rm tot}$ is the cloud total visible extinction.
It is also possible to introduce an illuminating star at a distance
$d_{*}$ from the cloud. The star is described by its spectral type,
corresponding to an average star radius and blackbody temperature.
It may be either on the same side as the observer ($d_{*}<0$) or
the opposite side ($d_{*}>0$). No star is set by $d_{*}=0$.

\placefigure{Fig_struct}

The gas column density follows from the relations $C_{\rm D}=\frac{N_{\rm{H}}}{E_{\rm{B-V}}}$
and $R_{\rm{V}}=\frac{A_{\rm{V}}}{E_{\rm{B-V}}}$,
where $N_{\rm{H}}$ is the total hydrogen column density in
${\rm \rm{cm}^{-2}}$ ($N_{\rm{H}}=N(\rm{H})+2\: N(\rm{H}_{2})$).
The standard galactic values are $C_{\rm D}=5.8\,10^{21}\,{\rm \rm{cm}^{-2}}\,{\rm mag}^{-1}$
\citep{Bohlin78,Rachford02} and $R_{\rm{V}}=3.1$. The relation
between the optical depth and the path length is:\[
l=\left(2.5\,\log_{10}{\rm e}\right)\,\frac{C_{\rm D}}{R_{\rm V}}\,\int_{0}^{\tau_{\rm max}}\frac{d\tau_{\rm V}}{n_{\rm H}(\tau_{\rm V})}\]
where $n_{\rm{H}}(\tau_{\rm{V}})$, in ${\rm cm^{-3}}$,
is the total hydrogen density at a visible optical depth $\tau_{\rm V}$.

\section{Numerical convergence\label{App_convergence}}

The steady state solution is reached only after a number of iterations
over the whole structure of the cloud. This comes from the necessity
to know in advance the optical depth in lines towards both sides of
the cloud in order to compute the thermal balance. This in turn requires
knowledge of column densities computed from abundances. Hence some
kind of initialisation is required, followed by iterations. Initialisation
is rather arbitrary.

Convergence of such a process is not guaranteed. However, one
finds that most physical quantities do converge in very few iterations
towards a value close to the final one so that most numerical problems
arise during the first or second iteration and are often cured by
skillful tuning of the initial guess.

\begin{figure}
\includegraphics[angle=270,scale=0.65]{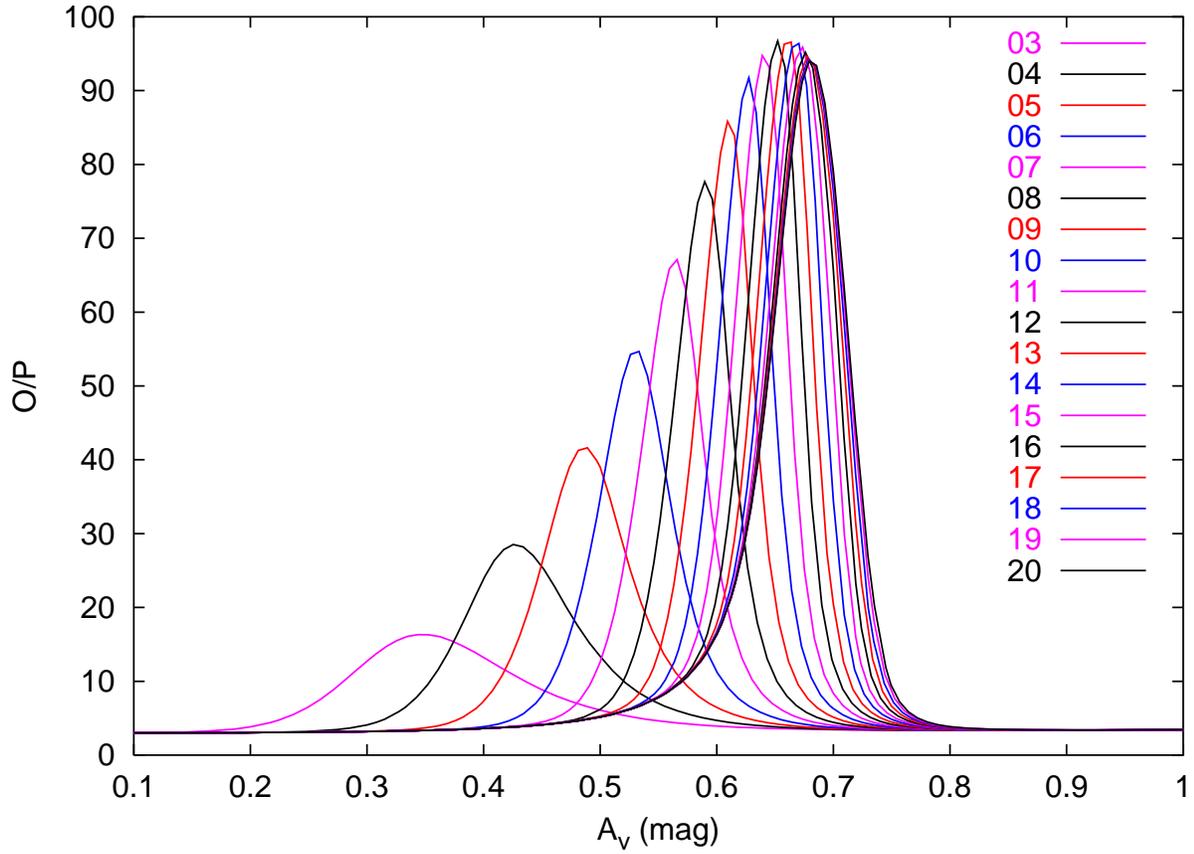}
\caption{Ortho to para ratio of ${\rm {H}_{2}}$ for a cloud model with $n_{\rm{H}}=10^{4}\,{\rm {cm}^{-3}}$
and $\chi=10^{4}$ as a function of the number of global model iterations.\label{Fig_convergence}}
\end{figure}

One quantity however converges significantly slower than any other
and thus controls the number of iterations: it is the position of
the ${\rm {H}/{\rm {H}_{2}}}$ transition which is sensitive to state
dependent photo-dissociation of ${\rm {H}_{2}}$. Self
shielding in lines depends on $J$ with a protection significantly
more efficient for ortho-${\rm {H}_{2}}$ compared to para-${\rm {H}_{2}}$.
The consequence is a peak in the ortho to para ratio occurring just
after the transition from atomic to molecular hydrogen, as already
pointed out by \cite{Abgrall92}. Iteration after iteration, that
transition is pushed further into the cloud as level column densities
are computed more accurately. The situation is illustrated on Fig.~\ref{Fig_convergence}
for a cloud model with $n_{\rm{H}}=10^{4}\,{\rm {cm}^{-3}}$
and $\chi=10^{4}$. 20 iterations were required to reach a satisfactory
stability. Less stringent physical conditions (i.e. usually a lower
radiation field) lead to convergence in less than 10 iterations.

\placefigure{Fig_convergence}

\section{FUV radiation field\label{App_UVfield}}

The basic physical quantity for radiation is the specific intensity
$I(\lambda)$, measured in ${\rm {erg}\,{\rm cm^{-2}\,{\rm s^{-1}\,{\rm sr^{-1}\,\mathring{A}^{-1}}}}}$.
Throughout most of the code, we use wavelengths expressed in ${\rm {\mathring{A}}}$
as the independent variable. The most useful derived quantities are
the mean intensity $J(\lambda)$ (in ${\rm {erg}\,{\rm cm^{-2}\,{\rm s^{-1}\,\mathring{A}^{-1}}}}$)
defined by\[
J(\lambda)=\frac{1}{4\pi}\,\int I(\lambda)\, d\Omega\]
and the energy density $u(\lambda)$ (in ${\rm {erg}\,{\rm cm^{-3}\,\mathring{A}^{-1}}}$)
defined by\[
u(\lambda)=\frac{1}{c}\,\int I(\lambda)\, d\Omega=\frac{4\pi}{c}\, J(\lambda)\]

If (and only if) the radiation field is isotropic, $I=J$. $I(\lambda)$
is the quantity to use to solve the radiative transfer equation. However,
most physical quantities depending on the radiation field can be computed
from $u(\lambda)$.

Our standard impinging radiation field is the value given by \cite{Draine78}.
Various incompatible expressions are found for that field, as discussed
in \cite{Kopp96}. We have chosen to use the formula of \cite{SD95}
which reads\[
I(\lambda)=\frac{1}{4\pi}\,\left(\frac{6.300\,10^{7}}{\lambda^{4}}-\frac{1.0237\,10^{11}}{\lambda^{5}}+\frac{4.0812\,10^{13}}{\lambda^{6}}\right)\]
where $\lambda$ is in ${\rm {\mathring{A}}}$, and $I(\lambda)$
in ${\rm {erg}\,{\rm cm^{-2}\,{\rm s^{-1}\,{\rm sr^{-1}\,\mathring{A}^{-1}}}}}$.
This expression is used from the Lyman cut-off up to the limit given
by \cite{Draine78} at $2000\,{\rm {\mathring{A}}}$. Longward, we
use\[
I(\lambda)=1.38243\,10^{-5}\,\lambda^{-0.3}\]

This field may be scaled by a factor $\chi$, and an additional radiation
from a nearby star taken as a diluted Black Body can also be introduced.
The star is characterised by its effective temperature $T_{\rm eff}$,
its radius $R_{*}$ and its distance to the cloud surface $d_{*}$.

It is common to measure the strength of the radiation field in reference
to \cite{Habing68}. We take the Habing standard value as $5.6\,10^{-14}\,{\rm {erg}\,{\rm cm^{-3}}}$
between $912\,{\rm {\mathring{A}}}$ and $2400\,{\rm {\mathring{A}}}$.
Note that this cut-off corresponds to $5.166\,{\rm {eV}}$, whereas
it is often given as $6\,{\rm {eV}}$. Although small, this difference
is one amongst the many inconsistencies between various codes. Here,
we \emph{define} the radiation scaling factor at some point by\[
G=\frac{1}{5.6\,10^{-14}}\,\int_{912}^{2400}u(\lambda)\, d\lambda\]
The {}``usual'' $G_{0}$ parameter is taken as the value of $G$
without cloud (in free space). Since $u(\lambda)$ comes from an integration
over $4\pi\,{\rm {sr}}$, this value differs from $G$ at the cloud
surface where half of the radiation is partly screened by the cloud
itself. For a semi-infinite cloud and an isotropic impinging radiation
field, taking into account back scattering of the radiation by dust,
$G_{\rm surface}$ is usually close to $0.54\, G_{0}$ (depending
on the dust characteristics). This may lead to (close to) a factor
of 2 of difference in the energy input of otherwise seemingly similar
models, and is a major source of discrepancies.

\section{Spherical Harmonics solution of the transfer equation\label{App_sph_harm}}

We follow \cite{Roberge83} for most of the development, but take
into account the fact that grain and/or gas extinction and scattering
properties may vary with position into the cloud. Here, we do not
take into account the possibility of embedded sources. The following
development applies to a plane parallel cloud with coherent scattering
so that the dependence on wavelength is omitted. In the radiative
transfer equations, Eq~(\ref{mathed:Eq_transfer}) and Eq~(\ref{mathed:Eq_redistrib}),
$\omega(\tau)=\frac{\sigma^{D}}{\kappa^{G}+\kappa^{D}+\sigma^{D}}$
is the troublesome contribution. This effective albedo includes absorption
by the gas in lines, and is thus smaller than the usual dust albedo
$\omega^{D}(\tau)=\frac{\sigma^{D}}{\kappa^{D}+\sigma^{D}}$. Boundary
conditions are (with $\tau_{max}$ the total slab optical depth at
$\lambda$):\[
\left\{ \begin{array}{cc}
I(\tau=0,\mu)=I^{-}(\mu) & \mu<0\\
I(\tau=\tau_{\rm max},\mu)=I^{+}(\mu) & \mu>0\end{array}\right.\]

If $\tau_{max}=\infty$, $I^{+}=0$. We expand $I$ and $p(\mu,\mu')$ in Legendre
polynomials $P_{l}$ by:\[
I(\tau,\mu)=\sum_{l=0}^{\infty}(2l+1)\, f_{l}(\tau)\, P_{l}(\mu)\]
\[
p(\mu,\mu')=\sum_{l=0}^{\infty}(2l+1)\,\sigma_{l}\, P_{l}(\mu)P_{l}(\mu')\]
\cite{FRR80,Roberge83} show that this leads to a infinite set of
equations whose general term is:\[
l\, f'_{l-1}+(l+1)f'_{l+1}=(2l+1)(1-\omega(\tau)\sigma_{l})\, f_{l}\]
Our expression differs from \cite{Roberge83} (Eq~9) by the explicit
$\tau$ dependence of $\omega$, so that this is \textbf{not} a constant
coefficient equation. We now stop the expansion at an odd value $L$,
and write it as a linear system:\[
A(\tau)\,{\bf {f'}={\bf {f}}}\]
From there, \cite{Roberge83} (Eq~11, 13, 14, 15, 16) hold, with
all coefficients $h_{l}$, $k_{m}$ and $R_{lm}$ now depending explicitly
on $\tau$. We still have $k_{-m}=-k_{m}$ and $R_{l,-m}=(-1)^{l}R_{l,m}$.
Letting ${\rm f}=R{\rm y}$, we have ${\rm f}'={\rm R}\,{\rm y}'+{\rm R}'\,{\rm y}$.
The derivative of ${\rm R}$ introduces a new term in the equation,
so that \cite{Roberge83} (Eq~16) is now:\[
{\bf {y'}={\bf {k}(\tau)\,{\bf {y}}}}+{\rm Q}\]
where ${\rm Q={\rm R}^{-1}\,{\rm R}'\,{\rm y}}$ does not decouple.
Each component $q_{m}(\tau)$ is a known linear combination of $y_{m}(\tau)$.
This is a linear equation, whose homogeneous part has a solution:\[
y_{m}(\tau)=C_{m}\,\exp\left(\int_{\tau_{m}}^{\tau}k_{m}(t)\, dt\right)\]
where the $2M$ constants $\tau_{m}$ may still be chosen arbitrarily
and $C_{m}=y_{m}(\tau_{m})$.

Adding the particular solution, one has:

\[
\begin{array}{l}
y_{m}(\tau)=\exp\left(\int_{\tau_{m}}^{\tau}k_{m}(t)\, dt\right)\,\times\\
\left\{ C_{m}+\int_{\tau_{m}}^{\tau}\,\exp\left(-\int_{\tau_{m}}^{s}k_{m}(t)\, dt\right)\, q_{m}(s)\, ds\right\} \end{array}\]

The integral involving $q_{m}$ is large only if both ${\rm R}'$
and $q_{m}$ are large, which occurs only where lines dominates the
absorption \textbf{and} are not yet saturated. A first order approximation
is obtained by considering this integral as a constant. Iterative refinements
are possible afterwards, but are not taken into account here. A more
accurate treatment will be the subject of another paper.

One sees that, provided the albedo $\omega(\tau)$ is known, a complete
solution is still computable, but requires numerical integrations.
Those expressions may thus be used in an iterative scheme. Back to
the original variables, we have now:\[
f_{l}(\tau)=\sum_{m=-M}^{+M}\, R_{lm}(\tau)\, C_{m}\,\exp\left(\int_{\tau_{m}}^{\tau}k_{m}(t)\, dt\right)\]
$m=0$ being omitted from the sum.

To avoid numerical instabilities, we also write:\[
f_{l}(\tau)=f_{l}^{+}(\tau)+\left(-1\right)^{l}\, f_{l}^{-}(\tau)\]
with\[
\left\{ \begin{array}{c}
f_{l}^{+}(\tau)=\sum_{m=1}^{M}R_{lm}(\tau)\, C_{m}\,\exp\left(-\int_{\tau}^{\tau_{\rm max}}k_{m}(t)\, dt\right)\\
f_{l}^{-}(\tau)=\sum_{m=1}^{M}R_{lm}(\tau)\, C_{-m}\,\exp\left(-\int_{0}^{\tau}k_{m}(t)\, dt\right)\end{array}\right.\]
where we have chosen $\tau_{m}=0$ if $\mu_{m}<0$, and $\tau_{m}=\tau_{\rm max}$
if $\mu_{m}>0$.

Constants $C_{m}$ are determined from boundary conditions by:\begin{equation}
\sum_{m=-M}^{M}C_{m}\, B_{im}=Q_{i}\label{mathed:boundary_cond}\end{equation}
with $B_{im}$ given by:

\begin{itemize}
\item $\mu_{i}<0\,;m<0$:\[
\sum_{l=0}^{L}(2l+1)\, P_{l}(\mu_{i})\, R_{lm}(0)\]

\item $\mu_{i}<0\,;m>0$:\[
\sum_{l=0}^{L}(2l+1)\, P_{l}(\mu_{i})\, R_{lm}(0)\,\exp\left(-\int_{0}^{\tau_{\rm max}}k_{m}(t)dt\right)\]

\item $\mu_{i}>0\,;m<0$:\[
\sum_{l=0}^{L}(2l+1)\, P_{l}(\mu_{i})\, R_{lm}(\tau_{\rm max})\,\exp\left(-\int_{0}^{\tau_{\rm max}}k_{m}(t)dt\right)\]

\item $\mu_{i}>0\,;m>0$:\[
\sum_{l=0}^{L}(2l+1)\, P_{l}(\mu_{i})\, R_{lm}(\tau_{\rm max})\]

\end{itemize}
and\[
Q_{i}=\left\{ \begin{array}{cc}
I^{-}(0,\mu_{i}) & \mu_{i}<0\\
I^{+}(\tau_{\rm max},\mu_{i}) & \mu_{i}>0\end{array}\right.\]

In the following, only the mean intensity $J_{\lambda}$ is needed.
This is given by $f_{0}$ which depends on $R_{0,m}=1$. So all coefficients
$R_{lm}(\tau)$ need not be kept during computation. One only needs
the coefficients $C_{m}$ and the integrated eigenvalues $\int_{0}^{\tau}k_{m}(t)\, dt$.

\section{Chemical processes\label{App_chemistry}}

Species (atoms and molecules), initial abundances and chemical reactions
to use in a run are given in a single file. The first part gives the
list of the species with their name, atomic composition, initial abundance
and enthalpy of formation in ${\rm {kcal}\,{\rm mol^{-1}}}$. These
enthalpies are used in the thermal balance to compute the variations
of enthalpy of the chemical reactions. 

The second part of the chemistry file is a list of the chemical reactions.
Chemical rates are computed from the three parameters $\gamma$, $\alpha$,
$\beta$ given for each reaction. Photo processes computed by direct
integration over the radiation field (i.e. photo-dissociation of ${\rm H_{2}}$,
${\rm {HD}}$, ${\rm {CO}}$ and its isotopes and photo-ionisation
of ${\rm {C}}$ and ${\rm {S}}$) must not be mentioned in this file.
Reactions on grains may be included, following \cite{LPRF95}. Most
of the listed reactions follow an Arrhenius law : 

\[
k=\gamma\,\left(\frac{T_{\rm K}}{300}\right)^{\alpha}\,\exp(-\beta/T_{\rm K})\;{\rm cm^{3}\,{\rm s^{-1}}}\]

with $k$ the chemical rate at a point of the cloud at the temperature
$T_{\rm K}$.

Here, we would just like to mention how we deal with some specific
reactions.

\subsection{Secondary photon processes}

\[
{\rm {X}+h\nu\,\rightarrow\, products}\]

\[
k=\gamma\,\zeta\,\left(\frac{T_{\rm K}}{300}\right)^{\alpha}\,\frac{n({\rm H_{2})}}{n({\rm {H})+n({\rm H_{2})}}}\,{\rm s^{-1}}\]

These photons are created deep into the cloud by electronic cascades
of ${\rm H_{2}}$ following excitation by electrons produced by cosmic
rays as described first by \citet{PT83}. Temperature dependence is
needed for ${\rm {CO}}$ only. Rates are taken from \citet{GLDH89} where an albedo of 0.5 is assumed.

\subsection{Radiative association}

\[
{\rm {X}+{\rm {Y}\,\rightarrow\,{\rm {XY}+h\nu}}}\]

\[
k=\gamma\,\left(\frac{T_{\rm K}}{300}\right)^{\alpha}\,\exp(-\beta/T_{\rm K})\;{\rm cm^{3}\,{\rm s^{-1}}}\]

These reactions are singled out from ordinary gas phase because it
is not possible (due to the escaping photons) to compute their contribution
to thermal balance from simple thermodynamic considerations.

\subsection{Endothermal reactions with $\rm{H}_{2}$}

\[
{\rm {X}+{\rm H_{2}\,\rightarrow\, products}}\]

\[
k=\gamma\,\left(\frac{T_{\rm K}}{300}\right)^{\alpha}\,\sum_{l=1}^{l_{\rm max}}n_{l}\,\exp(-(\beta-E_{l})/T_{\rm K})\;{\rm cm^{3}\,{\rm s^{-1}}}\]
where $n_{l}$ is the relative population of level $l$ of ${\rm H_{2}}$,
$E_{l}$ its energy, and $l_{\rm max}$ is the highest level of
${\rm H_{2}}$ such that $\beta-E_{l}>0$. Those reactions make the
implicit hypotheses that all internal energy of ${\rm H_{2}}$ may
be used to overcome an activation barrier or an endothermicity. 

\subsection{Photoreactions}

\[
{\rm {X}+h\nu\,\rightarrow\, products}\]

\[
k=\gamma\,\left(\chi^{-}\, e^{-\beta A_{\rm V}}+\chi^{+}\, e^{-\beta(A_{\rm V}^{\rm max}-A_{\rm V})}\right)\]

These reactions are used only for species whose destruction rate is
not computed directly by integration over the local radiation field.
$\chi^{\pm}$ are the scaling factors of the radiation field with respect
to that of Draine on the left and right side of the cloud respectively
(see Appendix~\ref{App_UVfield} and~\ref{App_cloud_struct} for
notations), and $A_{\rm{V}}^{\rm max}=2.5\,\log_{10}{\rm e}\,\tau_{\rm max}$
is the total cloud extinction. For a semi-infinite cloud, $\chi^{+}=0$.
Note that no factor of $\frac{1}{2}$ is needed to take into account
the fact that photons come from only a half space at each edge, since
this is already taken into account in the computation of $\chi^{\pm}$.

\subsection{Adsorption on grains\label{sub_chim_ads}}

\[
{\rm {X}+dust\,\rightarrow\,{\rm {X:}}}\]

\[
k=\gamma\,\left\langle \sigma n_{g}\right\rangle \,\overline{v}\]
where X: is a species adsorbed on grains, $<\sigma n_{g}>$ is the mean grain cross section per unit volume,
and $\overline{v}$ the mean particle velocity. Microphysics (including
the sticking coefficient) is {}``hidden'' in the term $\gamma$,
which is read from the chemical input file for most species. We use
specific expressions for ${\rm {H}}$ and ${\rm H_{2}}$:

\begin{itemize}
\item For ${\rm {H}}$, the default sticking coefficient is $\gamma=\sqrt{\frac{10}{\sup(10,T_{\rm K})}}$.
As $\overline{v}\propto\sqrt{T_{\rm K}}$, this leads to a constant
rate at high temperature. The consequences are discussed in Sect.~\ref{Sec_H2_form}.
\item For ${\rm H_{2}}$, the rate is tuned to give at most a single mono-layer
on the grain, as described in \cite{LeBourlot00}.
\end{itemize}

\subsection{Grain surface reactions}

\[
{\rm {X:}+{\rm {Y:}\,\rightarrow\, products}}\]

\[
k=\left(\frac{1}{t_{\rm hop}({\rm {X:})}}+\frac{1}{t_{\rm hop}({\rm {Y:})}}\right)F_{r}\]
where $F_{r}$ is the fraction of sites on the outermost layer occupied
by ${\rm {Y}}$, ${\rm {X}}$ is either ${\rm {H}}$ or ${\rm {D}}$,
and $t_{\rm hop}$ is the {}``hopping'' time of the particle from
one site to the next. See \citet[Appendix C]{LPRF95} for details
and the expression of $F_r$ . We use $t_{\rm hop}({\rm {H})=2\,10^{-11}\,{\rm {s}}}$.

Note that this expression supposes that the time to reach a specific
position varies as the distance to it and not as the square, as would
occur for a random walk on an infinite surface. It is also not valid
for reactions other than hydrogenation.

\subsection{Thermal evaporation from grains}

\[
{\rm {X:}\,\rightarrow\,{\rm {X}}}\]

\[
k=\gamma\,\sqrt{\frac{2k\beta}{m_{X}}}\,\exp\left(-\frac{\beta}{T_{\rm d}}\right)\]
where $\beta$ is the adsorption binding energy of ${\rm X}$ on the
surface, $\sqrt{\frac{2k\beta}{m_{\rm X}}}$ a vibration frequency
and $T_{\rm d}$ the dust temperature. see \cite{CT03} for details.

\acknowledgements{We thank the referee for his careful reading which improved the english considerably. The authors thank Pierre Hily-Blant for numerous discussions and
heavy testing of the code. Other contributions over the years are
too numerous to list in full, but are not forgotten. Support from
french CNRS's {}``PCMI'' program is gratefully acknowledged.}

\clearpage

\begin{deluxetable}{cccc}

\tablecaption{Model variables defined or calculated at each point in the cloud.\label{Tab_var}}
\tablewidth{0pt}
\tablehead{
\colhead{variable} & \colhead{Unit} & \colhead{Name} & \colhead{Comment}
}
\startdata
$T_{\rm K}$& ${\rm {K}}$& Kinetic temperature& Variable or parameter\\
$n_{\rm{H}}$& ${\rm cm^{-3}}$& Density (Hydrogen nuclei)& Variable or parameter\\
$n({\rm {X})}$& ${\rm cm^{-3}}$& Abundance of ${\rm {X}}$& \\
$n_{i}({\rm {X})}$& none& Population of level $i$ of ${\rm {X}}$& $\sum_{i}n_{i}({\rm {X})=1}$\\
$u(\lambda)$& ${\rm {erg}\,{\rm cm^{-3}\,{\rm \mathring{A}^{-1}}}}$& Radiative energy density& \\
${\bf \Lambda}$ \& ${\bf \Gamma}$& ${\rm {erg}\,{\rm cm^{-3}\,{\rm s^{-1}}}}$& Heating and cooling rates& See Sect.~\ref{Sec_Ther_Bal}\\
\enddata
\end{deluxetable}

\clearpage

\begin{deluxetable}{cccc}

\tablecaption{Astrophysical parameters : model definition.\label{Tab_model}}
\tablewidth{0pt}

\tablehead{
\colhead{Parameter} & \colhead{Units} & \colhead{Name or definition} & \colhead{Comment}
}
\startdata
$\chi$& 'Draine'& FUV radiation strength& see Appendix~\ref{App_UVfield}\\
$\zeta$& $10^{-17}\,{\rm s^{-1}}$& Cosmic ray ionisation rate& \\
$A_{\rm{V}}$& ${\rm {mag}}$& Extinction& Total cloud depth\\
$v_{\rm turb}$& ${\rm {cm}\,{\rm s^{-1}}}$& Turbulent Velocity& Should be {}``local{}``\\
$\delta_{\rm X}$& none& Depletion of atom ${\rm {X}}$& Relative to ${\rm {H}}$\\
$P$& ${\rm K\, cm^{-3}}$& Thermal pressure& Defined by $P=n\, T_{\rm K}$\\
       &                                   &                                 & with $n = n({\rm H}) + n({\rm H}_2) + n({\rm He})$ \\
\enddata
\end{deluxetable}

\clearpage

\begin{deluxetable}{cccc}

\tablecaption{Atomic and molecular parameters.\label{Tab_cte}}
\tablewidth{0pt}
\rotate

\tablehead{
\colhead{Constant} & \colhead{Units} & \colhead{Name} & \colhead{Comment}
}
\startdata
$k_{\rm XY}(T_{\rm K})$ & ${\rm cm^{3}\,{\rm s^{-1}}}$ & Chemical reaction rate coefficient & Elementary two body process\\
$\sigma_{\rm X}(E)$& ${\rm cm^{2}}$& Cross section& Photoionisation, dissociation, etc...\\
$A_{ij}$& ${\rm s^{-1}}$& spontaneous emission transition probability& Derived: $B_{ij}$, $B_{ji}$, $f_{ij}$\\
$q_{ij}(T_{\rm K})$& ${\rm cm^{3}\,{\rm s^{-1}}}$& Collisional excitation rate coefficients & Derived: $q_{ji}$\\
\enddata
\end{deluxetable}

\clearpage

\begin{deluxetable}{ccccc}

\tablecaption{Grain parameters.\label{Tab_grains}}
\tablewidth{0pt}

\tablehead{
\colhead{Parameter} & \colhead{Units} & \colhead{Name or definition} & \colhead{Comment} & \colhead{Typical value}
}
\startdata
$a_{\rm min}$& ${\rm {cm}}$& Lower size cut-off& MRN& $3\,10^{-7}$\\
$a_{\rm max}$& ${\rm {cm}}$& Upper size cut-off& MRN& $3\,10^{-5}$\\
$d$                    & $\rm {cm} $ & Mean distance between &  Le Bourlot  & $2.6\, 10^{-8}$\\ 
                           &                       & adsorption sites &                           &                           \\
$\alpha$& none& index& MRN& $-3.5$\\
$\omega$& none& dust albedo& fixed\tablenotemark{a}, M96& $0.42$\\
$g$& none& anisotropy factor $<\cos\theta>$& fixed\tablenotemark{a}, WD& $0.6$\\
$G_{r}$& none& $M_{\rm grain}/M_{\rm gas}$& fixed\tablenotemark{a}& 0.01\\
$\rho_{gr}$& ${\rm {g}\,{\rm cm^{-3}}}$& grain volumic mass& fixed\tablenotemark{a}& $3$\\
$R_{\rm{V}}$& & $A_{\rm{V}}/\rm{E}_{\rm{B-V}}$& See Appendix~\ref{App_cloud_struct}& 3.1\\
$C_{\rm D}$& ${\rm cm}^{2}\,{\rm mag^{-1}}$& $N_{\rm{H}}/\rm{E}_{\rm{B-V}}$& See Appendix~\ref{App_cloud_struct}& $5.8\,10^{21}$\\
$c_{3}$, $\gamma$, $y_{0}$& depend& Extinction curve, $2200\,\rm{Å}$ bump& FM& Galactic\\
$Q_{\rm abs}(a,\lambda)$& none& absorption coefficient& DL& \\
\enddata
\tablerefs{MRN: \cite{MRN77}, FM: \cite{FM86,FM88,FM90},
Le Bourlot: \cite{LPRF95}, 
DL: \cite{DL84} upgraded by \cite{LD93},
see \url{http://www.astro.princeton.edu/\textasciitilde{}draine/dust/dust/diel.html},
M96: \citet{Mathis96}, WD: \citet{WD01}.}
\tablenotetext{a}{Parameters quoted ``fixed'' are in fact dependent on grain composition. Modifications to the program for a more physical modelization are in progress. }
\end{deluxetable}

\clearpage

\begin{deluxetable}{ccccccc}

\tablecaption{Collisional processes included.\label{Tab_Coll}}
\tablewidth{0pt}

\tablehead{
\colhead{$\;$} & \colhead{Levels} & \colhead{${\rm {H}}$} & \colhead{${\rm {He}}$} & \colhead{${\rm H^{+}}$} & \colhead{${\rm H_{2}}$} & \colhead{${\rm e^{-}}$}
}
\startdata
${\rm {C}}$ & $^{3}P_{0}$, $^{3}P_{1}$, $^{3}P_{2}$, $^{1}D_{2}$, $^{1}S_{0}$ & (1) & (2) & (3) & (4) & (5)(6)\\
${\rm {O}}$ & $^{3}P_{2}$, $^{3}P_{1}$, $^{3}P_{0}$, $^{1}D_{2}$, $^{1}S_{0}$ & (1)(7) & (8) & (9) & (10) & (9)(11)\\
${\rm {S}}$ & $^{3}P_{2}$, $^{3}P_{1}$, $^{3}P_{0}$, $^{1}D_{2}$, $^{1}S_{0}$ & (1)(+) & (8)(+) & (9)(+) & (10)(+) & (9)(10)(+)\\
${\rm {Si}}$ & $^{3}P_{0}$, $^{3}P_{1}$, $^{3}P_{2}$, $^{1}D_{2}$, $^{1}S_{0}$ & (1)(\dag) & (2)(\dag) & (3)(\dag) & (4)(\dag) & (5)(6)(\dag)\\
${\rm C^{+}}$ & $^{2}P_{1/2}$, $^{2}P_{3/2}$, $^{4}P_{1/2}$, $^{4}P_{3/2}$, $^{4}P_{5/2}$ & (12) & - & - & (13) & (14)(15)\\
${\rm N^{+}}$ & $^{3}P_{0}$, $^{3}P_{1}$, $^{3}P_{2}$, $^{1}D_{2}$, $^{1}S_{0}$ & - & - & - & - & (6)\\
${\rm Si^{+}}$ & $^{2}P_{1/2}$, $^{2}P_{3/2}$, $^{4}P_{1/2}$, $^{4}P_{3/2}$, $^{4}P_{5/2}$ & (16) & - & - & (13)(\dag) & (17)\\
${\rm HCO^{+}}$ & $J=0\,\rightarrow\,20$ & - & - & - & (18) & (19)\\
${\rm {CO}}$ & $J=0\,\rightarrow\,30$ & (20) & (21) & - & (22) & -\\
${\rm {CS}}$ & $J=0\,\rightarrow\,20$ & - & - & - & (23) & (24)\\
${\rm H_{2}}$ & $J=0\,\rightarrow\,29$, $v=0\,\rightarrow\,14$ \tablenotemark{a} & (25) & (25) & - & (25) & -\\
${\rm {HD}}$ & $J=0\,\rightarrow\,9$ & (26) & (26) & - & (26) & -\\
\enddata
\tablecomments{
A ``-'' indicates that no collision is considered.
(+): Rates for ${\rm {S}}$ are taken from ${\rm {O}}$.
(\dag): Rates for ${\rm {Si}}$ are taken from ${\rm {C}}$, and some
rates for ${\rm Si}^{+}$ are taken from ${\rm C}^{+}$.
}
\tablenotetext{a}{
All ro-vibrational levels of ${\rm H_{2}}$ ground electronic state may
be included. Usually they are included up to a highest level ($l_{0}$)
chosen on physical ground.
}
\tablerefs{
(1)~\cite{LR77a}, (2)~\cite{LRR91,SF91,LHBCL02}, (3)~\cite{RL90},
(4)~\cite{SSSFJ91}, (5)~\cite{PA76}, (6)~\cite{Mendoza83}, (7)~\cite{FS83},
(8)~\cite{MF88}, (9)~\cite{CLMTMR80,Pequignot90}, (10)~\cite{JSSF92},
(11)~\cite{BBT98}, (12)~\cite{LR77b}, (13)~\cite{FL77}, (14)~\cite{LDHK85},
(15)~\cite{WB02}, (16)~\cite{Roueff90}, (17)~\cite{DK91}, (18)~\cite{Flower99},
(19)~\cite{FT01,ND89}, (20)~\cite{BYD02}, (21)~\cite{CBBD02},
(22)~\cite{Flower01}, (23)~\cite{TCGL92}, (24)~\cite{DPGPR77},
(25)~\cite{LPF99} and ref. therein, (26)~\cite{FLPR00} and ref. therein
}
\end{deluxetable}

\clearpage

\begin{deluxetable}{cc}
\tablewidth{0pt}

\tablecaption{Adopted gas phased abundances relative to $n_{\rm{H}}$ and adopted
parameters in the grid of models (see also Table~\ref{Tab_grains}).\label{Tab_paramdiff}}
\tablewidth{0pt}

\tablehead{
\colhead{Parameter} & \colhead{Value}
}
\startdata
${\rm He}$ & 0.1\\
${\rm C}$$^{(a)}$ & 1.3 (-4)\\
${\rm N}$$^{(b)}$ & 7.5 (-5)\\
${\rm O}$$^{(c)}$ & 3.2 (-4)\\
${\rm S}$$^{(a)}$ & 1.9 (-5)\\
${\rm Fe}$$^{(a)}$ & 1.5 (-8)\\
\tableline
$\zeta$($s^{-1}$)&  5.0 (-17)\\
$b\,({\rm km\, s}^{-1})$&  $2.0$\\
\enddata
\tablecomments{
Ref: (a) \citet{Savage96}, (b) \citet{MCS97}, (c) \citet{MJC98}\\
Figures in parentheses are powers of ten.
}
\end{deluxetable}

\clearpage

\begin{deluxetable}{ccccc}

\tablecaption{Edge destruction probabilities in ${\rm s}^{-1}$.\label{Tab_proba_edge}}
\tablewidth{0pt}

\tablehead{
\colhead{${\rm H}_{2}$} & \colhead{${\rm HD}$} & \colhead{${\rm CO}$} & \colhead{${\rm C}$} & \colhead{${\rm S}$}
}
\startdata
4.2 (-11)& 2.6 (-11)& 1.1 (-10)& 1.7 (-10)& 5.1 (-10)
\enddata
\tablecomments{
Model parameters:
$n_{\rm H}=100\,\rm{cm}^{-3}$, $\chi=1,$ $A_{\rm V}=1$.
Values in parentheses are powers of 10.}
\end{deluxetable}

\clearpage

\begin{deluxetable}{lcclcc}

\tablecaption{comparison of FGK approximation and exact radiative transfer.
\label{Tab_transf}}
\tablewidth{0pt}

\tablehead{
\colhead{$\;$} & \colhead{FGK} & \colhead{Exact} & \colhead{$\;$} & \colhead{FGK} & \colhead{Exact}
}
\startdata
${\rm H}$& 3.5(20)& 2.4(20)& $\rm{C}^{+}$& 2.4(17)& 2.4(17)\\
${\rm H}_{2}$& 7.6(20)& 8.1(20)& C& 1.0(15)& 1.6(15)\\
$T_{01}$& 65& 62& ${\rm CO}$& 4.6(13)& 9.3(13)\\
${\rm CH}$& 3.0(12)& 3.2(12)& ${\rm OH}$& 2.9(13)& 2.5(13)\\
\enddata
\tablecomments{
Column densities in $\rm{cm}^{-2}$ and $T_{01}$ in Kelvin
obtained with the FGK approximation and the exact radiative transfer
calculation up to $J=5$ (see Sect.~\ref{Sec_transfer}). For both
models, $n_{\rm{H}}~=~100\,\,\rm{cm}^{-3}$, $\chi~=~1$,
radiation field from both sides, $A_{\rm{V}}~=~1$ and the parameters
in Table~\ref{Tab_paramdiff} are used.}
\end{deluxetable}

\clearpage

\begin{deluxetable}{ccccccc}

\tablecaption{One and two sides results.\label{Tab_deuxf}}
\tablewidth{0pt}

\tablehead{
\colhead{$A_{\rm{V}}^{tot}$} &
\colhead{$\;$} & \colhead{0.2} & \colhead{0.5} & \colhead{1.0} & \colhead{5.0} & \colhead{7.0}
}
\startdata
${\rm H}$& 1& 1.7(20)& 2.6(20)& 3.2(20)& 4.8(20)& 5.5(20)\\
& 2& 1.9(20)& 2.9(20)& 3.5(20)& 5.0(20)& 5.7(20)\\
${\rm H}_{2}$& 1& 1.0(20)& 3.4(20)& 7.7(20)& 4.4(21)& 6.3(21)\\
& 2& 9.0(19)& 3.2(20)& 7.6(20)& 4.4(21)& 6.3(21)\\
$\rm{C}^{+}$& 1& 3.0(16)& 1.3(17)& 2.4(17)& 8.4(17)& 8.8(17)\\
& 2& 4.9(16)& 1.2(17)& 2.5(17)& 8.5(17)& 8.8(17)\\
${\rm C}$& 1& 9.2(13)& 3.7(14)& 1.5(15)& 3.7(17)& 8.0(17)\\
& 2& 4.8(13)& 2.1(14)& 1.0(15)& 3.7(17)& 8.0(17)\\
${\rm CO}$& 1& 8.8(12)& 3.0(13)& 8.6(13)& 2.1(16)& 5.4(16)\\
& 2& 2.7(12)& 1.3(13)& 4.6(13)& 1.9(16)& 5.2(16)\\
${\rm CH}$& 1& 4.4(11)& 1.8(12)& 4.8(12)& 6.4(13)& 1.2(14)\\
& 2& 1.5(11)& 8.3(11)& 3.0(12)& 5.6(13)& 9.9(13\\
\enddata
\tablecomments{
Comparison of column densities for 1 side and 2 sides models. Figures
in parentheses correspond to powers of 10. For all models, $n_{\rm{H}}=100\,\,\rm{cm}^{-3}$
and $\chi=1$. The second column gives the number of sides of the
model. For the one side model, column densities correspond to 2 times
the value at $A_{\rm V}/2$.}
\end{deluxetable}

\clearpage

\clearpage

\clearpage

\clearpage

\clearpage

\clearpage

\clearpage

\clearpage

\clearpage

\clearpage

\clearpage

\clearpage

\clearpage

\clearpage

\clearpage

\clearpage

\clearpage

\clearpage

\clearpage

\clearpage

\clearpage

\clearpage

\clearpage

\clearpage

\clearpage

\clearpage

\clearpage

\clearpage

\clearpage

\clearpage

\clearpage

\clearpage

\clearpage

\clearpage

\clearpage

\clearpage

\clearpage


\begin{thebibliography}{}
\bibitem[Abergel et al.(2003)]{Abergel03}Abergel, A., Teyssier, D., Bernard, J.P. et al. 2003, A\&A, 410, 577
\bibitem[Abgrall et al.(1992)]{Abgrall92}Abgrall, H., Le Bourlot, J., Pineau des For\^ets, G., Roueff, E., Flower,
D.R., \& Heck, L. 1992, A\&A, 253, 525
\bibitem[Abgrall et al.(2000)]{ARD00}Abgrall, H., Roueff, E., Drira, I. 2000, A\&AS, 141, 297
\bibitem[Andr\'e et al.(2004)]{Andre04}Andr\'e, M. K., Le Petit, F., Sonnentrucker, P., Ferlet, R., Roueff,
E., Civeit, T., D\'esert, J-M., Lacour, S., \& Vidal-Madjar, A. 2004,
A\&A, 422, 483 
\bibitem[Balakrishnan et al.(2002)]{BYD02}Balakrishnan, N., Yan, M., \& Dalgarno, A. 2002, ApJ, 568, 443
\bibitem[Bakes \& Tielens(1994)]{BT94}Bakes, E.L.O., \& Tielens, A.G.G.M. 1994, ApJ, 427, 822
\bibitem[Bates \& Spitzer(1951)]{Bates51}Bates, D. R. \& Spitzer, L.J. 1951, ApJ, 113, 441 
\bibitem[Bell et al.(1998)]{BBT98}Bell, K. L., Berrington, K. A., \& Thomas, M. R. J. 1998, MNRAS, 293,
L83
\bibitem[Black \& Dalgarno(1977)]{BD77}Black, J.H., \& Dalgarno, A. 1977, ApJS, 34, 405
\bibitem[Black(1987)]{Black87}Black, J.H., 1987, in ASSL Vol. 134, Interstellar processes, 731
\bibitem[Black \& van Dishoeck(1987)]{BvD87}Black, J. H., \& van Dishoeck, E. F. 1987, ApJ, 322, 412
\bibitem[Bluhm et al.(2003)]{Bluhm03}Bluhm, H., de Boer, K.S., Marggraf, O., Richter, P., \& Wakker, B.P.
2003, A\&A 398, 983
\bibitem[Bohlin et al.(1978)]{Bohlin78}Bohlin, R. C., Savage, B. D. \& Drake, J. F. 1978, ApJ, 224, 132
\bibitem[Burton et al.(1990)]{BHT90}Burton, M.G., Hollenbach, D.J., Tielens, A.G.G.M. 1990, ApJ, 365,
620
\bibitem[Burke \& Hollenbach(1983)]{BH83}Burke, J.R., Hollenbach, D.J. 1983, ApJ, 265, 223
\bibitem[Boiss\'e et al.(2005)]{Boisse05}Boiss\'e, P., Le Petit F., Rollinde, E., Roueff, E., Pineau des For\^ets
G., Andersson, B.G., Gry, C., \& Felenbok, P. 2005, A\&A 429, 509 
\bibitem[Cardelli(1994)]{Cardelli94}Cardelli, J.A. 1994, ASPC 58, 24
\bibitem[Casu(2003)]{Casu03}Casu S. 2003, PhD thesis
\bibitem[Cazaux \& Tielens(2003)]{CT03}Cazaux, S., \& Tielens, A.G.G.M. 2004, ApJ, 604, 222
\bibitem[Cecchi-Pestellini et al.(2002)]{CBBD02}Cecchi-Pestellini, C., Bodo, E., Balakrishnan, N., \& Dalgarno, A.
2002, ApJ, 571, 1015
\bibitem[Chambaud et al.(1980)]{CLMTMR80}Chambaud, G., Levy, B., Millie, P., Tran Minh, F., Launay, J.M., \&
Roueff, E. 1980, J. Ph. B, 13, 4205
\bibitem[Decamp \& Le Bourlot(2002)]{DLB02}Decamp, N. \& Le Bourlot, J. 2002, A\&A, 389, 1055
\bibitem[de Jong et al.(1980)]{dJBD80}de Jong, T., Boland, W., \& Dalgarno, A. 1980, A\&A, 91, 68
\bibitem[Dickinson et al.(1977)]{DPGPR77}Dickinson, A.S., Phillips, T.G., Goldsmith, P.F., Percival, I.C.,
\& Richards, D. 1977, A\&A, 54, 645
\bibitem[Draine(1978)]{Draine78}Draine, B. 1978, ApJS, 36, 595
\bibitem[Draine \& Lee(1984)]{DL84}Draine, B., \& Lee, H. 1984, ApJ, 285, 89
\bibitem[Dufton \& Kingston(1991)]{DK91}Dufton, P. L., \& Kingston, A. E. 1991, MNRAS, 248, 827
\bibitem[Federman et al.(1979)]{FGK79}Federman, S.R., Glassgold, A.E., \& Kwan, J. 1979, ApJ, 227, 466
\bibitem[Faure \& Tennyson(2001)]{FT01}Faure, A., \& Tennyson, J. 2001, MNRAS, 325, 443
\bibitem[Federman \& Shipsey(1983)]{FS83}Federman, S.R., \& Shipsey, E.J. 1983, ApJ, 269, 791
\bibitem[Federman et al.(1994)]{Federman94}Federman, S. R., Cardelli, J. A., Sheffer, Y., Lambert, D. L., \&
Morton, D. C. 1994, Ap J, 432, L139
\bibitem[Fitzpatrick \& Massa(1986)]{FM86}Fitzpatrick, E.L., \& Massa, D. 1986, ApJ, 307, 286
\bibitem[Fitzpatrick \& Massa(1988)]{FM88}Fitzpatrick, E.L., \& Massa, D. 1988, ApJ, 388, 734
\bibitem[Fitzpatrick \& Massa(1990)]{FM90}Fitzpatrick, E.L., \& Massa, D. 1990, ApJS, 72, 163
\bibitem[Flannery et al.(1980)]{FRR80}Flannery, B.P., Roberge, W.G., \& Rybicki, G.B. 1980, ApJ, 236, 598
\bibitem[Flower \& Launay(1977)]{FL77}Flower, D.R., \& Launay, J.M. 1977, J. Ph. B, 10, 3673
\bibitem[Flower(1999)]{Flower99}Flower, D.R. 1999, MNRAS, 305, 651
\bibitem[Flower(2001)]{Flower01}Flower, D.R. 2001, J. Ph. B, 34, 2731
\bibitem[Flower et al.(2000)]{FLPR00}Flower, D. R., Le Bourlot, J., Pineau des For\^ets, G., \& Roueff, E.
2000, MNRAS, 314, 753
\bibitem[Flower \& Pineau des For\^ets(1998)]{FPdF98}Flower, D. R., \& Pineau des For\^ets G. 1998, MNRAS, 297,1182
\bibitem[Gredel et al.(1989)]{GLDH89}Gredel, R., Lepp, S., Dalgarno, A., \& Herbst, E. 1989, ApJ 347, 289
\bibitem[Gry et al.(2002)]{GBNPHF02}Gry, C., Boulanger, F., Nehm\'e, C., Pineau des For\^ets, G., Habart,
E., \& Falgarone, E. 2002, A\&A, 391, 675
\bibitem[Habing(1968)]{Habing68}Habing, H. 1968, Bull Astr Inst Neth, 19, 421
\bibitem[Hollenbach \& Tielens(1999)]{HT99}Hollenbach, D. J., \& Tielens, A. G. G. M. 1999, Rev. Mod. Phys.,
71,173
\bibitem[Jaquet et al.(1992)]{JSSF92}Jaquet, R., Staemmler, V., Smith, M. D., \& Flower, D. R. 1992, J
Phys B, 25, 285
\bibitem[Jura(1975)]{Jura75}Jura, M. 1975, ApJ, 197, 575
\bibitem[Joulain et al.(1998)]{JFPdF98}Joulain, K., Falgarone, E., \& Pineau des For\^ets, G. 1998, A\&A, 340,
241 
\bibitem[Kaufman et al.(1999)]{KWHL99}Kaufman, M.J., Wolfire, M.G., Hollenbach, D.J., \& Luhman, M.L. 1999,
ApJ, 527, 795
\bibitem[Kaufman et al.(2006)]{Kau06} Kaufman , M.J., Wolfire, M.G., \& Hollenbach, D.J., submitted, ApJ
\bibitem[Kopp(1996)]{Kopp96}Kopp, M. 1996, Ph D Thesis
\bibitem[Kwan(1977)]{Kwan77}Kwan, J. 1977, ApJ, 216, 713
\bibitem[Lacour et al.(2005)]{Lacour05}Lacour, S., Andr\'e, M.K., Sonnentrucker, P., Le Petit, F., Welty, D.E.,
D\'esert, J.M., Ferlet, R., Roueff, E., \& York, D. 2005, A\&A, 430,
967
\bibitem[Laor \& Draine(1993)]{LD93}Laor, A., \& Draine, B. 1993, ApJ, 402, 441
\bibitem[Launay \& Roueff(1977a)]{LR77a}Launay, J.M., \& Roueff, E. 1977, A\&A, 56, 289
\bibitem[Launay \& Roueff(1977b)]{LR77b}Launay, J.M., \& Roueff, E. 1977, J. Ph. B, 10, 879
\bibitem[Lavendy et al.(1991)]{LRR91}Lavendy, H., Robbe, J.M., \& Roueff, E. 1991, A\&A, 241, 317
\bibitem[Le Bourlot et al.(1993a)]{LPRF93}Le Bourlot, J., Pineau des For\^ets, G., Roueff, E., \& Flower, D.R.
1993a, A\&A, 267, 23
\bibitem[Le Bourlot et al.(1993b)]{LPRS93}Le Bourlot, J., Pineau des For\^ets, G., Roueff, E., \& Schilke, P.
1993b, ApJ, 416, L87
\bibitem[Le Bourlot et al.(1995a)]{LPR95}Le Bourlot, J., Pineau des For\^ets, G., \& Roueff, E. 1995a, A\&A,
297, 251
\bibitem[Le Bourlot et al.(1995b)]{LPRF95}Le Bourlot, J., Pineau des For\^ets, G., Roueff, E., \& Flower, D.R.
1995b, A\&A, 302, 870
\bibitem[Le Bourlot et al.(1995c)]{LBPRDG95}Le Bourlot, J., Pineau des For\^ets, G., Roueff, E., Dalgarno, A., Gredel,
R. 1995c, ApJ, 449, 178
\bibitem[Le Bourlot et al.(1999)]{LPF99}Le Bourlot, J., Pineau des For\^ets, G., \& Flower, D. R. 1999, MNRAS,
305, 802
\bibitem[Le Bourlot(2000)]{LeBourlot00}Le Bourlot, J. 2000, A\&A, 360, 656
\bibitem[Le Petit et al.(2002)]{LRL02}Le Petit, F., Roueff, E., Le Bourlot, J. 2002, A\&A, 390, 369
\bibitem[Le Petit \& Roueff(2003)]{LPR01}Le Petit, F. \& Rouef, E. 2003, in {}``Dissociative Recombination
of Molecular Ions with Electrons'' 373, Ed. S. Gubermen, Kluwer Ac.
- Plenum Pub.
\bibitem[Le Picard et al.(2002)]{LHBCL02}Le Picard S.D., Honvault, P., Bussery-Honvault, B., Canosa, A., Laub\'e,
S., Launay, J.M., Rowe, B., Chastaing, D., \& Sims, I.R. 2002, J.
Chem. Phys., 117, 10109
\bibitem[Le Teuff et al.(2000)]{LTMM00}Le Teuff, Y.H., Millar, T.J., Markwick A.J., 2000, A\&AS, 146, 157
\bibitem[Lee et al.(1996)]{LHPRL96}Lee, H.-H., Herbst, E., Pineau des For\^ets, G., Roueff, E., Le Bourlot,
J. 1996, A\&A, 311, 690
\bibitem[Lennon et al.(1985)]{LDHK85}Lennon, D. J., Dufton, P. L., Hibbert, A., \& Kingston, A. E. 1985,
ApJ, 294, 200
\bibitem[Mathis et al.(1977)]{MRN77}Mathis, J., Rumpl, W., \& Nordsieck, K. 1977, ApJ, 217, 425
\bibitem[Mathis(1996)]{Mathis96}Mathis, J. 1996, ApJ, 472, 643
\bibitem[Mendoza(1983)]{Mendoza83}Mendoza, C. 1983, in {}``Planetary Nebulae'', IAU Symposium Num.
103, London, 143
\bibitem[Meyer et al.(1997)]{MCS97}Meyer, D. M., Cardelli, J. A., \& Sofia U. J. 1997, ApJ 490, L103
\bibitem[Meyer et al.(1998)]{MJC98}Meyer, D. M., Jura, M., \& Cardelli, J. A. 1998, ApJ, 493, 222
\bibitem[Monteiro \& Flower(1988)]{MF88}Monteiro, T.S., \& Flower, D.R. 1987, MNRAS, 228, 101
\bibitem[Morton(1975)]{Morton75}Morton, D.C. 1975, ApJ 197, 85
\bibitem[Nehm\'e et al.(2005)]{NLGBP05}Nehm\'e, C., Le Bourlot, J., Gry, C., Boulanger, F., \&
Pineau des For\^ets, G. 2005, A\&A, submited
\bibitem[Neufeld \& Dalgarno(1989)]{ND89}Neufeld, D.A., \& Dalgarno, A. 1989, Phys. Rev. A, 40, 633
\bibitem[Oka et al.(2003)]{Oka03}Oka, T., Thorburn, J. A., McCall, B. J. et al. 2003, ApJ, 582, 823
\bibitem[Patriarchi et al.(2001)]{PMPB01}Patriarchi, P., Morbidelli, L., Perinotto, M., Barbaro, G. 2001, A\&A,
372, 644
\bibitem[Patriarchi et al.(2003)]{PMPB03}Patriarchi, P., Morbidelli, L., Perinotto, M., \& Barbaro, G. 2003,
A\&A, 410, 905
\bibitem[Pequignot \& Aldrovandi(1976)]{PA76}P\'equignot, D., \& Aldrovani, S.M.V. 1976, A\&A, 50, 141
\bibitem[P\'equignot(1990)]{Pequignot90}P\'equignot, D. 1990, A\&A, 231, 499, erratum in P\'equignot, D. 1996,
A\&A, 313, 1026
\bibitem[Pety et al.(2005)]{PTFGRAHC05}Pety, J., Teyssier, D., Foss\'e, D., Gerin, M., Roueff, E., Abergel,
A., Habart, E., \& Cernicharo, J. 2005, A\&A, 435, 885
\bibitem[Prasad \& Tarafdar(1983)]{PT83}Prasad, S. S., Tarafdar, S. P. 1983, ApJ, 267, 603
\bibitem[Rachford et al.(2001)]{Rachford01}Rachford, B. L., Snow, T. P., Tumlinson, J. et al. 2001, ApJ, 555,
839
\bibitem[Rachford et al.(2002)]{Rachford02}Rachford, B. L., Snow, T. P., Tumlinson, J. et al. 2002, ApJ, 577,
221
\bibitem[Roberge(1983)]{Roberge83}Roberge, W.G. 1983, ApJ, 275, 292
\bibitem[Roellig et al.(2005)]{roellig05}Roellig M. et al. 2005, in preparation
\bibitem[Roueff(1990)]{Roueff90}Roueff, E. 1990, A\&A, 234, 567
\bibitem[Roueff et al.(1996)]{RLBP96}Roueff E., Le Bourlot J., Pineau des Forets G., 1996, in \char`\"{}Dissociative
Recombination : Theory, Experiment and Applications \char`\"{}, Zajfman
D., Mitchell J.B.A., Schwalm D., Rowe B.R. eds, World Scientific,
Singapore, 11
\bibitem[Roueff \& Le Bourlot(1990)]{RL90}Roueff, E., \& Le Bourlot, J. 1990, A\&A, 236, 515
\bibitem[Savage \& Sembach(1996)]{Savage96}Savage, B. D., \& Sembach, K. R. 1996, ARA\&A, 34, 279
\bibitem[Schroder et al.(1991)]{SSSFJ91}Schroder, K., Staemmler, V., Smith, M.D., Flower, D.R., \& Jaquet,
R. 1991, J. Ph. B, 24, 2487
\bibitem[Stephens \& Dalgarno(1973)]{SD73}Stephens, T. L.; Dalgarno, A. 1973, ApJ 186, 65
\bibitem[Sternberg \& Dalgarno(1995)]{SD95}Sternberg, A., \& Dalgarno, A. 1995, ApJS, 99, 565
\bibitem[Sternberg \& Dalgarno(1989)]{SD89}Sternberg, A., \& Dalgarno, A. 1989, ApJ, 338, 197
\bibitem[Staemmler \& Flower(1991)]{SF91}Staemmler V., \& Flower D.R. 1991, J. Ph. B, 24, 2343
\bibitem[Störzer \& Hollenbach(2000)]{SH00}Störzer, H., \& Hollenbach, D. 2000, ApJ, 539, 751
\bibitem[Teyssier et al.(2004)]{Teyssier04}Teyssier, D., Foss\'e, D., Gerin, M., Pety, J., Abergel, A., \& Roueff,
E. 2004, A\&A 417, 135
\bibitem[Tielens \& Hollenbach (1985a)]{Tielens85a} Tielens, A.G.G.M., \& Hollenbach, D. 1985a, ApJ 291, 722
\bibitem[Tielens \& Hollenbach (1985b)]{Tielens85b} Tielens, A.G.G.M., \& Hollenbach, D. 1985b, ApJ 291, 747
\bibitem[Turner et al.(1992)]{TCGL92}Turner, B.E., Chan, K.W., Green, S., \& Lubovich, D. 1992, ApJ, 399,
114
\bibitem[Tumlinson et al.(2002)]{Tumlinson02}Tumlinson, J., Shull, J.M., Rachford, B.L. et al. 2002, ApJ 566, 857
\bibitem[van Dishoeck(1988)]{evD88}van Dishoeck, E., in \char`\"{}Rate Coefficients for Astrochemistry,
1988, T.J. Millar and D.A. Williams (eds), Kluwer Academic Publishers,
49
\bibitem[Wells(1999)]{Wells99}Wells, R.J. 1999 JQSRT, 62, 29
\bibitem[Weingartner \& Draine(2001)]{WD01}Weingartner, J.C., \& Draine, B.T. 2001, ApJ, 548, 296
\bibitem[Wilson \& Bell(2002)]{WB02}Wilson, N. J., \& Bell, K. L. 2002, MNRAS, 337, 1027
\end{thebibliography}
\end{document}